\documentclass[10pt,journal,comsoc]{IEEEtran}
\ifCLASSOPTIONcompsoc
  \usepackage[nocompress]{cite}
\else
  \usepackage{cite}
\fi
\ifCLASSINFOpdf
\else
\fi
\usepackage{graphicx}
\usepackage{booktabs}
\usepackage{multirow}
\usepackage{float}
\usepackage{graphics}
\usepackage{forest}
\usepackage{bbding}
\usepackage{amsfonts}
\usepackage{array}
\usepackage{tabu}
\usepackage[utf8]{inputenc}
\usepackage{color, colortbl}
\definecolor{lighgray}{gray}{0.9}
\usepackage{pgfplots}
\pgfplotsset{compat=newest}
\begin{document}
%
\title{An Overview of Attacks and Defences on Intelligent Connected Vehicles}
%
%
%


%
\author{Mahdi~Dibaei,
        Xi~Zheng,~\IEEEmembership{Member,~IEEE,}
        Kun~Jiang,~\IEEEmembership{Member,~IEEE,}
        Sasa~Maric,
        Robert~Abbas,~\IEEEmembership{Member,~IEEE,}
        Shigang~Liu,~\IEEEmembership{Member,~IEEE,}
        Yuexin~Zhang,~\IEEEmembership{Member,~IEEE,}
        Yao~Deng, 
        Sheng~Wen,~\IEEEmembership{Member,~IEEE,}
        Jun~Zhang,~\IEEEmembership{Member,~IEEE,}
        ~Yang~Xiang,~\IEEEmembership{Senior Member,~IEEE,}
       and ~Shui~Yu,~\IEEEmembership{Senior Member,~IEEE,}
\thanks{}
\thanks{M. Dibaei, X. Zheng (Corresponding Author), and Y. Deng are with the Department of Computing, Macquarie University, Sydney, NSW, Australia. E-mail:
dibaeimahdi@yahoo.com, james.zheng@mq.edu.au (Corresponding Author), yao.deng@hdr.mq.edu.au.

K.Jiang is with the China State Key Lab of Automotive Safety and Energy, Tsinghua University, Beijing, China. E-mail: jiangkun@tsinghua.edu.cn.

S. Maric and R.Abbas are with the School of Engineering, Macquarie University, Sydney, NSW, Australia. E-mail:
sasa.maric@students.mq.edu.au,robert.abbas@mq.edu.au.

S.Liu,Y.X.Zhang, S.Wen, J.Zhang, Y.Xiang are with the School of Software and Electrical Engineering, Swinburne University of Technology, Hawthorn, VIC, Australia. E-mail:shigangliu@swin.edu.au,yuexinzhang@swin.edu.au,
swen@swin.edu.au,junzhang@swin.edu.au,yxiang@swin.edu.au.

S. Yu is with the Faculty of Engineering \& Information Technology, University of Technology Sydney, Sydney, NSW, Australia. Email: Shui.Yu@uts.edu.au
}

}

\IEEEtitleabstractindextext{%
\begin{abstract}

Cyber security is one of the most significant challenges in connected vehicular systems and connected vehicles are prone to different cybersecurity attacks that endanger passengers’ safety. Cyber security in intelligent connected vehicles is composed of in-vehicle security and security of inter-vehicle communications. Security of Electronic Control Units (ECUs) and the Control Area Network (CAN) bus are the most significant parts of in-vehicle security. Besides, with the development of 4G LTE and 5G remote communication technologies for vehicle-to-everything (V2X) communications, the security of inter-vehicle communications is another potential problem. After giving a short introduction to the architecture of next-generation vehicles including driverless and intelligent vehicles, this review paper identifies a few major security attacks on the intelligent connected vehicles.  Based on these attacks, we provide a comprehensive survey of available defences against these attacks and classify them into four categories, i.e. cryptography, network security, software vulnerability detection, and malware detection. We also explore the future directions for preventing attacks on intelligent vehicle systems. 

\end{abstract}

}

\maketitle

\IEEEdisplaynontitleabstractindextext

%
\IEEEpeerreviewmaketitle

\section{Introduction}
\label{sec:introduction}


The past decade has seen a rapid development of vehicular systems in a wide variety of aspects. The complexity of current vehicular systems, with a dramatic increase in the use of electronic systems and wireless technology, has changed the traditional concept of security in the automotive industry. Moreover, the growing interest in the development of Vehicular Ad hoc Networks (VANETs) and Intelligent Transportation Systems (ITS) has involved new security challenges and vulnerabilities. However, long-established computer security policies are not followed by the industry standards for in-vehicle and vehicular communications because of hardware constraints and differences in network configuration \cite{patsakis2014towards,cheah2018building}.

Previous reports have illustrated highly practical wireless attacks on core functions of vehicles and disengaging the engine and the brakes \cite{forbes}, \cite{wierd}, \cite{guardian}, \cite{helpnetsecurity}. For instance, by hijacking the steering and brakes of a Ford Escape and a Toyota Prius, Miller and Valasek indicated that a vehicle system is not just a simple machine of glass and steel \cite{forbes}. In contrast, it is important to realize that it is also a hackable network of computers. In 2015, 1.4 million vehicles were a subject of a recall by Chrysler because hackers could remotely take the control of a jeep’s digital system over the internet \cite{wierd}. In another report, a team of hackers remotely hijacked a Tesla Model S from a distance of 12 miles \cite{guardian}. In a recent study, researchers have found 14 vulnerabilities in the infotainment system in several BMW's series \cite{helpnetsecurity}. Overall, these cases support the view that security in intelligent vehicular systems becomes essential and must be addressed in order to protect them.

At the present time, successful cybersecurity attacks on vehicles are mainly due to sharing information and wireless communications that increase the susceptibility of vehicles to different malicious attacks. Consequently, information privacy, data privacy, securing data exchange including input and output data as well as protecting Electronic Control Units (ECUs) inside the vehicle systems are among the most significant security and privacy issues for the intelligent vehicles \cite{MCAFEE}.

In this paper, the term security encompasses attack, defence, or vulnerability. With this in mind, some significant survey and review studies related to the issue are mentioned below. These works are limited only in attacks or vulnerabilities while not focusing on defences. Mokhtar and Azab \cite{mokhtar2015survey}, Sakiz and Sen \cite{sakiz2017survey}, and Hasrouny et. al. \cite{hasrouny2017vanet} have focused on security attacks of VANETs. In all the above-mentioned papers security attacks have been focused and categorized, however, security defence mechanisms have not been classified. Moreover, although they have performed some good exploratory work on the network vulnerabilities in vehicular systems, they have largely missed the in-vehicle vulnerabilities (e.g. vulnerabilities of ECUs, software vulnerabilities). A survey by Bernardini et al. \cite{bernardini2017security} covers security vulnerabilities in internal vehicle communications including the ECU and in gateways including the On-Board Diagnostics (OBD), Tire Pressure Monitoring System (TPMS), electrical charging system, Remote Keyless System (RKS), and infotainment system. However, it should be noted that they have not mentioned about defence mechanisms and techniques.

Despite a few published survey papers that cover the security attacks or security vulnerabilities in vehicles, far too little attention has been paid to the defences. We aim at presenting a detailed analysis of security attacks and challenges in intelligent vehicle systems and their possible related defences. Currently, security defences in in-vehicle and inter-vehicular communications have not been given enough attention and to the best of our knowledge, no previous study has classified current defences against cyber security attacks on vehicles. 

For the purpose of paper selection, a search of IEEE Xplore Digital Library, Google Scholar, ScienceDirect, and Springer databases with different combinations of following queries was performed:
vehicle, security attack, VANET, intelligent transportation system, EEA, LiDAR, ECU, TPMS, OBD, RKS, CAN, MOST, machine learning, SVM, Markov chain model, malware, cryptography, symmetric encryption, asymmetric encryption, intrusion detection, software vulnerability detection, mutation testing, lightweight authentication, V2X, LiFi, 3GPP, software defined security, deep learning, intelligent, CNN, LSTM.

Search results were papers only in English. We screened titles and abstracts of the papers manually and we selected the papers considering their quality. In this way, we focused on papers from Q1 journals based on SCIMAGO INSTITUTION RANKINGS or Conference A* and A papers based on Core Conference Portal and we included related high-quality papers as more as possible. As a matter of fact, we have included other papers where we have not found related Q1, A*, or A papers. Moreover, as old papers may not relevant today, we just included papers in the past 10 years. The classification of our solutions is based on the research categories from those papers we reviewed and attacks on the intelligent  vehicle systems.

This review paper classifies the main current defence mechanisms to cryptography,  network security, software vulnerability detection, and malware detection. Moreover, we propose new research directions in these four defence categories:  Light-weight authentication (cryptography), 3GPP and software defined security (network security), deep learning (software vulnerability and malware detection). The major motivation that led us to carry out this work is the ever-expanding gap between security attacks and existing safety measures.

\begin{figure}
\begin{centering}
\includegraphics[width=1 \columnwidth]{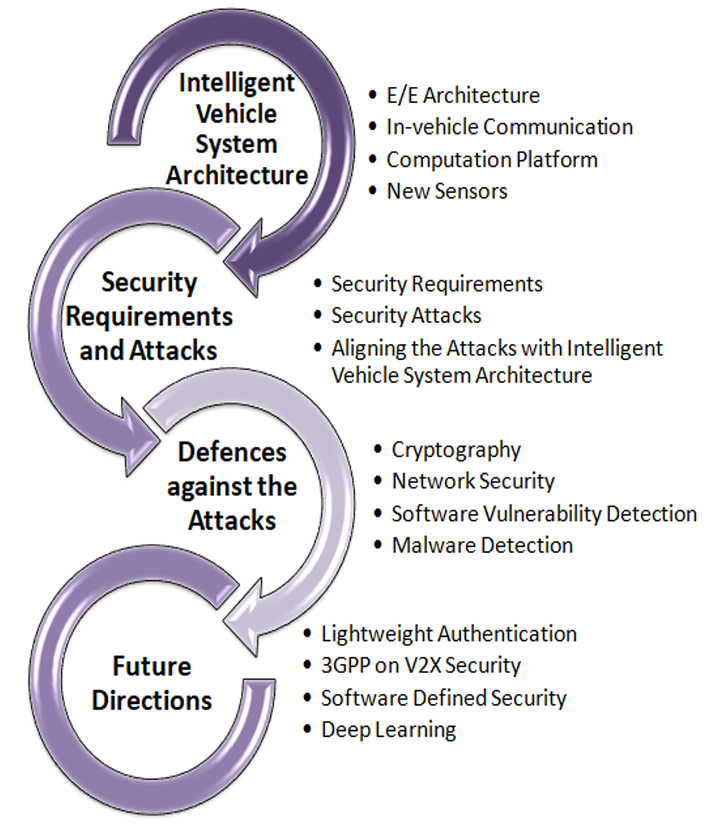}
\par\end{centering}
\caption{Key structures of our technical contribution}
\label{structure}
\end{figure}

The key structures of the technical content in the survey paper is illustrated in Figure \ref{structure}. It taxonomizes the security of intelligent  vehicles according to the following attributes: (i) Intelligent Vehicle System Architecture: makes an overview on the Electronic/Electrical architecture of intelligent vehicle systems and presents an overall view of Electronic/Electrical architecture as well as the in-vehicle communication network, computation platform, and new sensors in intelligent vehicles. (ii) Security Requirements and Identified Attacks: discusses security requirements for vehicular systems in four categories (authentication, integrity, privacy, and availability) and presents a classification of attacks on vehicles and vehicular networks including denial-of-service attacks, distributed denial of service attacks, black-hole attacks, replay attacks, Sybil attacks, impersonation attacks, and also more identified attacks on components of intelligent vehicles (iii) Defences against the attacks: refers to the list of existing novel techniques including cryptography, signature-based detection, anomaly-based detection, software vulnerability detection, and malware detection which can be used to deal with security challenges in automotive systems. (iv) Future Directions: shows the possible areas (e.g. lightweight authentication, software defined security, deep learning) for further studies. Future research should, therefore, concentrate on the investigation of these solutions.

This study makes the following contributions:
\begin {itemize}
\item{Detailed description of the intelligent vehicle system}
\item{Explaining attacks associated with the intelligent vehicle system}
\item{Defences against the attacks and possible research direction}

\end{itemize}

In particular, this review paper aimed to examine four research questions: 
\begin {itemize}
\item{What is the state of the art of vehicle systems? }
\item{What are unique research challenges in securing vehicle systems?}
\item{ What are the main solutions and their pros and cons? }
\item{ And what are promising solutions to improve security?}
\end{itemize}

The remainder of this paper has been divided into four sections:
Section~\ref{sec:modernVehicleSystemArchitecture} gives an overall review of the state-of-the-art intelligent vehicle systems architecture.
Section~\ref{sec:AttacksChallenges} discusses security attacks experienced and challenges currently faced by intelligent vehicle systems.
In Section~\ref{sec:solutions} we highlight the best practices to deal with these security challenges in intelligent vehicle systems.
Section~\ref{sec:futureDirections} discusses some promising future directions to address those security challenges in intelligent vehicle systems.
Limitations and threats to the validity of this study are discussed in Section~\ref{sec:ValidityDiscuss}.
Finally, Section~\ref{sec:concl} summarises our findings.



\section{Intelligent Vehicle System Architecture}
\label{sec:modernVehicleSystemArchitecture}


The series production of high-level intelligent connected vehicle (ICV) is an active research topic in the automotive industry. Many intelligent driving functionalities have been installed on passenger cars, such as lane keeping assistance (LKA), lane departure warning (LDW) and other assistance systems. It is sure that the high-level intelligent vehicle should be able to accomplish all these functionalities. However, it is not feasible to integrate all these intelligent assistance systems by simply put them together, as the traditional Electrical/Electronic Architecture (EEA) was not designed for supporting so many intelligent functionalities. Especially, the required abilities of data acquisition and processing are beyond the limit of traditional EEA. The next-generation EEA which could support the high-level ICV is the key to ICV’s series production. The next-generation EEA needs fundamental advancement in three parts, which is the overall structure design, the in-vehicle/ inter-vehicle communication network, and the computation platform.

\begin{figure}
\begin{centering}
\includegraphics[width=0.9\columnwidth]{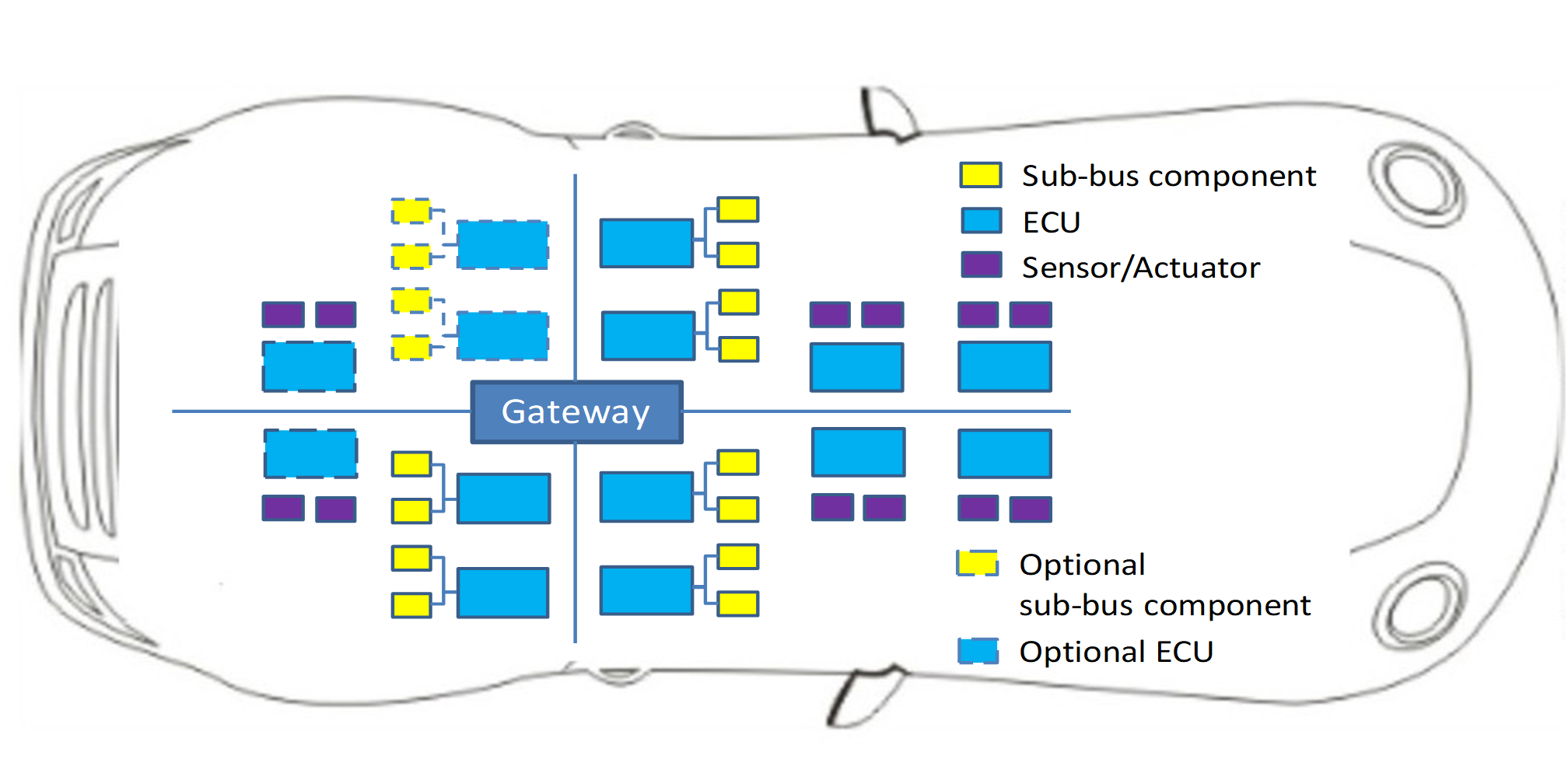}
\par\end{centering}
\caption{Gateway-based E/E architecture}
\label{gateway-based}
\end{figure}
\subsection{The Overall E/E Architecture of Intelligent Vehicle}

The topology design of overall architecture is fundamental to improve the performance of EEA. The main task of topology design is to ensure the data flow on the network matches the need of each node. As shown in Figure \ref{gateway-based}, the traditional EEA topology is based on the controller area network (CAN). Due to the characteristics of CAN, every node in the network must share the bandwidth with each other. The bandwidth is like a bottleneck that limits the data processing ability of each ECU on the network. The core problem of traditional EEA is the lack of space for high computation power unit, which is necessary for intelligent driving. The topology of next-generation EEA should specify where the complex computation is realized and how the huge amount of data is transferred.     
\begin{figure}
\begin{centering}
\includegraphics[width=0.9\columnwidth]{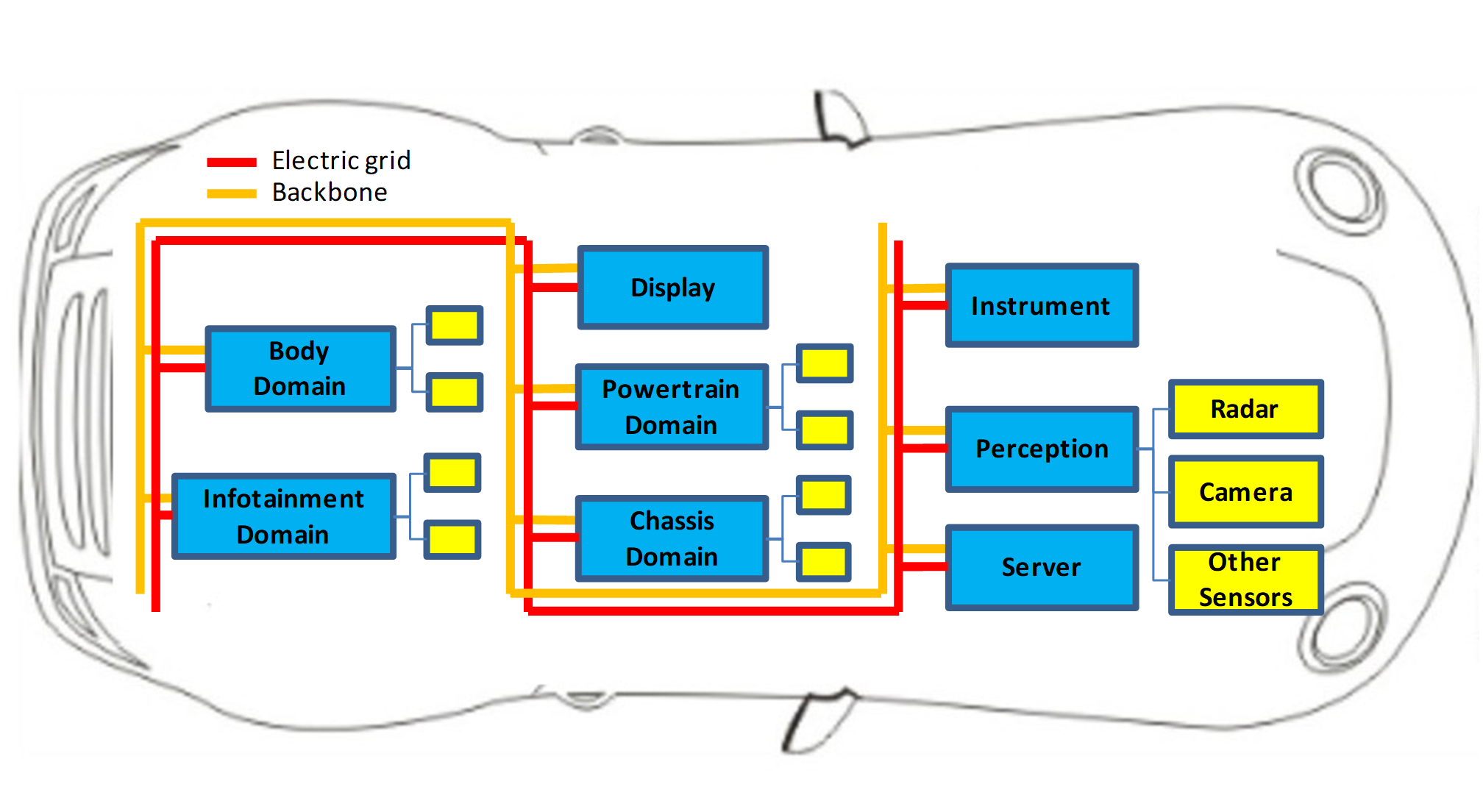}
\par\end{centering}
\caption{Domain-based E/E architecture}
\label{domain-based}
\end{figure}

One feasible approach is the domain-based topology which has been recently applied in production vehicles. Its concept is to divide the autonomous driving system into several domains. Its main difference with the traditional EEA is the occurrence of domain ECU, which is the core computation platform of each domain. The vehicle components can be classified into different domains according to their functionalities. Usually, the sensors and actuators those can be shared by different functionalities would be grouped as one domain. For example, the commonly used domains are the infotainment domain, the chassis and safety domain. The domain-based EEA can be illustrated by Figure \ref{domain-based}. The domain-based topology has advantages over the traditional one. First of all, it can support more complex intelligent driving functions, as each domain ECU has more power in both communication and computation. The domain ECU can be directly connected to sensors in the domain without the problem of sharing bandwidth. It is also a computation platform to integrate related simple control functions into a complex behavior control function\cite{haas2016cross}. Furthermore, the distributed computation strategy of domain-based topology has the advantage of being more compatible with traditional EEA system. The domains can be relatively independent to each other, while only transmit necessary information to other domains. The data flow within the domain will not occupy the bandwidth and other resources of the backbone. 
\begin{figure}
\begin{centering}
\includegraphics[width=0.9\columnwidth]{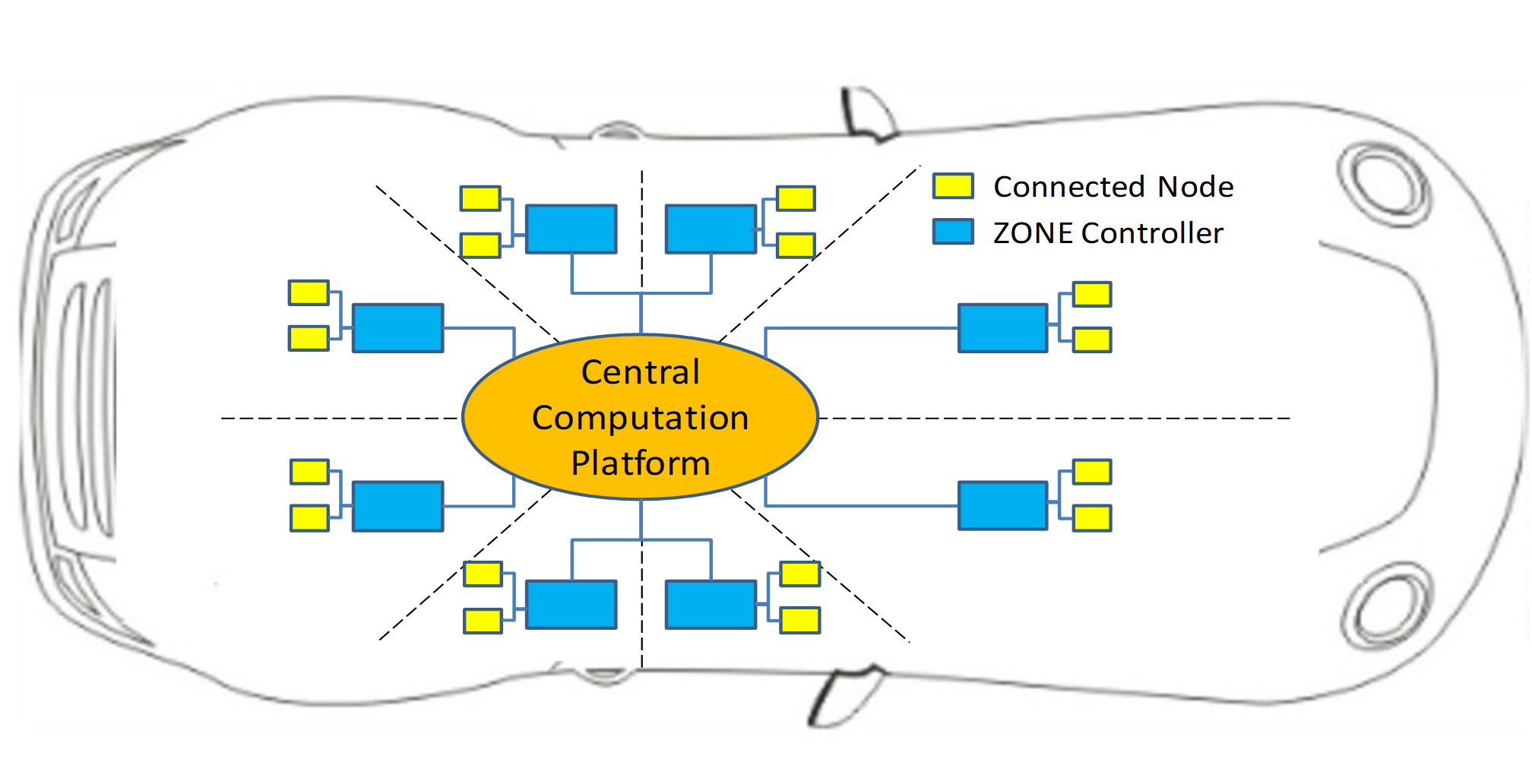}
\par\end{centering}
\caption{Centralized E/E Architecture}
\label{centralized}
\end{figure}

The centralized architecture is another approach for the EEA of the next generation. In a centralized architecture, most of the computation tasks are realized in the central computation entity, as illustrated in Figure \ref{centralized}, rather than distributed in different functional domains. Most of the components should be connected to the central computation entity, which could access all sensors and actuators. The benefit of a centralized topology is the ability of realizing complete sensor fusion. In theory, when the central computation entity could combine more information, it has the potentials to make a better decision and execute better behavior. However, a centralized topology has higher demands on the data communication capacity. The centralized EEA needs to group the components into different sub-networks according to their physical placement or the network properties to improve the efficiency of communication. The controller of the sub-network is called a zone controller in \cite{brunner2017automotive}.

\subsection{The In-Vehicle Communication Network of Intelligent Vehicles}

As mentioned above, one of the most significant challenges for the next-generation EEA is managing the high-speed communication among a
vehicle's electronic components with a limited cost. The most successful communication network in the current automotive industry is the controller area network (CAN) protocol. The CAN protocol is developed by BOSCH and it has been the most widely used standard in the field of vehicle hardware communication since its publication in 1986 \cite{zeng2016vehicle}. Compared to other network technologies, CAN
has two outstanding advantages: cost efficiency and flexibility. A variant of CAN is CAN with
flexible data rate (CAN-FD) \cite{afsin2017c}, \cite{hartwich2012can} with a
 bandwidth of up to 8 Mb/s \cite{bosch2012can}. CAN is a multi-master
network in which every node could equally and independently receive
and broadcast information. With this characteristic, CAN is almost
a plug-and-play system: new ECUs or diagnostic tools can be easily
connected to the network without special modification of the network.
Nevertheless, it also makes the communication system vulnerable to
attacks.

In many subsystems of vehicles, different kinds of specialized communication networks are developed for specific automotive applications. Each of these networks has its own advantages and will possibly still exist in the vehicle for a long time, but they are unlikely to be employed as the backbone communication network in the EEA of the next generation. A LIN (Local Interconnect Network) permits a low-cost and flexible wire harness, and can be easily
implemented without special support requirements. however, the bandwidth capacity of LIN is only 20 kb/s. It is usually used in the switches and motors that roll windows and control seats. FlexRay protocol was designed to support the use of fully electric/electronic systems for performing vehicle's safety-critical functions including \textquotedblleft brake-by-wire\textquotedblright{}, \textquotedblleft suspension-by-wire\textquotedblright{}, \textquotedblleft steer-by-wire\textquotedblright{}, and in general \textquotedblleft x-by wire\textquotedblright{} \cite{flexray2010flexray}. With a built-in mechanism of time synchronization,
FlexRay can ensure real-time communication between safety-critical
components with little time delay. Media Oriented Serial Transport
(MOST) is another in-vehicle network. MOST was developed to support  infotainment devices and related applications in vehicles \cite{Engelmann2010most}, \cite{grzemba2011most}, \cite{zeeb2001optical}. It employs plastic optical fibers as its physical layer, so the network is isolated from EMI (Electro Magnetic Interference), preventing problems like buzzing sounds
in the infotainment system.

A promising candidate for the backbone communication network is the automotive ethernet\cite{matheus2017automotive}, \cite{hank2012automotive}, \cite{ABI2014ethernet}. Though ethernet is not a new idea for data communication, it still needs a great amount of modification and research to be utilized by vehicles. It was until 2013 that the first application of AE appeared in production vehicles, when the BMW X5 used AE for connection of the onboard cameras. The comparison between AE and other networks is shown in Table I. The main advantages of AE are as follows: (1) Larger bandwidth. Currently, the bandwidth capacity of the AE protocol is 100 Mbps and in the near future, it will be increased to 1 Gbps. (2) Improved security. An IP-based routing method is employed by the Ethernet, thus it prevents one compromised ECU to perform malicious attacks on the whole communication system. Moreover, the switches in ethernet can manage the information flow in the network, and avoid hi-jacked ECUs flooding overload data into the network.

\begin{table}
\caption{Specifications of common vehicle buses \cite{yang2018intelligent}}
\centering{}\includegraphics[width=0.9\columnwidth]{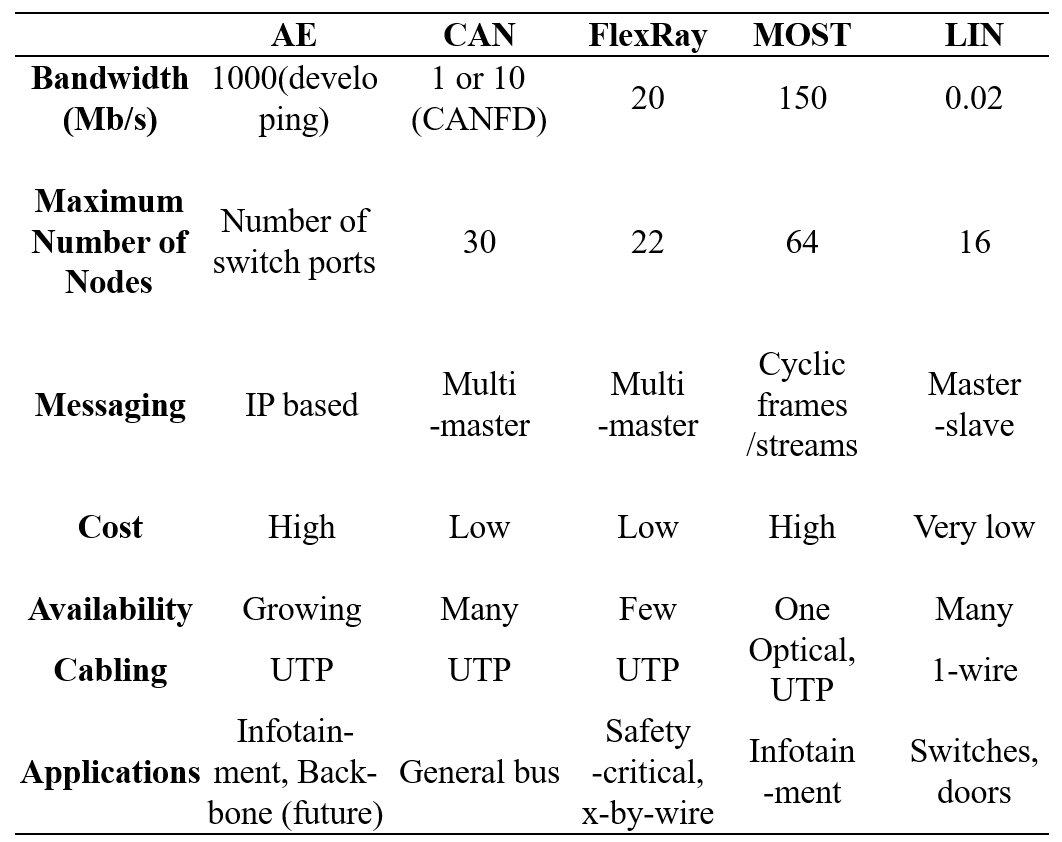}
\end{table}

The emergence of various wireless communication technologies enables the development of cooperative communication. In particular, the breakthrough of 4G LTE and 5G remote communication technologies, and the development of DSRC (Dedicated Short Range Communications) supported V2X communications.
The next important step was accident-free driving based on inter-vehicle communications and Cooperative ITS (Intelligent Transportation Systems)\cite{weiss2011v2x}. Cooperative perception-based V2X provides an exciting opportunity for developing more reliable target recognition and tracking in the field of view (FOV) \cite{obst2014multi}. Furthermore, it can also provide an accurate  perception of occluded objects \cite{kim2015multivehicle}, \cite{liu2014vehicle}, \cite{kim2015impact} or objects outside the FOV by multimodal sensor data fusion \cite{luthardt2017efficient}, \cite{leung2012decentralized}, \cite{yang2018intelligent}.

\subsection{The Computation Platform for Intelligent Vehicle}

Traditionally, simple controllers such as MCUs and DSPs were well established in vehicles for data processing of various functions including taillights on and off \cite{shenjiang2010design}, air-conditioning\cite{Xu2014automotive},
powertrain\cite{Gan2004study}, etc. Meanwhile, DSPs can execute more complicated applications,
like onboard multimedia systems \cite{Yu2011dsp} and driver assistance
functions\cite{Yu2009dsp-based}, \cite{Yu2007design} which require high integration and
great processing capacity.

In high-level ICV, more than hundreds of millions of lines of code are expected to be executed by the processors to realize intelligent algorithms including sensor fusion and deep learning. Therefore, a powerful computation platform with better hardware and software design is urgently needed. Both GPU and FPGA are believed to have wide application in the automotive industry in near future. 
A GPU is specialized in massively parallel computation, and thus it is very good at image processing\cite{lindholm2008nvidia}, which makes it ideal in self-driving vehicles for computational complex systems such as obstacle detection system and collision avoidance system. Another option is Field-programmable gate arrays (FPGAs) which are suitable for parallel computing and have less energy consumption.

The software system is another indispensable task of a computation platform.
The software used in the automotive industry has its own requirements. OSEK/VDX is a joint project that is developed by the European automotive industry. The aim of this project is to develop a real-time operating system for automotive applications \cite{Liu2002osek}.
Another important project is JASPAR (Japan Automotive Software Platform and ARchitecture) established in 2004 by Japan, and well-known corporations including Toyota, Nissan, and Honda are among its member companies. It should be mentioned that one major drawback of OSEK/VDX and JASPAR is that they fail to take the reusability and transferability demanded by the modern automotive electronics industry into account. The AUTOSAR (AUTO-motive Open System ARchitecture) standard is developed to separate application software from the associated hardware, and thus save development costs \cite{guettier2016standardization}. However, AUTOSAR still needs further development to support complicated perception algorithms and AI applications.

Although general operating systems such as Linux and Android support highly complex algorithms, the major problem is that they cannot be used as the automotive embedded software. It is necessary to develop a software platform combining the advantages
of both an automotive software system and a general operating system\cite{aly2017consolidating}.
Currently, AUTOSAR as a global partnership for developing automotive software is standardizing AUTOSAR adaptive platform. In particular, providing a stable programming interface as well as supporting Ethernet-based E/E architectures are the two major objectives of this software platform \cite{furst2016autosar}.
In software development, software update and security are two primary concerns.
For an autonomous driving vehicle, it is necessary to update its software
even after it has been sold, just like a smartphone. Update through
over-the-air can bring lots of convenience and benefits to both consumers
and manufacturers. The security during updates is quite important
and is becoming a hot research topic\cite{sagstetter2013security}.

\subsection{New Sensors in Intelligent Vehicles}

\begin{figure}
\begin{centering}
\includegraphics[width=0.9\columnwidth]{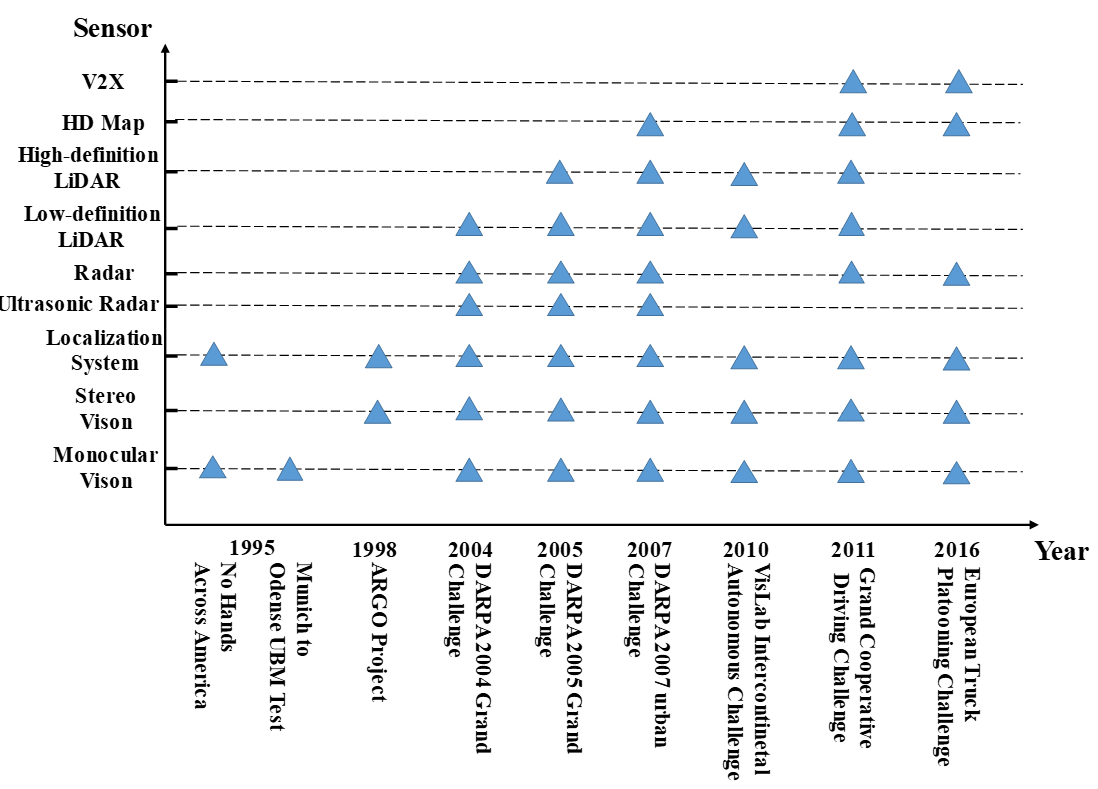}
\par\end{centering}
\caption{Mainstream sensors used in autonomous driving competitions and projects}
\label{Mainstream}
\end{figure}

In order to achieve full observations of both the vehicle's own state, the
surroundings and even the situation beyond the visual range, the intelligent
vehicle needs to be equipped with many new sensors. By comparing the
sensors used in autonomous driving competitions \cite{jochem1995no}, \cite{maurer1996compact}, \cite{bertozzi1998vision}, \cite{broggi2000architectural}, \cite{campbell2010autonomous} as shown in Figure \ref{Mainstream}, we can see the current trends in perception
technology. The fusion of multi-sensors is widely accepted as an essential
method to ensure perception robustness. The sensor fusion for high-level ICV mainly refers to the following sensors: LiDAR, Radar and intelligent cameras.

\begin{itemize}
\item \textbf{LiDAR:} LiDAR which stands for Light Detection and Ranging enables self-driving cars to observe the world. In fact, it is achieved by utilizing laser light pulses. High-definition LiDAR provides a 360-degree field of view with more than 16 laser channels. Regarding rotation mechanisms, LiDAR may be classified into three main categories: mechanical LiDAR, semisolid-state LiDAR, and solid-state LiDAR.

\item \textbf{Radar:} Millimeter-wave radar is capable of penetrating nontransparent materials such as smoke, dust, snow, and fog. In other words, the main advantage of Millimeter-wave radar is its capability to handle small size, all-weather, and long detection distance. However,  low horizontal resolution and low lateral detection accuracy are the most significant limitations of Millimeter-ware radars \cite{johnson2008development}, \cite{wang2014bionic}. Due to these shortcomings, the fusion of Millimeter-wave radars with other sensors with the aim of increasing the accuracy of object perception system is required. One solution is the fusion of Millimeter-wave radar and Monocular camera \cite{han2016frontal}, \cite{wang2016road}, \cite{kato2002obstacle}.

\item \textbf{Intelligent visual sensors:} A monocular visual system and stereo vision system are the main intelligent visual sensors in intelligent vehicles. They are utilized in order to achieve semantic segmentation of the driving environment\cite{johnson2008development},
target detection and tracking \cite{song2014robust}, ranging \cite{dagan2004forward}, \cite{park2014robust}, driver distraction and fatigue detection \cite{dong2013driver}, and so on \cite{bertozzi2000vision}, \cite{tawari2014looking}. AI technology, such as deep learning, is deeply integrated into the visual sensors to provide more accurate detection results. However, visual sensors are unstable in changing light conditions. What is worse, the AI algorithm can be attacked leading, to a false detection result.

\end{itemize}



\section{Security requirements and identified attacks}
\label{sec:AttacksChallenges}
In recent years we have seen an increased amount of research related to vehicular networks, and fully automated vehicles are fast becoming a reality. They have a tremendous potential to increase efficiency and safety for their occupants and they have already been implemented in trials in a number of locations around the United States and throughout Europe \cite{WorldCar}, \cite{Europe}. Understandably, as with any new technology, there is a certain amount of hesitation regarding self-driving vehicles; this has been further enhanced by recent incidents in California and Arizona in the United States \cite{crashOne}, \cite{crashTwo} where self-driving vehicles have been involved in incidents with pedestrians. This has prompted a number of companies like Uber to slow down or suspend the deployment of self-driving vehicles \cite{uber}, in an attempt to optimize operation and restore public confidence. In addition to unpredictable accidents, VANETs are susceptible to a number of malicious attacks and intrusions, where users are going to cause damage to intentionally the vehicle or its occupants.
\newline
\indent VANETs' dynamic nature and their need for real-time communications mean that attacks are particularly effective and often have detrimental effects. Consequently, awareness of potential threats and developing mitigation methods to proactively mitigate attacks is required. A wide range of attacks on intelligent vehicle systems have been discovered  in recent years. This is in addition to the existing attacks that occur in traditional networks. As a matter of fact, cybersecurity has been a growing area of concern for many governments and corporations around the world and, as we become increasingly dependent on technology, cybersecurity becomes a central issue in the design stage. Since VANETs are envisioned to operate, at least in part, over existing architecture, the taxonomy of attacks that VANETs face becomes extensive.
\newline
\indent In 2016 a group of Chinese security researchers from Keen Security Lab discovered a method to hack a Tesla controller area network (CAN) bus, which can be found in almost all intelligent cars and controls indicators and brakes \cite{tesla}. They were able to remotely access the central control unit and adjust the mirrors, lock the doors, manipulate the dashboard and even apply the brakes. This was reported to Tesla who quickly provided an update for its vehicles. However, this event clearly revealed that there is a real issue with outdated software being used. Recently, another team from Keen Security Lab discovered 14 vulnerabilities in BMW cars \cite{bmw}. They discovered that they were able to use a backdoor to gain access to the telematics control unit as well as the CAN bus. Similar to Tesla, BMW's response was to roll out upgrades for the affected models. These were made available over an air connection or for customers at the BMW dealerships. Similarly, researchers in the Netherlands discovered a method to get around RFID-based key immobilizers \cite{hash2}, which have been used as a primary security feature by many automotive manufacturers since 1996. The authors in \cite{hash2} used a method that bypasses the cryptographic authentication, while it can be conducted in less than six minutes with no specialized hardware.

The next section discusses some system level requirements for vehicular networks. The subsequent section details the various types of attacks that intelligent vehicles are susceptible to.

\subsection{Security Requirements}

Successful, safe and secure implementation of intelligent vehicular systems is dependent on designing and developing an extensive security framework. Therefore, vehicular systems must abide by strict security requirements. Identification of appropriate security requirements in the early stages of conceptual design and development plays a key role in ensuring that vehicles and occupants will remain safe and secure at all times. Throughout the literature, authentication, integrity, privacy, and availability are among the most significant prerequisites needing to be provided by a security system \cite{de2018driverless}. In this section, we discuss these four categories as key requirements for successful and secure integration of vehicular systems. Studying security requirements will provide deep knowledge of security attacks, security vulnerabilities, and security defences.

\subsubsection{Authentication}
Authentication is one of the key security requirements of any communication systems. In fact,  it is a requirement for verifying the identity of participants in a communication and protecting sensitive information and critical data by preventing all unauthorized access \cite{stavrou2010survey}. Authentication in vehicular systems is an important attribute that needs to be considered carefully in the early stages of system design and implementation. It means that data/information can only be accessed by authorized users. In essence, only the intended parties should be able to have access to the message and retrieve its original contents. To fulfill the authentication requirement, it is essential that key management and distribution be efficient and accurate.

\subsubsection{Integrity}

The integrity of a communication system refers to the validity of data between the sender and the receiver. The primary requirement of integrity within a communication system is that the received data is accurate and is not altered maliciously \cite{stavrou2010survey}. In vehicular networks, it is essential to be able to validate that the message has not been corrupted during transmission by degradation factors such as noise and fading, as well as deliberately by an attacker. To achieve this goal, error detection and correction codes must be implemented.

\subsubsection{Privacy}
In the Vehicle-to-Vehicle (V2V) and Vehicle-to-Infrastructure (V2I) modes of communication, where vehicles employ different techniques for sharing information (e.g. information about their geographical location) and creating a cooperation-oriented environment among vehicles and RSUs, the shared information can be used maliciously to track users \cite{safi2018cloud}. Hence, privacy is another significant challenge in intelligent vehicle systems, and sensitive information must be protected by intelligent cars \cite{bernardini2017security}.

\subsubsection{Availability}

In vehicular networks, improving the chance of getting information by all targeted vehicles is of great importance. VANETs are highly dynamic and the network must be able to respond in real time. Continuous availability is difficult to achieve in normal operating conditions and it becomes even harder when we consider that updates and patches will also have to be used at some stages. Thus, it is important that replication is considered in the early stages of design for increasing content availability and decreasing delivery cost. When there is a fault or temporary outage in one part of the network, it is important that network operations continue and that the vehicles be unaware of any problems. It is essential that services be available at all times. Consequently, the required redundancy for this purpose must be implemented properly \cite{silva2016geo}.

\subsection{Security Attacks}
In addition to the requirements described above, in order for vehicles to be securely and successfully deployed they must be dynamic and responsive to threats. Threats to security must be mitigated whenever possible; in other words, a proactive approach to threats should be a key requirement that must be met. However, since it is impossible to predict all possible threats to the network, reactive approaches must be effective and deployed quickly and efficiently. It is important that users experience as little disruption as possible as a result of an attack. The following sections present a classification of attacks that affect vehicles.

\subsubsection{Denial of Service Attacks (DoS)}
Denial-of-service-attacks (DoS) are well known and have been used extensively to disrupt network operations for many years. Denial-of-service-attacks involve flooding a host with an enormous amount of information in an attempt to overload it, effectively preventing it from receiving or processing information coming in from legitimate users. Denial-of-service-attacks are very effective at disrupting networks' operations, but they are computationally expensive to execute. As such DoS attacks have evolved over the years. DoS attacks have been optimised by targeting specific nodes that are seen as core elements on the network. In addition, common control channels have also been identified as key targets. If an attacker takes out the common control channel in the network, it essentially takes out the method of communication between users. This is an optimum strategy that allows for maximum results with minimum effort. In VANETs, the primary target for attackers would be the road-side unit (RSU). RSUs are a core component in VANETs as they authenticate, manage and update users and their information. Therefore, a successful attack on an RSU can have a detrimental effect on network operations. The simplest method to combat denial of service attacks is to block the attacker's IP address.
\subsubsection{Distributed Denial-of-service-attacks (DDoS)}
In traditional denial-of-service-attacks, a single attacker attacks a node or a channel using a single IP address, usually from a single computer. This often places a huge burden on the attacker’s resources. As a result, attackers often use multiple IP addresses in a distributed attack, reducing the resource burden. Often an attacker will plant Trojans in unsuspecting users and use their resources to perform DDoS attacks. Distributed DoS attacks are even harder to mitigate and combat because the incoming messages can come from a large number of users. Therefore, it becomes useless to simply block a single IP address. It becomes difficult to distinguish between attackers and legitimate users that have been infected. Similar to a DoS attack, distributed DoS attacks can be performed on both RSUs and other vehicles on the network.
\subsubsection{Black-Hole Attacks}
Black-hole attacks are common in communication systems. In a black-hole attack, an attacker instead of forwarding packets to their destination drops them, creating a hole where no packets are able to move through the network. This type of attack can have serious implications on network performance and routing. If an attacker is in a critical path between two groups of users, where no other paths exist, a black-hole attack effectively means that the two groups are not able to communicate with each other and become isolated. Black-hole attacks can be mitigated in a number of ways. One way is to introduce reputation-based routing, where paths that are chosen have a reputation associated with them. If the path does not deliver the packets the reputation is decreased and a path with the highest reputation is used. Deep learning and machine learning can also be implemented in networks to accurately predict which paths are going to be safe. These methods all require additional overhead on the network. A variant of a black-hole attack is called a grey-hole attack. In a grey-hole attack, the attacker only drops a percentage of packets. Attackers do this in order to avoid detection. This makes it difficult for mitigation methods to distinguish between dropping packets on purpose and packets that are dropped or corrupted during transmission even when all users on the network are legitimate.
\subsubsection{Replay Attacks}
Replay attacks are somewhat related to black-hole attacks. However, in black-hole attacks the sender intentionally sends out packets through the attacker, thinking they are legitimate, and will relay the packets to their destination. In a replay attack, the sender does not know that there is a node in the middle intercepting packs \cite{attacks}. Replay attacks are a variant of man-in-the-middle attacks. In man-in-the-middle attacks, the attacker sits between the receiver and the transmitter and intercepts packets as they are transmitted. Unlike in a black-hole attack (where packets are dropped) packets are quickly retransmitted making it appear as though nothing happened. In a replay attack, the attacker captures packets on their way to the destination and manipulates them to attack the network. In a classic scenario, an attacker intercepts a message with a password attached to it, later using the same password to authenticate themselves and impersonate a node. In VANETs, replay attacks often target communication between the vehicle and the road-side unit (RSU). If an attacker intercepts a message between an RSU and a vehicle which contains the encryption key or password they would be able to authenticate themselves at a later time. Man-in-the-middle and replay attacks are difficult to mitigate effectively because it is almost impossible to know when you are under attack. Since in most cases attackers are highly mobile and do not alter the packets in any way. Mitigation methods consist of the implementation of a strong encryption method, using virtual private networks and using time-delay variation.
\subsubsection{Sybil Attacks}
Sybil attacks, or pseudospoofing attacks as they are sometimes known, involve a user creating a large number of pseudonymous identities \cite{sybilTwo},\cite{sybilThree}. Traditionally, Sybil attacks have been used in peer-to-peer networks where the large number of users allows the attacker to have a greater influence on the network. In cases where resources are equally distributed, a Sybil attack would allow the user to have access to a large portion of the overall resources. In 2014 an attack on TOR (a peer-to-peer sharing site) was detected where a group of users flooded the TOR network for several months \cite{TorAttack}. In VANETs Sybil attacks can be used with a similar intention. However, Sybil attacks can also be used to route traffic in a certain direction, for example, when an attacker creates a large number of pseudonymous identities at certain locations. The increase in the number of users at a certain location indicates that there is severe congestion at that location which would force other vehicles to change their own routing to avoid the congested area. In VANETs, when Sybil attacks are performed with the assistance of global positioning systems (GPS) spoofing attacks (where an attacker attempts to appear to be at a location where they are not) would allow the attacker to ensure that they have a congestion-free route. A congestion-free route would be created because all other vehicles would attempt to route around problem areas. In peer-to-peer networks, this type of attack is difficult to mitigate since peer-to-peer networks rely heavily on the anonymity of users. The most effective mitigation methods are identification and authentication based methods using cryptology or public encryption.
\subsubsection{Impersonation Attacks}
Impersonation-based attacks are common in many real-life scenarios. The simplest example of an impersonation is when one person pretends to be another to gain an advantage (monetary or material) or conceal their real identity. In communication systems, users are motivated by similar reasons. Impersonation attacks can be used to gain access to resources or classified information that has restricted access. In this scenario users often have access to login details and passwords of authorised parties, which they use to impersonate authorised users. Impersonation attacks are therefore often a result of man-in-the-middle attacks, where users intercept packets travelling through the network. This is sometimes described as a phishing attack, since users often intercept random packets looking for one with authentication information (much like what one would do while fishing). In VANETs malicious nodes would impersonate road-side units in an attempt to trick users into divulging their authentication details. After authentication information has been acquired it can be used to access classified information or even as authentication with other parties. Attackers could also impersonate other vehicles to gain an advantage. For example, as an attacker might choose to impersonate an emergency vehicle, that would give them a higher priority within the network and would lead to less congestion. A number of methods have been proposed to mitigate against impersonation attacks in communication systems. Methods based on encryption, localization, and clustering can be used to mitigate the effects of impersonation attacks.
\subsubsection{Malware}
Malware and spyware have been designed around since the early days of the internet. Malicious nodes insert specialized malicious software (malware) within legitimate software. When a user downloads and installs the software they unknowingly also install malware on their system. Malware often refers to a number of malicious software agents such as Spyware, Adware, and Trojans. The primary goal of Malware software is to collect information on the host,  often looking for bank-account details or confidential information. Another common method that attackers use to infect machines is through false updates. Bogus update requires are sent to users that intentionally introduce malware in their system. Indeed, malware has great detrimental effects on VANETs. Since VANET networks are highly dynamic and will be changed and updated often, vehicles must ensure that updates and information that they are receiving coming from a trusted source. If it is not, and they become infected, they are at a risk of losing personal information and in some cases having critical malfunctions. The simplest method for mitigation of Malware attacks is the introduction of a firewall which is able to filter malicious messages from legitimate ones. However, additional methods are sometimes needed, as attacks have been known to find methods around firewalls \cite{firewall}. In addition to firewall protection, reputation-based schemes are often introduced to ensure that only messages from trusted parties are accepted.
\subsubsection{Falsified-Information Attack}
The spreading of falsified information is again commonly used in communication systems. A Sybil can be seen as an example of a falsified-information attack. In a similar way, attackers can spread falsified information about the congestion on roads to effectively force other drivers to diverge to alternate routes. They can also create congestion by neglecting to report congestion or accidents on the road. This form of attack is commonly used because it is computationally inexpensive and can have a high impact because of the distributed nature of VANETs. If an attacker is able to convince a single vehicle, that vehicle would unknowingly become an attacker as it would propagate falsified information to the next vehicle. This form of attack is often combated using reputation-based schemes that reward drivers that send out legitimate information and punish drivers that send out falsified information.
\subsubsection{Timing Attacks}
Time synchronization is a key aspect in VANETs. Vehicles move in and out of networks very rapidly which introduces the need for real-time updates and information exchange between both RSUs and vehicles. Since time-critical message exchange is critical, any delay in messages can cause serious problems. Timing attacks are similar in many ways to black and grey hole attacks. However, instead of dropping all or a portion of the packets a malicious node adds a time slot to introduce intentional delay. This causes major issues, especially in autonomous vehicles where a delay in time-sensitive information can cause a major accident.

\subsection{Aligning the Attacks with Intelligent Vehicle System Architecture}
This part of the paper aligns above-mentioned attacks with different components of the intelligent vehicle system architecture identified in Section~\ref{sec:modernVehicleSystemArchitecture}. Table \ref{tab:component} lists security attacks on different components in intelligent vehicles including overall E/E architecture, communication network, computation platform, and new sensors. An intelligent vehicle with its components is shown in Figure \ref{component}. Below, we demonstrate the identified attacks on these components. 

\begin{table*}

\caption{Security Attacks on Components Of Intelligent Vehicles  }
\centering
\renewcommand{\arraystretch}{1.40}
\begin{tabular}{p{4 cm} c  c  c c c c c c c }
\toprule
                                                          &DoS          &DDoS             &Black-Hole          &Replay         &Sybil          &Impersonation                &Malware                      &   Falsified-Information   &  Timing        \\
\midrule%
Overall E/E architecture                    &                 &                       &                         &\checkmark  & \checkmark  &  \checkmark                   &                                   &                                     &                      \\
Communication network                    &\checkmark& \checkmark   & \checkmark       & \checkmark&\checkmark&                                      &\checkmark                 &                                    &  \checkmark      \\
Computation platform                       & \checkmark&                       &                          &                   &                  &                                       & \checkmark                 &                                    &                         \\
New sensors                                      &                 &                      &                          &                    &                 &      \checkmark                   &                                  &  \checkmark                  &                            \\

\hline
\end{tabular}
\label{tab:component}
\end{table*}

\begin{itemize}
\item \textbf{Overall E/E architecture:} As mentioned earlier, the in-vehicle networks (CAN, LIN, FlexRay, and MOST) are vulnerable to different cybersecurity attacks. Through an on-board diagnostics (ODB) port or a USB port attackers can stop the engine or brakes of a vehicle and cause a fatal car crash \cite{pan2017cyber}. Replay attack and impersonation attack on CAN bus are reported in \cite{markovitz2017field}. Nilsson et al. \cite{nilsson2009first} simulated spoofing (Sybil) attack on the FlexRay bus by creating and injecting diagnostic messages. Another instance is the E/E architecture of Electrical Vehicle (EV). In recent years, the use of EVs, that can be recharged from an external source of electricity, has been dramatically increased. The architecture of EV charging-station systems makes it possible for information exchange between EV and electrical vehicle supply equipment (EVSV) that may be used for payment systems for public charging stations. Consequently, EVs are subject to cybersecurity attacks by the charging-plug interface. The International Electrotechnical Commission (IEC) has defined the communication protocol between a charging station and an EV by the ICE 61851 and ISO 15118 standards \cite{leszczyna2018review}.
\item \textbf{Communication network:} Wi-Fi, WiMAX, Long-Term Evolution (LTE), Near-Field Communication (NFC), and Dedicated Short-Range Communications (DSRC) are among the available communication standards and technologies for Vehicle-to-vehicle (V2V) and Vehicle-to-infrastructure (V2I) data communications. Vehicle-to-everything (V2X) or inter-vehicle communications are wireless, and security is considered one of the most significant challenges of V2X technology \cite{dey2016vehicle}. Moreover, audio and video players, automotive navigation systems, USB and Bluetooth connectivity, carputers, audio control, hands-free voice control and in general infotainment systems have increased security concerns about potential remote car hacking. In reviewing the literature, Sybil attack \cite{zaidi2016host}, black-hole attack \cite{sedjelmaci2015accurate}, Dos attack, DDoS attack, replay attack, and timing attack \cite{cui2018review} on the communication network of intelligent vehicle systems were found. In additions, V2V communications are susceptible to malware attack \cite{wei2018virus}.
\item \textbf{Computational platform:} Malware can penetrate into the software systems of intelligent vehicles. Additionally, DoS attacks can be launched in order to disrupt processing ability of a vehicle \cite{petit2015potential}.
\item \textbf{New sensors:} The tire-pressure monitoring system is a warning system for measuring the air pressure of tires by pressure sensors or monitoring individual wheel rotational speeds and warning the driver when tires are under-inflated. A TMPS notifies the driver when a vehicle’s tire pressure is low. Under those circumstances, a security issue related to TPMS is that a vehicle may be tracked using existing sensors along the roadways \cite{petit2015potential}. Another instance is Remote Keyless System (RKS) or smart key that is most widely used as electronic authorization system in order to controls access to the vehicle. Sensors in the vehicle are able to sense the received signal from the remote key. Along with this growth in using smart keys, however, there is increasing concern over their security vulnerabilities. The most compelling evidence is a surveillance video released by West Midland police department in Birmingham, England in 2017 that shows two hackers exploiting keyless technology to steal a Mercedes-Benz \cite{fussel2017birmingham}. Besides, In \cite{eiza2017driving} a falsified-information attack on LiDAR system of a vehicle is reported.  
\begin{figure}
\begin{centering}
\includegraphics[width=0.9\columnwidth]{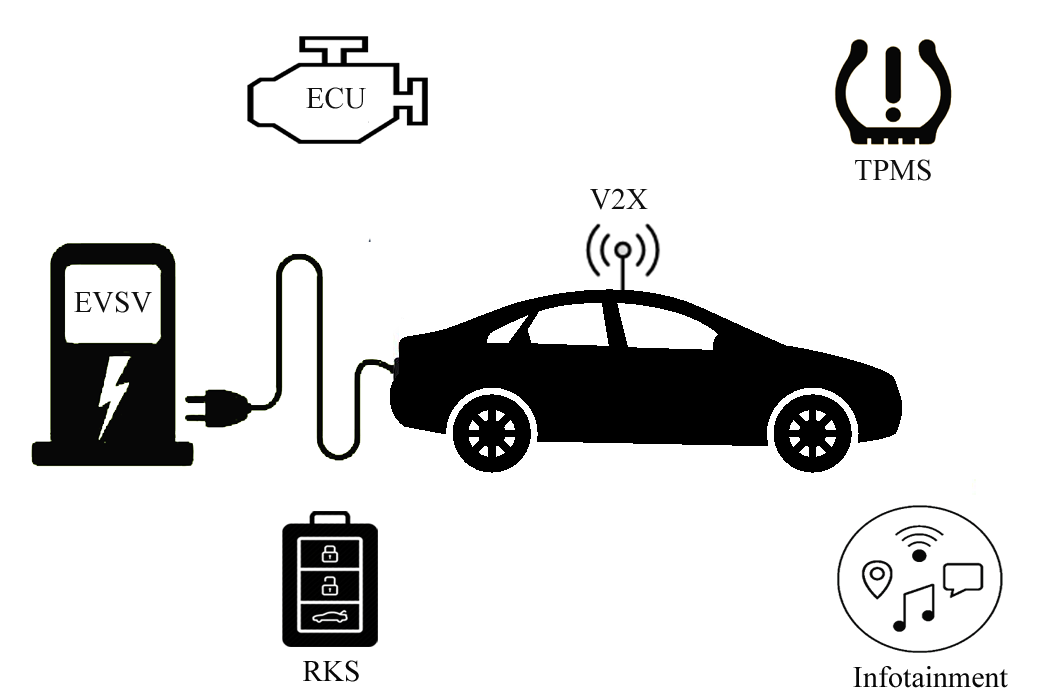}
\par\end{centering}
\caption{Typical components of intelligent vehicles}
\label{component}
\end{figure}
\end{itemize}

All the aforementioned requirements and attacks lead to the conclusion that securing intelligent vehicles is of great importance. Thus, security solutions, mechanisms, and techniques should be used to deal with these attacks. In the next section, we will present some key findings and our analysis.

\section{Existing Defences Against The Attacks}
\label{sec:solutions}
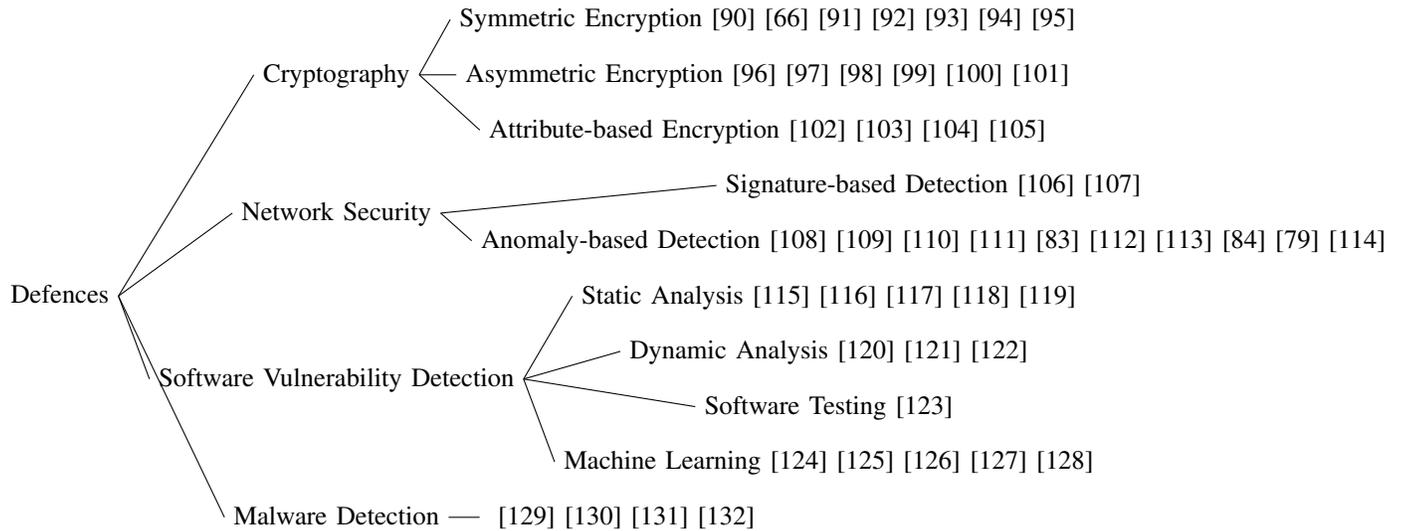
\begin{figure*}[h]

\begin{centering}
       \begin{forest}
            for tree={
                calign=center,
                grow'=east,
                parent anchor=east, child anchor=west,
                text height=1.4ex, text depth=0.2ex 
            }
            [ Defences [Cryptography[Symmetric Encryption \cite{2flip}  \cite{tesla}  \cite{RAISE}\cite{huang2011pacp} \cite{ecdsa} \cite{sakmp} \cite{gkmpan}][Asymmetric Encryption \cite{ppaa} \cite{Calandriello}   \cite{ppgcv}  \cite{tack}      \cite{GSIS}  \cite{sraac}][Attribute-based Encryption    \cite{att4} \cite{abacs}     \cite{att5}     \cite{att6}    ]]
            [Network Security[Signature-based Detection      \cite{bissmeyer2010intrusion}    \cite{tomandl2014rest}             ][Anomaly-based Detection \cite{cho2016fingerprinting}  \cite{martynov2007design} \cite{sedjelmaci2014new}     \cite{song2016intrusion}     \cite{zaidi2016host}   \cite{leeotids2017OTIDS}    \cite {yu2013detecting}   \cite{sedjelmaci2015accurate}       \cite{markovitz2017field}    \cite{zhang2018distributed}]]
              [Software Vulnerability Detection [Static Analysis  \cite{tice2014enforcing} \cite{dahse2014static} \cite{ding2017efficient} \cite{castro2006securing}  \cite{ayewah2008using}]  [Dynamic Analysis  \cite{godefroid2008automated}   \cite{newsome2005dynamic} \cite{clause2007dytan}][Software Testing  \cite{le2012genprog}][Machine Learning \cite{shin2011evaluating}  \cite{perl2015vccfinder} \cite{zhou2017automated}   \cite{shar2015web}   \cite{grieco2016toward}]] 
              [Malware Detection [  \cite{fan2016malicious} \cite{huda2016hybrids}   \cite{huda2017defending}      \cite{mirza2017cloudintell}]]
          ]]
        \end{forest}
\par\end{centering}
\caption{Existing defences against the attacks}
\label{existing-defences}
\end{figure*}

In this section, we walk through a variety of existing defences (Figure\ref{existing-defences}) which can be used as best practices to deal with the security attacks identified in Section~\ref{sec:AttacksChallenges},  and analyze the pros and cons of these defences. Table \ref{tab:defences} lists all security defences presented in this section and associates them with security requirements and security attacks.

\begin{table*}[htbp]  %
\label{tab: security solutions}
\caption{comparison of Security Defences  based on security attacks and security requirements}
\centering

\renewcommand{\arraystretch}{1.40}

\setlength{\tabcolsep}{10pt}
\begin{tabular}{c l c c c c c }

\toprule
\rowcolor{lightgray}
Category& Security solutions                                           &main mitigated attacks &   Authentication                          &  Integrity                                           &   Privacy &  Availability \\
\midrule%
\multirow{15}{*}{\rotatebox[origin=c]{90}{Cryptography}}&2FLIP \cite{2flip}       &         DoS                                                                   &\checkmark                     &  \checkmark                                                           &           \checkmark         &         Medium        \\
&TESLA \cite{tesla}                                             &   DoS, packet injection    &        \checkmark             &              \checkmark                                              &       \checkmark               &       Medium         \\
&RAISE \cite{RAISE}                                           &    Impersonation     &\checkmark                     &   \checkmark                                                          & \checkmark                     &     Medium           \\
&PACP \cite{huang2011pacp}                             &    Eavesdropping, replay, impersonation   &         \checkmark   &                                &      \checkmark                &          Medium     \\
&ECDSA \cite{ecdsa}                                            &All malicious attacks&\checkmark  &&&Medium\\
&SA-KMP \cite{sakmp}                                          & DoS, replay, impersonation    &  \checkmark                                &        \checkmark                                                     &      \checkmark                 &    Medium          \\
&GKMPAN \cite{gkmpan}                                      &Eavesdropping, replay  &  \checkmark                        &                                                             &                      &     Limited        \\
&PPAA \cite{ppaa}                                               &  Sybil  &   \checkmark                               &                                                              &  \checkmark                   &         N/A      \\
&Calandriello et al. \cite{Calandriello}                     &               DoS, jamming  & \checkmark                                       &          \checkmark                                                    &         \checkmark             &  Limited             \\
&PPGCV \cite{ppgcv}                                              &    Collusion        &        \checkmark                              &                                                             &          \checkmark            &    Limited         \\
&TACKs \cite{tack}                                                  &   Eavesdropping, Sybil, correlation     & \checkmark                                   &                                                             &          \checkmark           &     Limited         \\
&GSIS \cite{GSIS}                                                   &  DoS    &            \checkmark                    &                                        \checkmark                      &               \checkmark      &        Limited         \\
&SRAAC \cite{sraac}                                              &   Unauthorized access    &          \checkmark                   &                                                              &                      &  Limited              \\
&DABE \cite{att4}                                                   &   Collusion   &   \checkmark                                      &                                                              &                      &        Limited        \\
&ABACS \cite{abacs}                                              &   Collusion   &              \checkmark                         &                                                              &                      &     Limited           \\
&Xia et al. \cite{att5}                                              &   Collusion, replay   &       \checkmark                               &                                                              &          \checkmark            &         Limited       \\
&Bouabdellah et al. \cite{att6}                               & Black-hole   &  \checkmark                                        &                                                              &                      &       Limited         \\
\arrayrulecolor{gray}\hline
\multirow{15}{*}{\rotatebox[origin=c]{90}{Network Security} \newline \rotatebox[origin=c]{90}{}}
&Bi{\ss}meyer et al. \cite{bissmeyer2010intrusion}                        &     Sybil         &               \checkmark                        &                               \checkmark                               &            \checkmark          &        Medium        \\
&REST-Net  \cite{tomandl2014rest}                                                &    Impersonation, falsified information    &    \checkmark                                &                              \checkmark                                &                      &         N/A       \\
&CIDS \cite{cho2016fingerprinting}                                                 & DoS, masquarade       &                                  &                         \checkmark                                     &                      &           Limited     \\
&Martynov et al.\cite{martynov2007design}                              &      DoS   &              &                                   &                                                              &                  Good                   \\
&IDFV \cite{sedjelmaci2014new}                                                     &    Selective forwarding, black-hole   &                            &                                                              &                      &     Medium           \\
&Song et al. \cite{song2016intrusion}                                             & Message injection       &                          &                                                              &                      &   N/A             \\
&Zaidi et al. \cite{zaidi2016host}                                                     &       Sybil, falsified information       &             \checkmark         &                                                              &                      &       Good         \\
&OTIDS \cite{leeotids2017OTIDS}                                              &         DoS, impersonation, fuzzy    &                     &                                                              &                      &       N/A         \\
&PES  \cite {yu2013detecting}                                         &     Sybil                    &                                         &               \checkmark                                            &  \checkmark            &      Medium           \\
&AECFV \cite{sedjelmaci2015accurate}                           &      Black-hole, worme hole, Sybil                  &                                          &                        \checkmark                                   &        \checkmark      &     Limited              \\
&Markovitz et al. \cite{markovitz2017field}                     &         Falsified information                    &                                        &                                                           &              &              Meduim      \\
&PML-CIDS \cite{zhang2018distributed}                          &          DoS, probing, unauthorized access                 &                                        &                                                           &         \checkmark     &           Medium        \\

\arrayrulecolor{gray}\hline
\multirow{13}{*}{\rotatebox[origin=c]{90}{Software Vulnerability} \newline \rotatebox[origin=c]{90}{Detection}}&Tice et al. \cite{tice2014enforcing}            &     Control-flow             &                         &             \checkmark                                      &    \checkmark            &           Good       \\
&Dahse and Holz \cite{dahse2014static}                          &         XSS, remote code execution                &           \checkmark               &                                                                     &     \checkmark          &       Good             \\
&PITTYPAT\cite{ding2017efficient}                                   &       Control-flow                &         \checkmark                 &                     \checkmark                              &        \checkmark       &        Good             \\
&DFI \cite{castro2006securing}                                        &    Buffer overflow                  &            \checkmark               &                      \checkmark                            &       \checkmark      &     Good                \\
&FindBugs \cite{ayewah2008using}                                  &      Buffer overflow                &                                             &                              \checkmark                    &                               &         Medium          \\
&Generational Search \cite{godefroid2008automated}                  &  Malware (Bug)      &                                            &                                 \checkmark                    &         \checkmark    &         N/A                \\
&TaintCheck \cite{newsome2005dynamic}                                     &   Overwrite     & \checkmark                         &        \checkmark                                         &    \checkmark        &  Medium                   \\
&Dytan \cite{clause2007dytan}                                                      &   Control-flow, data-flow, overwrte     &                                            &                     \checkmark                                       &     \checkmark          &      Good           \\
&GenProg \cite{le2012genprog}                                                     & DoS, overflow       &       \checkmark                    &                        \checkmark                                    &               &        Medium         \\


&Shin et al. \cite{shin2011evaluating}                                     &    Malware (Bug)             &                                       &                           \checkmark    &    \checkmark           &      Good          \\
&Perl et al. \cite{perl2015vccfinder}                                        &     Malware (Bug)               &                                       &                                      \checkmark                     &           \checkmark    &          Good         \\
&Zhou and Sharma \cite{zhou2017automated}                       &   DoS              &                                       &                                          \checkmark                 &         \checkmark      &         Good          \\
&Shar et al. \cite{shar2015web}                                              &   Injection, file inclusion             &                                       &                                         \checkmark                   &     \checkmark           &         Good         \\
&VDiscover \cite{grieco2016toward}                                    &    Malware            &                                       &                            \checkmark                                &        \checkmark        &            Good      \\

\arrayrulecolor{gray}\hline
\multirow{4}{*}{\rotatebox[origin=c]{90}{Malware} \newline \rotatebox[origin=c]{90}{}\newline \rotatebox[origin=c]{90}{Detection}}&MSPMD \cite{fan2016malicious}                             &      Malware                           &                                      &                                              \checkmark               &        \checkmark       &          Limited          \\
&MRMR-SVMS \cite{huda2016hybrids}                             &    Malware                    &   \checkmark                 &                                           \checkmark                  &              \checkmark    &          Limited         \\
&Huda et al. \cite{huda2017defending}                           &    Malware                   &                                       &                                                    \checkmark         &       \checkmark           &         N/A           \\
&CloudIntell \cite{mirza2017cloudintell}                            &   Malware                    &                                       &                                       \checkmark                      &            \checkmark       &     Variable             \\
\hline

\end{tabular}
\label{tab:defences}
\end{table*}

\subsection{Cryptography}
This section provides an overview of cryptography-based algorithms used to enhance security for vehicular networks. Encryption has been used by humans long before computer systems
to hide messages from unwanted parties. As computer networks have developed encryption has become an essential part in ensuring that the integrity of information remains intact.
In intelligent vehicular systems, encryption is an essential key to ensure safety. The section that follows outlines a number of existing security algorithms based on symmetric key encryption,
asymmetric key encryption, and attribute-based encryption.

\subsubsection{{Symmetric Encryption}}
\begin{figure}
\begin{centering}
\includegraphics[width=0.8\columnwidth]{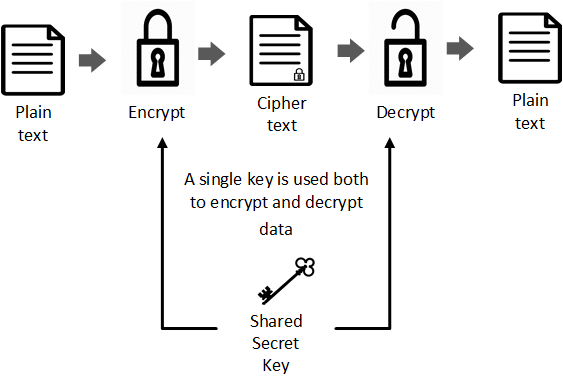}
\par\end{centering}
\caption{Symmetric encryption}
\label{Symmetric}
\end{figure}

Symmetric-key cryptography is used in a wide range of applications. In symmetric-key cryptology, a single key is used both to encrypt and decrypt data, as shown in Figure \ref{Symmetric}.
It is known to be less secure than asymmetric cryptography, which uses one key for encryption and another, separate, key for decryption. In symmetric-key encryption, it is
essential that a secure channel is established so that keys can be exchanged safely. If this channel is compromised or the key is mistakenly shared with the attacker,
that attacker would have full access to the network. Traditionally, Symmetric keys were seldom used in point-to-point communication. They were primarily used in retrieval
situations, where data is stored in a database at a central location. However, they gained popularity because they are simple, and much faster than asymmetric. The number of keys
needed for symmetric encryption is far less. This is primarily because that they use a single key and they usually have smaller key sizes, which significantly reduces overhead.

\indent In \cite{2flip} a decentralized method is proposed to authenticate vehicles using hash functions. The Two-Factor Lightweight Privacy preserving
(2FLIP) algorithm works in two steps to provide fast and accurate authentication. In the first step, a thematic device is used to identify the
driver using fingerprints or face recognition. The key strength of using biological password is nonrepudiation. In other words, driving evidence or sent messages are undeniable. This is denoted as a biological password and is needed along with the certificate authority (CA), for user
authentication. The second step is decentralization of the CA, which means that constant transmission of the CA is
not needed, increasing overall efficiency. The authors propose that the new method presents a large improvement in terms of computational
complexity as opposed to existing schemes. The results of performance evaluation indicate that in 2FLIP, computation cost has been reduced 100-1000 times and communication overhead has been decreased between 55\% and 77\%. The reduction in overhead makes this method highly practical. A disadvantage of this algorithm is that all new
drivers would have to be subjected to an authentication phase to add them to the list of authenticated drivers.
\newline
\indent A broadcast authentication protocol called the Timed Efficient Stream Loss-tolerant Authentication (TESLA) is proposed in \cite{tesla}.
In the TESLA scheme, a user sends out packets encrypted with a public key. Initially, the receiver is not able to authenticate the packets.
The packet is put into a buffer and the receiver waits for the next packet to authenticate the current packet. The next packet is sent along with the key for the previous packet.
The receiver is able to authenticate the first packet. If the authentication is successful the packet is kept and if not it is discarded.
Packets are kept, until they are all authenticated. TESLA uses one-way hash functions and loose time synchronization between the sender and the receiver.
Time synchronization does not have to be accurate as long as the receiver has an idea of the upper bound of the time it takes for the packet
to reach them. In the TESLA method packets are split up and the key for the current packet is attached to the next packet. The receiver is
only able to authenticate a packet after it has received the next packet. One of the disadvantages of this method is that, with a high volume of
packets coming through, the size of the buffer at the receiving end must be able to accommodate all incoming packets. This method is popular
because of its simplicity. It does not need two-way authentication and it is not bound by tight synchronization. An improvement designed to implement
solutions to these deficiencies is developed in \cite{teslaone} called the Tesla++. Similarly to the original algorithm, Tesla++ broadcasts authentication
keys in a symmetric manner. However, instead of storing all messages and their corresponding MACs,
the receiver only stores the MAC. Therefore, there is a large reduction in the amount of memory that is needed at the receiver. However, much like the original algorithm,
Tesla++ still has high overhead and is prone to DoS attacks.
\newline
\indent The RSU-aided message authentication scheme (RAISE) is a novel approach to vehicle authentication using symmetric functions
\cite{RAISE}. In this method, the RSU is responsible for all authentication. This algorithm assumes that messages coming from the RSU do not need to be verified,
which leads to a significant decrease in the total overhead of the network. The RSU is responsible for distribution of a short hash key which is
used as an authentication method for inter-vehicle communication.  In addition, the RSU must also assign unique identification numbers for each
vehicle within the network, which introduces another level of security into the network. Through simulations, the authors were able to prove that
the overhead of RAISE is significantly less than group-based and public-key infrastructure schemes (PKI). One drawback of this method is its reliance
on the integrity of the RSU. If an attacker were able to impersonate an RSU they would gain total control of the network. Additionally, because of the
reliance of the algorithm on a central node (the RSU), DoS attacks can have significant effects on the network if they are aimed at the RSU.
\newline
\indent Pseudorandom authentication can also be used to authenticate a vehicle, both in vehicle-to-vehicle authentication and vehicle-to-roadside
unit (RSU) authentication. In \cite{huang2011pacp} a pseudorandom method is used to authenticate users at the RSU. Each user is sent out a pseudorandom that no other
vehicle on the network knows. This keeps the identity of each user hidden and known only by the RSU. Pseudonymous authentication with conditional
privacy (PACP) is used to generate the pseudorandom keys. When a user registers at the motor-vehicle department they are handed a ticket with a unique
ID which is used to authenticate them and generate a pseudorandom key. Although this scheme is effective, it has very high overhead and is prone to
identity theft, as malicious nodes could possibly intercept or copy the ticket acquired at the motor-vehicle department.  A similar scheme is presented where the RSU sends out a large number of pseudorandom keys and matching certificates. These messages are sent to all users on the network.
Then when a user wants to transmit they would do so by using the pseudorandom key. When the message reaches the receiver it can be authenticated using
the corresponding certificate. Similarly to the previous method, this algorithm needs a large amount of overhead.
\newline
 \indent In \cite{ecdsa} an elliptic-curve digital signature algorithm (ECDSA)-based scheme is proposed. A typical ECDSA scheme works in three stages: the key generation, signature generation, and signature verification stages. In order for a vehicle to be authenticated, they must be able to generate a valid signature, using a public key, whereas the signature generation phase has a low computational complexity associated with it. The highest computational complexity lies in the signature-verification phase, where the receiver must verify that the signature is legitimate from a large list of possible signatures. The method proposed in \cite{ecdsa} introduces a scheme that implements an ECDSA verification engine that is able to verify up to 27000 signatures per second. This presents a significant reduction of latency within the network. A latency of 37 microseconds for a single signature verification and an efficiency of 24.5 sGE were achieved, which is a significant improvement when compared to previous methods.
\newline
\indent A secure and authenticated key-management protocol (SAKMP) is presented in \cite{sakmp}. SAKMP is a distributed key-management protocol that assigns public
keys to users based on their geographic location. The key is generated using a function that ensures that each key is unique and dependent on the ID of
the user and their location. The location of the users is obtained using GPS. In order for the algorithm to generate unique keys, a 3D position must be established.
The key is generated using the x, y, and z coordinates. The algorithm is only concerned with secure communication between the RSU and the OBUs. It would be difficult
to implement this algorithm in vehicle-to-vehicle communication because of its dependence on location and fairly complicated key generation function. Much like the other symmetric-encryption
based methods, storing a large number of keys has a very large overhead, especially for vehicle-to-vehicle communication.
\newline
\indent A scalable and efficient group rekeying protocol based solely on symmetric-key cryptography called GKMPAN is presented in \cite{gkmpan}.
GKMPAN attempts to exploit the properties of ad hoc networks to propagate group keys from one host to another. This reduces the risk
of eavesdropping and ensures that the key is only distributed to trusted users. In order to keep each group key secret, prior to the
deployment of the ad hoc network a key encryption key is distributed for delivering keys safely. Keys are distributed and updated at
intervals to ensure a higher level of security. The current key can be decoded by a user even if they missed previous keys. This makes
the algorithm stateless. This method is simple and effective. Since group keying is used, some of the overhead of having to generate a
key for each user is alleviated. This means that the algorithm proposed in \cite{gkmpan} is highly scalable. However, since only a single
key of keys is generated, if an attacker was to obtain it they would be able to get all subsequent keys. Since keys are passed on through
the network, as the network becomes large, severe bottlenecks could form.

\subsubsection{{Asymmetric Encryption}}
\begin{figure}
\begin{centering}
\includegraphics[width=0.8\columnwidth]{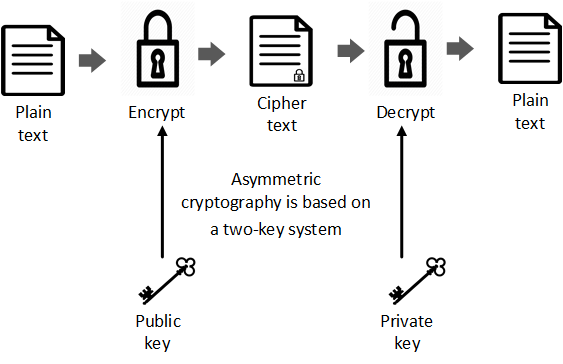}
\par\end{centering}
\caption{Asymmetric encryption}
\label{Asymmetric}
\end{figure}

Asymmetric cryptography is based on a two-key system (Figure \ref{Asymmetric}). A user sends out a public key that is distributed to every member of the network.
This key is used by other users on the network to send encrypted messages to the reference user. It is not computationally feasible to
decrypt the message without a private key which only the original node has. Asymmetric cryptology is more secure than Symmetric cryptography
because the private key is kept from the public. No secret channel is necessary for the exchange of keys because the public key acts as a
secure communication method. Asymmetric cryptology also generates significantly fewer keys. For example, asymmetric keys only need 2n secret
keys, where n is the number of users on the network, whereas, symmetric keys would need nx(n+1)/2 keys in order to securely communicate.
One of its disadvantages over symmetric cryptography is that it is much slower because of the harder mathematical problems associated with
encryption and decryption using separate keys. The keys must also be longer in order for asymmetric cryptography to be effective.

The following section describes a number of algorithms that use asymmetric cryptography to establish secure communications between vehicles, infrastructure, and humans.
\newline
\indent A Peer-to-Peer Anonymous Authentication (PPAA) method proposed in \cite{ppaa} makes the assumption that all clients and servers are equal as peers,
unlike conventional schemes that assume authentication must only be done for the client and never the server. This method ensures that both the server
and the client must be authenticated to achieve maximum security. PPAA introduces a novel system in which peers; are pseudonymous to individual peers,
although users are able to authenticate each other if they have previously been in contact they remain anonymous. The algorithm uses a centralized node
called a General Manager (GM) whose sole purpose is efficient setup and registration of the nodes. It is important to note that this scheme cannot
validate more than one node at a time, which can cause problems in large networks where many nodes are transmitting concurrently. Since this scheme
relies on a peer-to-peer framework users remain anonymous, but a new session must be started each time a user wants to communicate with another
user. This could cause unwanted overhead.
\newline
\indent In \cite{Calandriello} a pseudonym-based authentication method is used to secure communication in VANETs. A hybrid scheme is proposed that uses group
signatures to generate on-the-fly pseudonym keys which, in a similar way to the algorithm presented in \cite{ppaa}, allow users to remain anonymous within
the network. In this scheme, a user registers within a group, at which point they are given the group public key that can be used to authenticate their messages.
This method has low computational complexity and allows the users to quickly authenticate their messages using a group dynamic. Through the use of group public keys,
the authors have reduced the amount of storage that is needed to execute this method. However, the use of group keys could lead to serious breaches in the
security framework. If an adversary was able to gain access to the group keys they would be able to authenticate their own messages.
\newline
\indent A privacy-preserving group communication scheme for VANETs (PPGCV) is proposed in \cite{ppgcv}. The algorithm works in two phases. In the first phase, each user on
the network is given a pool of keys which is randomly distributed. These keys are used for Key Encrypting Keys (KEKs). To ensure that users within the group
can communicate, a group key is also established which can be used to change the pool of keys in case they are compromised. If a single user on the network is
compromised the central authority assumes that all keys are compromised. This scheme has a comprehensive method for key relocation and as such has the advantage
of being robust and hard to predict. But it does add overhead to the network. In addition, during the key reallocation users are left with no encryption methods
and cannot transmit data, further decreasing network efficiency. This method also assumes that users can keep track of which keys have been compromised which puts
an additional burden on them. If this is not managed properly then revoked keys can be used by attackers.
\newline
\indent A VANET key-management scheme based on Temporary Anonymous Certified Keys (TACKs) is introduced in \cite{tack}. In this method, users are grouped
according to their location. Users that are in close proximity are given a single group public. Regional Authorities (RAs) are appointed within each
group to distribute certificates. A TACK is a short-term certificate that is acquired from the RA. The TACK is used for signing messages; it is a
method of authenticating each vehicle within the group. When a vehicle enters a new geographical area or after some set period of time the TACK
expires and a new one is issued. This ensures that attackers are not able to associate any particular key with any particular user. The new key is
generated randomly by each vehicle after their previous key has expired. The new key along with the group user key is sent out for authentication by the
RA. The RA authenticates the key and updates its internal records accordingly.  This scheme's main advantage is the minimal overhead associated with it,
especially at the OBU. Most of the computational complexity and authentication is done at the RA. A disadvantage is that the scheme is very infrastructure-dependent
 and cannot be used in a distributed environment. Temporary Authentication and Revocation Indicator (TARI), an algorithm based on TACK, was proposed in \cite{tari}.
It is based on the same security principles as TACK. TARI also uses group signatures that are dependent on the geographical location of the OBUs. TARI has a different
method of authentication. It uses an AI to authenticate OBUs after they have received a message. Each user is verified within their own group. Its primary advantage over
TACK is that it uses symmetric-key cryptography, which significantly reduces its overhead. However, it suffers from the same drawbacks at TACK, which makes it
highly dependent on a centralised topology.
\newline
\indent A method called Group Signature and Identity (ID)-based Signature (GSIS) is used to tackle security and conditional privacy in vehicular
networks \cite{GSIS}. GSIS proposes to use another two-step process to ensure a high level of security. It groups vehicles into clusters based
on their location. Then, using a group structure each vehicle is able to securely communicate with each other in a safe and secure manner.
Users outside the cluster are ignored. A significant reduction in computational complexity is achieved, as only a single key has to be stored.
This could present a problem: if an attacker were to acquire the public key they would be treated as a part of the group. Communication
between the road-side unit and the cars is achieved using ID-based cryptography. Each message is sent out with a digital signature called an ID.
The ID numbers for the road-side units are used as public keys, whereas, the licence plates of the vehicles are used as their public keys. The
main contribution of this paper is a lightweight accurate algorithm that is able to ensure a high level of authentication for vehicular
communication. The drawback of this method is its susceptibility to man-in-the-middle-attacks, especially within the cluster configuration
where vehicle-to-vehicle communication is conducted.
\newline
\indent Secure revocable anonymous authenticated inter-vehicle communication (SRAAC) is an encryption method designed to make autonomous car networks safer and more reliable \cite{sraac}.
SRAAC has three distinct stages. The first is an OBU Registration, then a Certificate Issuance, Anonymity Revocation and finally the actual communication including Signature and Verification
of messages. The CA sends out multiple certificates to a single OBU, ensuring that even if one is compromised the OBU can use the other keys. The certificates are generated and distributed
blindly, making this method completely anonymous. A tag is kept for every set of certificates that are sent. The primary reason for this is to be able to revoke the rights of the user if they
become compromised.  This method has overhead and enables users to be completely anonymous. However, other nodes in the network would find it difficult to know which users have been misbehaving,
making it difficult to isolate malicious nodes.

\subsubsection{{Attribute-based Encryption}}

\begin{figure}
\begin{centering}
\includegraphics[width=0.9\columnwidth]{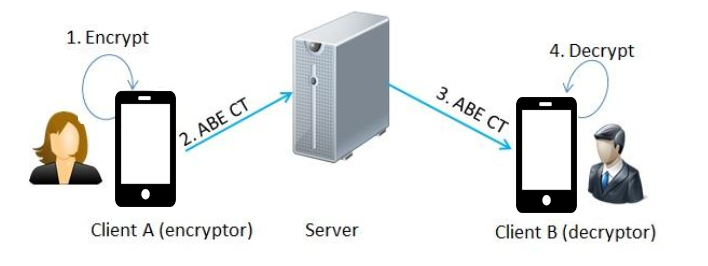}
\par\end{centering}
\caption{Architecture for a typical ABE \cite{asim2014attribute}}
\label{ABE}
\end{figure}

Attribute-based encryption (ABE) is a form of encryption that uses specific attributes (or a set of attributes) to encrypt data, as shown in Figure \ref{ABE}. As such, in order to decrypt the data one must
 have a satisfactory configuration/combination of attributes \cite{att1}. Attribute-based encryption was first introduced in 2005 by Sahai and Walters in \cite{att2}. They
  presented attribute-based encryption as an application of Fuzzy Identity based Encryption. This was later expanded in \cite{att3}, where the authors present a general
 framework for attribute based cryptology, seen as a more flexible alternative to the rigid traditional public-private key cryptography. Instead of
 using fixed public and private keys, encryption is done using specific attributes. The attributes are taken from a pool which includes an entire library. Only users or groups
 of users that have attributes, that are the same as the ones chosen, are able to decrypt the message.

\begin{table*}[htbp] %
\caption{Cryptographic  defences }
\centering
\begin{tabular}{p{3 cm} p{5 cm}  p{4cm} p{4.0cm}  }
\toprule
Method                                                                   & Key idea                                                              & Advantages                                                                            & Disadvantages   \\
\midrule%
 2FLIP \cite{2flip}                                                  &Uses decentralized CA and biological password & Reduces message delay, low message loss ratio & Telematics devices are required for all vehicles \\
TESLA \cite{tesla}                                                &A loose time synchronous authentication between sender and receiver & Simplicity &High overhead, prone to DoS attack \\
RAISE \cite{RAISE}                                               &RSU-based authentication &Scalability, low overhead&Not resilient against the RSU compromisation, prone to DoS attack\\
PACP \cite{huang2011pacp}                                 &Motor vehicle department gives unique ID to the vehicles for authentication & Scalability &Prone to identity theft, high overhead\\
ECDSA \cite{ecdsa}                                                &Provides an implementation of ECDSA for a fast signature verification&Low latency&Proper implementation of ECDSA is difficult\\
SA-KMP \cite{sakmp}                                             &Uses geographic information of the vehicle for key generation &Robust against DoS attack&Overhead of storing a large number of keys \\
GKMPAN \cite{gkmpan}                                        &Utilizes the nodes of the network for a hop-by-hop key propagation &Overhead of key generation is distributed&Does not preserve the privacy of the nodes \\
PPAA \cite{ppaa}                                                  &Considers authentication of not only clients but also servers &Secret handshake&Not suitable for a large-scale network, overhead \\
Calandriello et al. \cite{Calandriello}                     &Combines pseudonym-based approach with group signatures &Reduces overhead&serious breaches in the security framework \\
PPGCV \cite{ppgcv}                                               &A probabilistic key distribution approach  & Preserving the privacy of the nodes, robust &Overhead of key relocation \\
TACKs \cite{tack}                                                 &Utilizes short-lived keys that are certified by regional authorities &Reduces overhead especially at the OBU& Low speed is the main disadvantage of asymmetric encryption\\
GSIS \cite{GSIS}                                                   &Integrates group signature and identity-based signature schemes  &  High level of authentication&Heavy verification procedure for large verification lists \\
SRAAC \cite{sraac}                                             &Uses blinded certificate issuance for providing anonymity&Anonimity, non-repudiation &Overhead of a message certification by multiple servers, low speed\\
DABE \cite{att4}                                                   &Allows users to change the attributes that are associated with their private keys &Reduces overhead  &Synchronization, not applicable to unpredictable attribute changes \\
ABACS \cite{abacs}                                              &Utilizes attribute-based encryption for emergency services & Flexible and scalable access control&attribute revocation mechanism is a key challenge of ABE\\
Xia et al. \cite{att5}                                              &  Uses CP-ABE scheme for multi-hop multi-media data transmission in VANETs&Privacy-preserving, access control &It is heavily dependent on the structure of the network  \\
Bouabdellah et al. \cite{att6}                               &Combines CP-ABE and trust management scheme for multi-hop V2V communication &Anonymity, access control&Overhead of calculating and storing trust and reputation values\\\bottomrule
\end{tabular}
\label{tab:cryptography}
\end{table*}

This form of encryption is quickly becoming very popular because of its flexibility, effectiveness, and efficiency.
\indent In \cite{att4} a Secure, Selective Group Broadcast in Vehicular Networks using Dynamic Attribute-Based Encryption is presented. In traditional attribute-based methods,
key generation is based on a combination of certain attributes. The attributes selected depend on the policy of the sender. This again depends on what user/group of users the sender
intended the message for. When an attribute expires or needs to be replaced, the entire set of attributes must be replaced. This causes considerable overhead and delay. This is
especially a problem in vehicular based networks, where the number of users within a network is highly dynamic. The authors in \cite{att4}, propose an algorithm that uses attribute-based
 cryptology where each attribute is treated independently, as opposed to a set. When a single attribute elapses or needs to be replaced for some other reason, it is replaced independently
and without the need to change the other attributes. The main advantage of this method is the significant reduction in the overhead of the network. However, synchronization between users
about which attribute is relevant and which has been changed presents a problem in practical implementation. A similar scheme is presented in \cite{att1point1}, where the authors present
an algorithm that looks to dynamically add and remove attributes without affecting the rest of the access control policy tree. In \cite{att1point1} a fading function is introduced to each
attribute, making attributes dynamic and independent. Although this algorithm presents the problem and solution in a slightly different framework it has very similar advantages and disadvantages.
\newline
\indent In \cite{abacs}, an Attribute-Based Access-Control System (ABACS) is proposed to enable improved efficiency of emergency vehicles over VANETs. When an emergency occurs it is
important for the emergency vehicle (EV) to be able to get to the site of the emergency as quickly and efficiently as possible. The EV must be able to communicate efficiently with roadside
units (RSUs). RSUs must be able to identify which EVs are close by and which can respond to the emergency the fastest. Road-side units broadcast a message encrypted using attribute encryption
that uses attributes such as location, type of emergency vehicle (depending on the emergency this would be a police car, an ambulance or a fire truck) and event type. The emergency vehicles
that have these attributes are the only ones that are able to decrypt the message. When an emergency vehicle encrypts the message they get the relevant information and are able to respond.
The algorithm presented by the authors offers high security and reduces the overhead since only a single broadcast message has to be sent.
\newline
\indent In \cite{att5} an adaptive multimedia data-forwarding method is proposed for privacy preservation in vehicular ad hoc networks. This paper puts forward a scheme that reduces the overhead
placed on on-board units in vehicles. It does this by allowing road-side units to perform a large portion of the overall encryption. Unlike conventional schemes, this paper presents a framework
for not only short messages, but also multimedia applications such as social media. It does this by ensuring that the overhead is spread between the on-board unit and the road-side unit. Therefore,
it is important for vehicles to effectively pick which RSU they are going to involve in the dissemination process. Since decryption takes time it is important that the vehicle remains within the
transmission range of the RSU, otherwise they will receive partial or incomplete information. Attribute-based encryption is used by both the RSU and the OBU to ensure that users with the appropriate attributes
are the only ones that receive the message.  This method reduces the computational overhead to the RSUs, which enables them to be smaller and cheaper to implement. It relies heavily on the structure of the
network. In large networks, there is a possibility that the RSU would be overwhelmed with large quantities of traffic, causing a bottleneck in the network.
\newline
\indent A distributed multi-hop algorithm is proposed in \cite{att6}, where the authors present a protocol that can be utilized when no direct link between the vehicle and the road-side unit exist.
Much work has gone into RSU to OBU communications, but situations where OBUs are out of the range of RSUs are yet to be dealt with effectively. The primary issue with relaying information across the network
is that malicious users between the source and destination could have a large impact \cite{mamaghani2018security,radhappa2018practical}. To tackle this a scheme using attribute-based encryption is employed to ensure that only users that have the right attributes
receive and are able to read the message, and a reputation-based function is used to ensure that the messages are passed over the safest possible path. The framework for the reputation/trust function is poorly defined.
Additional storage and overhead are added in order to calculate the reputation and then store the reputation of each secondary user on the network. However, this scheme is not independent of a centralized
topology. OBUs are able to communicate in a distributed manner, which would be quite effective in more remote areas or when the RSU is under attack.
\newline
\indent A fine-grained privacy-preserving protocol is introduced in \cite{att7}, which is used to allow service providers to offer certain services to certain vehicles within the network. The algorithm uses
 attribute-based encryption to ensure that only vehicles that are authorised are able to access the services that are offered by the service providers. Different attributes allow users to access different
 services. To further add security, a secret sharing scheme is proposed to enforce the fine grained access control requirements. The algorithm also allows vehicles to remain anonymous by using pseudonyms as
 unique ID-based signatures. This algorithm is well defined, and it is very effective in ensuring that service providers only allow certain users to access their services. It also allows for vehicles to remain
 anonymous. However, there are some concerns that the algorithm would have a high overhead in practical situations, especially in large networks.

\subsubsection{{Summarising  Cryptographic Defences}}

It is necessary here to compare cryptography-based algorithms used to enhance the security of intelligent vehicules. Table \ref{tab:cryptography} gives a comparative view of above-mentioned cryptographic defences and focuses on their key idea, advantages, and disadvantages. 
Cryptography in VANETs is key to providing safety and security for users and service providers. However, many of the existing cryptographical standards and practises are inadequate for the new generation of VANETs. Current cytological standards are often overly complicated and place a high computational burden on users. They are seldom suitable for real time high speed applications that VANETs are inevitably becoming. Latency or a delay in communications between the vehicle and the road-side unit could cause serious accidents for users. It is therefore key that cytological standards are lightweight, but secure. It must be noted, that security is paramount in VANETs and even though real time applications require low latency, security must also be considered as a priority. Therefore, the application of cryptology algorithms in VANETs must consider the tradeoff between the security of the network and the user, against, application based parameters that enable low latency and delay. 
\newline
\indent To solve these problems we propose a number of solutions throughout this paper, these include but are not limited to new 3GPP standards, software defined networks, light authentication and block chaining (BC). In the previous sections, we discussed a number of lightweight authentications methods. In the sections that follow we discuss 3GPP standards and software defined radios. Another promising solution to the problems faced by VANETs is block chaining. As an illustration, blockchain is a distributed data structure that can manage financial transactions without the need to a centralized authority. In other words, a genuine copy of digital ledger is shared among the parties. Besides, in order to validate new transactions, public-key cryptography is utilized for providing multi-signature protection \cite{block1}. In \cite{block1} an IoT based block chaining method is discussed. IoT requires similar attributes from security protocols as VANETs, they both require low latency and computational complexity as well as a high level of security. It is concluded that blockchain enhances the security of authentication and authorization and also provides a strong defence against IoT security attacks such as IP spoofing. This is primarily due to their high level of security and scalability \cite{block1}. 
\newline
 \indent In \cite{block3} the authors present a framework for a lightweight algorithm that is secure and has low overhead. It is claimed that the fundamental security goals of confidentiality, integrity, and availability are considered and delivered using the approach presented in the paper. Significant reductions in overhead were achieved and confirmed through simulation of a variety of scenarios. In \cite{block2} a similar algorithm is presented that is lightweight and preserves all the security features of traditional blockchain algorithms. They proposed an architecture that uses distributed trust to reduce the block validation processing time. The experimentation and trials were conducted in a smart home setting which has similar goals and constraints of VANETs. Simulation of the proposed framework indicates that it has low packet overhead and low processing overhead.

\subsection{Network Security}\label{sec:Network Security}
Intelligent vehicles require cooperation from other devices and sensors to perform communications. These communications are implemented between the Controller Area Network
(CAN) and the Electronic Control Units (ECUs), and security mechanisms have not been considered in these settings at all. The CAN and ECUs are valuable targets for adversaries.
 For example, vehicles can connect to wired devices using USB, CD, wireless such as 3G, 4G, WiFi and smartphone, and all of these make the car become an open system.   Therefore, it is
 very necessary to invent suitable countermeasures to relieve the security risks in the intelligent car. Considering that  Intrusion Detection Systems (IDS) are the most closely countermeasure and
  the most reliable approach \cite{dhaliwal2018effective}, \cite{zaidi2016host}, \cite{butun2014survey} in terms of protecting vehicular networks or traditional computer networks, this section reviews related work
   using IDS in intelligent cars.
\begin{figure}
\begin{centering}
\includegraphics[width=0.9\columnwidth]{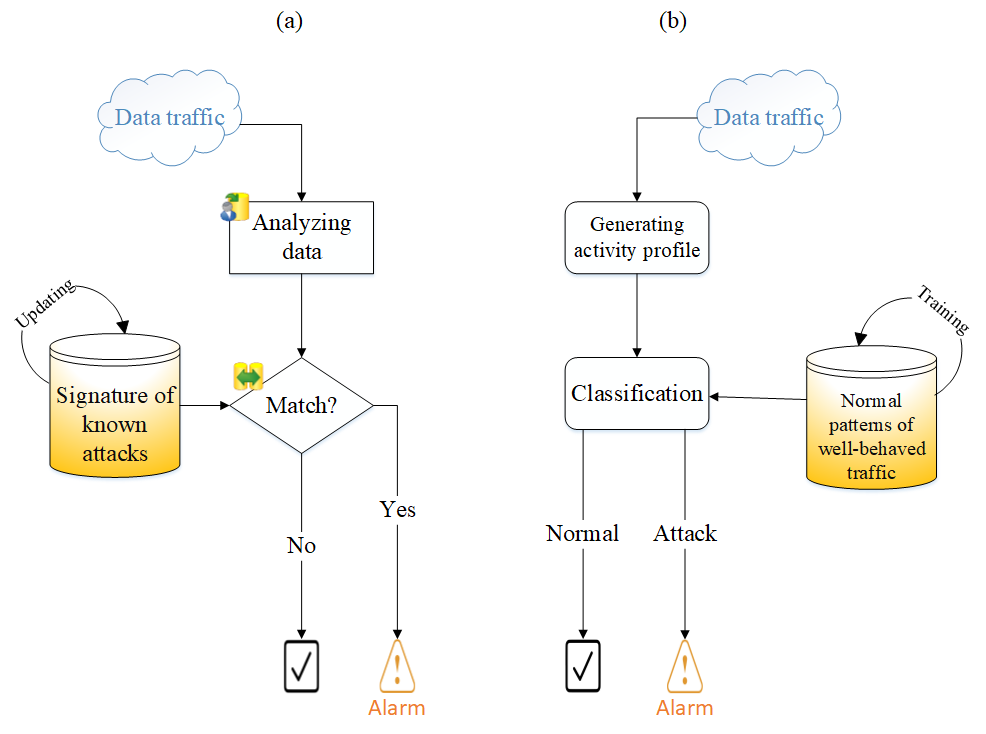}
\par\end{centering}
\caption{(a) Signatute-based detection; (b) Anomaly-based detection}
\label{intrusion}
\end{figure}
As shown in Figure \ref{intrusion}, there are two main classes of IDS including signature-based detection and anomaly-based detection. Recently published works \cite{miller2014survey}, \cite{sun2017attacks},
\cite{sharma2018survey} have discussed possible ways of securing vehicles from remote attack by assuming the defense strategy as a network intrusion detection problem. Recently, 3GPP has working in verifying that 5G
systems are able to utilise 256-bit symmetric cryptography mechanisms inherited from legacy 4G systems. The handover from one system architecture to the next must remain seamless.
As a matter of fact, legacy security visibility and configurability functionality are developing with technology evolution and in the future, devices will be more reactive and flexible to various security configurations. 

\subsubsection{{Signature-based Detection}}
This method first stores various existing signatures of known attacks in a database for retrieving and making a comparison. Then, it detects the intrusion attack by comparing oncoming
cases from the Internet of Vehicle with existing signatures of known attacks in store.

Bi{\ss}meyer et al. \cite{bissmeyer2010intrusion} developed a signature-based IDS that utilizes a plausibility model for vehicle movement data. The proposed scheme is able to
detect a single fake vehicle even if it uses a valid movement. Two kinds of attackers can be detected using the proposed algorithm: 1) a fake congestion attack; 2) a denial of congestion attack.

Tomandl et al. \cite{tomandl2014rest} introduced a novel IDS called REST-Net for Vehicular Ad hoc Networks (VANETs) to check fake messages. Different from previous solutions,
REST-Net uses a dynamic engine to analyze and monitor the data, and it achieves very high detection rates and adaptive warning levels in case drivers are interrupted. It is also
implemented with a concept which is used for recalling the fake message as long as an attacker is identified.

One of the disadvantages is that it usually causes high false-negative rates when facing unknown or new attacks. Another disadvantage is that signature-based detection fails to detect
intrusions with the development of onboard applications. For example, signature-based detection may be invalid sometimes as more and more extra devices such as sensors are integrated into vehicles.

\subsubsection{{Anomaly-based Detection}}
Anomaly-based detections predefines the baseline of normal cases, then new types of attack can be identified once they are observed to have abnormal information
 beyond the baseline \cite{sedjelmaci2014efficient}.

Cho and Shin \cite{cho2016fingerprinting} developed a Clock-based IDS (CIDS) for intrusion detection. CIDS constructs a baseline of ECUs' clock behaviors based on the thus-derived
fingerprints which are extracted from the intervals of periodic in-vehicle messages. Then, CIDS employs Cumulative Sum to detect any abnormal shifts (i.e. signs of intrusion)
in the identification errors. Experiments showed that CIDS could achieve a low false-positive rate of 0.055\%.

Martynov et al.\cite{martynov2007design} developed a software-based light-weight IDS based on properties selected from the signal database. Then, the authors
studied the message cycle time and the plausibility of the messages, and introduced two anomaly-based methods for IDS. Experiments were conducted in terms of
both simulation and real-world scenarios. Experimental results demonstrate that some malicious events such as injection of malformed CAN frames, unauthorized CAN
frames as well as Denial-of-Service (DoS) attacks can be recognized by the proposed IDS.

Sedjelmaci and Senouci \cite{sedjelmaci2014new} proposed a novel intrusion detection framework for a vehicular network (IDFV) utilizing detection and eviction
techniques. IDFV is implemented in two detection agents: a local intrusion-detection module and global intrusion-detection-module. Experiments demonstrate that IDFV exhibits
 a very high detection rate with more than 98\% and a low false-positive rate of lower than 1.3\%.


Song et al. \cite{song2016intrusion} proposed a light-weight intrusion-detection strategy by analyzing the time intervals of CAN messages. The authors first experimentally
show the difference between time intervals of messages in the normal status and in the under-attack status. Then, experiments were conducted based on the CAN messages from the
cars made by a famous manufacturer. The results show the effectiveness of the proposed method.

Zaidi et al. \cite{zaidi2016host} presented a statistical-technique-based IDS for anomalies and rogue-nodes detection using a traffic model. The proposed IDS can work independently
without relying on any infrastructure. In order to use the proposed mechanism, network message congestion is controlled to avoid broadcast storms. Experiments show that the proposed
IDS can keep the network working even if up to 40\% of nodes are malicious.

Lee et al. \cite{leeotids2017OTIDS} studied the offset ratio and time interval of message response performance, and proposed an intrusion-detection method. The proposed method
assumes that the receiver node will respond to the remote frame immediately once a particular identifier is transmitted. It also assumes that the response performance should be
different between an attack-free state and an attack state. In order to enhance the overall performance and accuracy of the proposed strategy, a novel algorithm is also proposed
to monitor the change of in-vehicle nodes. OTIDS can achieve very good performance without modifying the CAN protocol. Moreover, it can not only identify message injection
attacks and impersonating node attack but also can detect the types of messages in the injection attacks.


Yu et al. proposed a Presence Evidence System (PES) \cite{yu2013detecting}. PES is a statistical method for detecting Sybil attacks in VANETs. The authors have considered signal-strength distribution analysis
of vehicles to estimate their physical position because position verification is regarded as one of the best methods for the detection of Sybil attacks. When a claimer node broadcasts a
beacon message at a beacon interval for the purpose of neighboring discovery, an estimated position will be calculated for the claimer. The main idea is to improve estimating the position of a vehicle by
using a random-sample consensus (RANSAC)-based method. It should be mentioned that the RANSAC algorithm is a well-known learning method in the field of computer vision for outliers detection. The main
limitation of the PES, however, is that it cannot detect all Sybil attacks.

AECFV is an intrusion detection mechanism introduced by Sedjelmaci and Senouci \cite{sedjelmaci2015accurate}. In fact, AECFV includes
intrusion-detection systems at three levels: (1) cluster members level; (2) cluster heads (CH) level; (3) Road-Side Units (RSU) level. It should be noted that, along with a rule-based decision technique
and a trust-based scheme, AECFV makes use of a Support Vector Machine (SVM) as a machine-learning method to detect anomalies at the cluster level. As a matter of fact, in machine learning, SVMs are supervised
learning models that are used for solving classification problems, not to mention that SVMs are a major area of interest in the field of anomaly detection.  Furthermore, feature extraction, the training process, and
the classification process are the three main components of the SVM in the proposed model. Moreover, the detection mechanism of AECFV against different and various types of attacks on vehicular ad hoc networks has been
discussed. AECFV is expected to suit scenarios such as a lightweight communication overhead as well as fast attacks detection. However, inasmuch as AECFV needs to implement IDS on lots of vehicles as cluster
members, it causes a high overhead in large-scale vehicular networks. In the same way, other studies by Sharma and Kaul \cite {sharma2018hybrid}, Wahab et al. \cite{wahab2016ceap}, and Sedmalci et al.
\cite{sedjelmaci2017hierarchical} use SVMs for intrusion detection of vehicular networks.


\begin{table*}[htbp] %
\caption{ NEtwork security defences }
\centering
\begin{tabular}{p{3cm} p{5cm}  p{4cm} p{4cm}  }
\toprule
Method                                                                   & Key idea                                                              & Advantages                                                                            & Disadvantages \\
\midrule%
Bi{\ss}meyer et al. \cite{bissmeyer2010intrusion} &Uses position information of vehicles to detect message forgery&Additional hardware (radar, lidar, camera) are not required&Accuracy of GPSs, movements of vehicles, data transmission delays\\
REST-Net  \cite{tomandl2014rest}                         &Utilizes data plausibility checks to analyze and detect an attacker's fake messages&High detection rate, high adaptability&Ineffective against unknown attacks \\
CIDS \cite{cho2016fingerprinting}                         &Estimates clock skew of CAN messages to detect intrusion&Evaluation in real conditions, low false positive rate&CIDS cannot detect irregular time sequence attacks\\
Martynov et al.\cite{martynov2007design}   &Simulates anomaly-based detection against DoS attacks on wireless sensors&Protects nodes against unknown attacks&The simulation is performed on a limited number of fixed nodes \\
IDFV \cite{sedjelmaci2014new}                      &Uses rule-based, learninig-based, and trust-based techniques to detect malicious nodes &High detection rate, low false positive rate&Overhead of using many techniques\\
Song et al. \cite{song2016intrusion}              &analyzes the frequency of CAN messages to detect message injection and DoS attack&Lightweight&Cannot guarantee protection against other types of attacks\\
Zaidi et al. \cite{zaidi2016host}                      &Vehicles collect and analyze traffic information of other vehicles to train the IDS &High accuracy, low overhead&Dose not consider data communication attacks\\
OTIDS \cite{leeotids2017OTIDS}                    &Considers offset and time intervals of CAN messages to detect three types of attacks &accurate, low detection time&IDS cannot detect attacks with irregular \textit{remote frames} \\
PES  \cite {yu2013detecting}                                 &Uses RANSAC to improve estimating physical position of vehicles  & Robust estimation                                                                   & RANSAC can only estimate one model for a particular dataset  \\
AECFV \cite{sedjelmaci2015accurate}                   & Uses support vector machine at cluster heads                              & High detection rate, low false positive                                   & High overhead \\
Markovitz et al. \cite{markovitz2017field}              & Uses TCAM for CAN packets classification                                     & CAN packets are easy to represent as TCAMs, adaptable               & Not tested against different attack scenarios \\
PML-CIDS \cite{zhang2018distributed}                  & Uses DVP approach to decentralize a centralized machine learning approach & Decreases overhead, provides a certain degree of privacy, scalable    &Computational complexity \\\bottomrule
\end{tabular}
\label{tab:network}
\end{table*}


Recent developments in the field of machine learning have also led to a renewed interest in designing intelligent IDSs for in-vehicle anomalies. One study by Markovitz et al. \cite{markovitz2017field}
has involved designing a domain-aware anomaly detection system for the CAN traffic bus in which ternary content-addressable memories (TCAMs) have been used for detecting anomalies in CAN bus network
traffic. It must be mentioned that TCAMs are special types of high-speed memories that modern switches and routers use them for fast route lookup and packet classification. At first, in the learning
phase, TCMA learns how to classify CAN packets into three categories: constant, multi-value, counter/sensor. Then, in the testing phase, the TCMA classifier detects irregular messages that do not match
the trained model. The authors have evaluated the proposed scheme by simulated CAN bus traffic and also by real traffic data. It should be mentioned that TCMA is implementable in both
software and hardware.

Zhang et al. \cite{zhang2018distributed} proposed the idea of collaboration of vehicles in vehicular networks in order to manage a machine-learning scheme against malicious nodes.
The authors argued that privacy is a serious concern for the proposed approach, PML-CIDS, because vehicles may exchange sensitive information. PML-CIDS consists of
different parts: The pre processing engine is responsible for collecting and preprocessing data. The local detection engine is a logistic-regression classifier which is responsible for
intrusion detection by analyzing the preprocessed data and determining malicious activities. The P-CML engine is responsible for updating the classifier. The main philosophy of PML-CIDS
is decentralizing a centralized machine-learning approach. For solving this problem, a distributive optimization method called Alternating Direct Method for Multipliers (ADMM) has been
used to decentralize regularized empirical-risk-minimization (ERM) algorithms to achieve distributed training of large datasets. Moreover, PML-CIDS employs a privacy-preserving scheme of
regularized ERM-based optimization called dual-variable perturbation (DVP) which perturbs the dual variable of each vehicle at every ADMM iteration. It should be noted that PML-CIDS is
a distributive approach and decreases the overhead. In fact, with the huge size of training data, a centralized machine-learning approach can lead to communications overhead. Moreover, PML-CIDS
preserves privacy because vehicles do not have direct data communication \cite{zhang2017dynamic}.


The disadvantages of anomaly-based detection are: 1) it may cause high false-positive rates; 2) it is usually hard to prepare proper metrics to determine the baseline. However, it
is expected that data analysis techniques can help to improve the performance in the future.

\subsubsection{{Summarising  Network Security Defences}}
In this part of the paper, most popular network security defences of vehicular networks are compared. Table \ref{tab:network} lists network security defences and compares their merits and demerits.


\subsection{Software Vulnerability Detection}

\subsubsection{{Static Analysis}}
Static analysis is a set of program analysis methods to check and verify properties of program code without the need to execute it \cite{emanuelsson2008comparative,zheng2017security}.
It is helpful for finding structure errors and security vulnerabilities in a program so as to ensure the quality and security of software. Static analysis could detect software
run-time errors that are difficult to find by testing, including resource leaks, illegal operations like illegal arithmetic expressions, dead code, program termination problem, unreachable, and unusable data \cite{emanuelsson2008comparative}. Currently, static analysis methods are also applied to extract features of code for software
vulnerability detection using machine learning, which will be discussed in the following.

There are several common techniques of static analysis for securing software systems. Lexical analysis is a technique to scan the source code and transfer it to a token stream then
match the token stream against vulnerable constructs to find potential vulnerabilities \cite{mcgraw2004software}. Control-flow analysis is a technique to build the control-flow graph (CFG) of the program.
A node of the graph represents a basic code block and a directed edge represents a program execution path. Control-flow analysis could
illustrate the logic of the program and evaluate whether the program could terminate normally and which part of the code is dead. It is also the foundation of some other analysis
techniques. Control-flow integrity (CFI) is a technique that uses the normal CFG to detect attacks that aim to control the system by changing the normal behavior of software \cite{abadi2005control}.
Data flow analysis is a technique to collect run-time information about data in programs \cite{wogerer2005survey}. It traverses the CFG to record the initialize and reference blocks of variables
and their related data to construct data flow. The data flow could be used to classify the vulnerable variables, input and code segments. Taint analysis is a technique to identify whether variables
that accept input from an external user interface over the data flow could cause critical implications on the system.

There are several research projects on applying static analysis for detecting software vulnerabilities and defense attacks. \cite{tice2014enforcing} implemented the integration of CFI with the production
compilers GCC and LLVM. In their experiments, the proposed CFI mechanisms could protect 95\% to 99.8\% of indirect function calls of C++ programs. \cite{dahse2014static} proposed using static analysis
approaches to detect vulnerabilities in web applications. The authors refined the taint analysis by adding a persistent database store (PDS) to collect all data to detect second-order vulnerabilities
like second-order cross-site scripting and SQL injection. The experiment shows the proposed taint analysis identifying 159 previously unknown vulnerabilities. \cite{ding2017efficient} proposed a stronger
CFI called path-sensitive variation of CFI that could block more attacks. \cite{castro2006securing} proposed an implementation of data flow integrity enforcement that could detect attacks and errors of
C and C++ programs with low overhead. FindBugs \cite{ayewah2008using} is an open-source static analysis tool to look for bugs of a java program. Google uses the program to check the source code and build
code-review policies.

The main advantage of static analysis is that it does not execute the code so it has fast execution and high efficiency. On the other hand, for a large-scale software system, it is hard for a
human to review the code and find potential bugs carefully. The static analysis is able to detect vulnerabilities and errors automatically. However, a high false-positive rate is an important disadvantage
of static analysis, which means that it may report many vulnerabilities that are actually not vulnerable. Now some research projects try to combine static analysis and machine-learning techniques to solve the problem.

\subsubsection{{Dynamic Analysis}}
In contrast to static analysis, dynamic analysis depends on running the program to examine whether it has errors and vulnerabilities~\cite{zheng2017real}. The main two techniques of dynamic analysis are fuzzing and dynamic taint
 analysis.
 Fuzzing is a kind of automated approach that involves sending invalid or random data to the program and monitoring whether the program would crash or be in an error condition \cite{sutton2007fuzzing}.
 During this process, potential software vulnerabilities could be detected. Dynamic taint analysis is a technique that aims to analyze marked information flow when the program is executed, and 
this method could detect most of the software vulnerabilities \cite{ji2018coming}.

Fuzzing is a proactive method to find software flaws. In addition, it is also applied to quality-assurance processes by monitoring the product lifecycle and product maturity \cite{takanen2008fuzzing}.
There are three categories of fuzzing. The first one is black-box fuzzing that does not analyze the program and just generates random data as input. In the research of Godefroid et al.
\cite{godefroid2008automated}, the author used black box fuzzing tools to find more than 20 previously unknown vulnerabilities of a Windows software. White-box fuzzing is another one that
needs to analyze the program to get information about the program first. It could generate more specific input but the efficiency is not good. Combined the advantages of black-box and white-box
fuzzing, grey-box fuzzing was proposed, which is a technique that could get information of the internal structure about the program without program analysis.

Dynamic taint analysis was proposed by James Newsome and Dawn Song \cite{newsome2005dynamic}. High detection rate for the overwrite attacks and low false positive are the most important advantages of
the technique. The program value that is from a taint source is called tainted. Dynamic taint analysis could track how taint flows during the execution of the program to detect potential vulnerabilities.
There are two challenges of dynamic taint analysis including over-tainting and under-tainting. Over-tainting means the marked tainted value is not from a tainted source and under-tainting means some
information flow was missed \cite{schwartz2010all}. A precise dynamic taint analysis system should avoid those problems. Now Dytan is a general dynamic taint analysis framework that is used widely to
detect software vulnerabilities \cite{clause2007dytan}.

\subsubsection{{Software Testing}}
\label{sec:Softwaretesting}
Comparing with static analysis, dynamic analysis is more accurate to find vulnerabilities because it could analyze run-time information. Another advantage is that dynamic analysis usually does not need
source code. However, dynamic analysis has less code coverage. Currently, the hybrid analysis that combines static analysis and dynamic analysis is an available approach to find software vulnerability
and secure software system better.



Intelligent vehicles including aircraft, airplane, and car have million lines of code (Table \ref{tab:size}) and software is responsible for many safety-critical functions of the vehicle.

\begin{table}

\caption{software size in intelligent vehicles}
\centering
\begin{tabular}{p{2cm} p{2cm} c}
\toprule
Manufacturer                                                                   & Model                                                              & Software Size (lines of code)               \\
\midrule%
Boeing                                                                             & 787                                                                   & 14 million                          \\
Lockheed                                                                        &F-22                                                                    & 8 million                                        \\
Mercedes-Benz                                                               &S series                                                                       &20 million                                        \\
Ford                                                                                & GT                                                                       &10 million                                         \\
Ford                                                                                & CES 2016                                                             & 150 million \\
\hline
\end{tabular}
\label{tab:size}
\end{table}

\begin{figure}
\begin{centering}
\includegraphics[width=0.9\columnwidth]{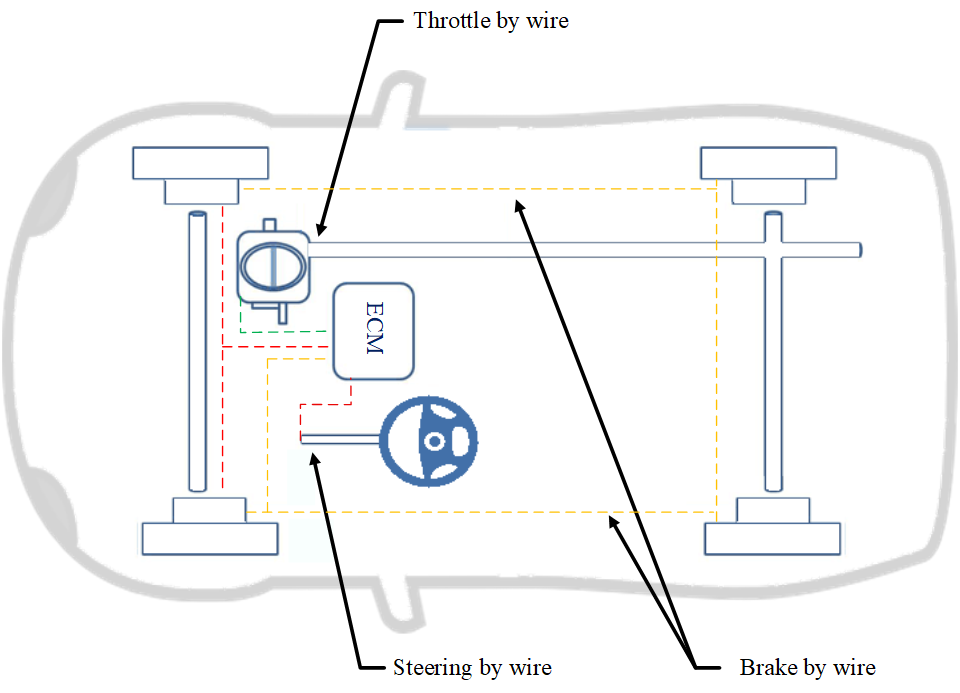}
\par\end{centering}
\caption{X by wire}
\label{Xbywire}
\end{figure}

Drive-by-wire, brake-by-wire, suspension-by-wire and in general X-by-wire (Figure \ref{Xbywire}) in automotive industry refers to the use of fully electric/electronic systems for performing vehicle functions, such as braking or steering, instead of mechanical or hydraulic systems \cite{sinha2011architectural}. Since these systems are highly safety-critical, they must comply with safety standards such as ISO 26262. Established traceability is one of the requirements of these standards to ensure well tested and safe resulting systems \cite{maro2018software}.
In this part of the paper, we will review mutation testing as one of the major software testing methods. As a matter of fact, this technique has been used in many industrial practices for testing software and securing intelligent vehicle systems. 
Mutation testing is a powerful software testing for evaluating and improving software programs. For this purpose, small defects are artificially introduced to the original program to create mutants and then, test cases are applied into the original program as well as the mutants. The outputs are compared and when the results for the original program and mutant program are different, the mutant is detected (killed). A test case considered a good one if it can distinguish between the behavior of the original version and the mutants \cite{deng2017mutation}.

\begin{table*}[htbp] %
\caption{Software vulnerability detection defences }
\centering
\begin{tabular}{p{3cm} p{5 cm}  p{4cm} p{4.0cm}  }
\toprule
Method                                                                   & Key idea                                                              & Advantages                                                                            & Disadvantages   \\
\midrule%
Tice et al. \cite{tice2014enforcing}  & Uses VTV and IFCC techniques to prevent unintended control transfers in the CFG&Low overhead&It is not practical when attackers chain edges together\\
Dahse and Holz \cite{dahse2014static}&Static analysis of storage writings and readings to detect second order vulnerabilities&Static analysis is fast for finding and fixing vulnerabilities&Static analysis cannot pinpoint all run-time troubles\\
PITTYPAT\cite{ding2017efficient}          & Implements a path-sensitive CFI that transfers control only to valid target points&increases precision of CFI mechanism, acceptable run-time overhead&It is dependent on specific hardware, not portable \\
DFI \cite{castro2006securing}                      &Creates runtime definition table (RDT) to guarantee data-flow integrity  &Detects a broad class of attacks&It makes the programs slower\\
FindBugs \cite{ayewah2008using}                            &Reviewes Findbugs as a static analysis tool in Google's testing and code review process &Simplicity&Does not guarantee detection of complicated vulnerabilities\\
Generational Search \cite{godefroid2008automated} &Combines mutation-based fuzz testing approach with symbolic execution &Covers all paths of the program&Overhead of path explosion in symbolic execution\\
TaintCheck \cite{newsome2005dynamic}& labels data from untrusted sources and tracks them during program execution time&High detection rate, low false positive rate&Over-tainting, under-tainting\\
Dytan \cite{clause2007dytan}    &Considers not only data-flow tainting but also control-flow tainting&Generality, high detection rate, low false positive rate &Over-tainting, under-tainting\\
GenProg \cite{le2012genprog}&Uses genetic algorithm approach in order to find program bugs &Low overhead, low false positive rate&Convergence is not guaranteed\\
Shin and Williams \cite{shin2011evaluating}   &Explores the relationship between three software product metrics and vulnerability&Provides an estimation of vulnerable locations&Does not guarantee software vulnerability detection\\
Perl et al. \cite{perl2015vccfinder} &Employs SVM classifier in order to detect suspicious commits&Generality, scalability, explainability&Difficulty of choosing good kernel functions for SVMs\\
Zhou and Sharma \cite{zhou2017automated}           &Uses a group of machine learning classifiers to detect unidentified vulnerabilities&independent of programming languages, Scalability& Low recall rate\\
Shar et al. \cite{shar2015web}   &Uses supervised and semi-supervised learning to detect web application vulnerabilities& Highly scalable&The complexity of feature selection\\
VDiscover \cite{grieco2016toward}    &Trains three machine learning classifiers to detect memory corruption vulnerabilities&Scalability&Low accuracy \\\bottomrule

\end{tabular}
\label{tab:software}
\end{table*}

GenProg \cite{le2012genprog} is a genetic method for automatic software repair that takes a program with a defect and a set of failing negative and passing positive test cases to produces a 1-minimal repair. The main theoretical premise behind using a genetic approach is to maximize a fitness function which evaluates the acceptability of a program variant. It should be noted that new program variations (variants) play the role of the chromosomes. Genetic representation of the proposed genetic algorithm is an abstract syntax tree which includes all of the statements in the program and a weighted path which is a sequence of \textless statement, weight\textgreater\ pairs. Moreover, crossover and mutation operators are utilized for creating new variants of the mating pool. Based on the fitness values of variants, the selection function selects the next generation members.

\subsubsection{{Machine Learning}}
In this part of the paper, a number of significant and well-known machine-learning approaches for software vulnerability detection are reviewed and compared with each other. Traditional approaches to software vulnerability detection are based on expertise. The work is tedious and achieves a high false-negative rate \cite{li2018vuldeepecker}. There are two available ways to apply machine-learning and deep-learning methods for software vulnerability detection: Supervised learning approaches create a vulnerability prediction model and a vulnerable code pattern recognition system using a labeled dataset \cite{ghaffarian2017software}. For different methods, the key difference is the feature selection. For the software vulnerability prediction model, software metrics that are degrees of some property that is relative to the software are used as features to train machine-learning and deep-learning models. For a software vulnerability-pattern-recognition system, features are extracted from software source code using traditional static and dynamic program analysis methods.

Shin and Williams \cite{shin2011evaluating} investigated a software-vulnerability-prediction model using the complexity of software code, code-churn and developer-activity metrics as a feature set. They proposed 28 software metrics including 3 code-churn metrics, 11 developer-activity metrics, and 14 complexity metrics as features to train a logistic-regression model. Those metrics are extracted from each program
code file. Therefore, the prediction model could predict which file is vulnerable. They conducted experiments on two projects and found that models using the development-history metrics could achieve over
70\% recall and less than 25\% false-positive rate so they concluded that development-history metrics are strongly relevant to vulnerability prediction.

Perl et al. \cite{perl2015vccfinder} explored using commit information metrics as features to predict which commits of open-source projects are vulnerable. They proposed that a commit submitted by a new
contributor more likely has vulnerabilities, bigger commits may have vulnerabilities and commits that modify more developers' work could introduce vulnerabilities. They extracted relative metrics from
commits to train an SVM software-vulnerability-prediction model called VCCFinder. They compared their model with a traditional static-analysis tool FlawFinder. The result showed that, when the two tools
achieved the same recall, VCCFinder had 60\% precision while FlawFinder had only 1\% precision.

Zhou and Sharma \cite{zhou2017automated} investigated both commits and bug-reports metrics for vulnerability-prediction models. They chose commit messages as commits metrics and title, description,
comments, comment number, attachment number, labels, created date, and last edited date as bug-reports metrics. word2vec was applied to transfer those metrics to vector features. They trained SVM,
random forest, Gaussian naive Bayes, K-nearest neighbors, AdaBoost and gradient boosting models and used a K-fold stacking algorithm to ensemble those models to improve the result. The experiment
result showed the model achieved an 83\% precision rate and a 74\% recall rate.

Shar et al. \cite{shar2015web} proposed a vulnerability pattern-recognition system to find vulnerable code that may cause SQL injection, cross-site scripting, remote code execution and file inclusion
in a PHP web application. They focused on code attributes named input validation and sanitization (IVS) attributes on the software function. Static program analysis was applied to find features that
could indicate whether a function needed to receive inputs from different entries such as File and client HTTP requests.  control flow graph (CFG), program dependence graph (PDG) and system dependence
graph (SDG) were applied to find features that could characterize program functions and operations. For instance, the PHP function str\_re- place('<', ' ', \$input) removes HTML tags from the input.
In fact, the presence of HTML tags in \$input could cause XSS, the function has a security property that filters HTML tags to  prevents XSS. Then a feature called ‘HTML-tag’ was proposed to indicate
whether the program function filters HTML-tags. Finally, the paper proposed 32 attributes related to input validation and function operation as features. If the function contains an attribute, the
corresponding feature value is 1. Therefore, the features are vectorized as the input of machine-learning models. Random forest was applied as the machine-learning classifier of the vulnerability-recognition
system. Results showed that the system achieved the best performance on finding SQL injection (92\% recall and 82\% precision rate) and worst performance on remote-code-execution prediction
(64\% recall and 55\% precision).

Grieco et al. \cite{grieco2016toward} explored using machine learning to detect memory-corruption vulnerability of C programs. They applied static and dynamic analysis on binary code rather than source
code. Static analysis was used to extract features relative to the call sequence of standard C libraries, and dynamic analysis was used to capture features that show the behavior of a program when it
runs along sequential calls to the C standard library. Logistic regression, random forest and Multiple Layers Perceptron (MLP) were applied in their experiments. The result showed that the model using
random forest trained by dynamical features achieved the best performance (55\% true positive rate and 17\% false-positive rate).

\subsubsection{{Summarising Software Vulnerability Detection Defences}}
Table \ref{tab:software} briefly illustrates the comparison of software vulnerability detection defences in vehicular systems.

The main advantage of the software-vulnerability-prediction model is that it does not care about the specific software code and its implementation but just software metrics that are easier to collect and
analyze. Any software metrics could be studied, whether they are related to software vulnerability so it is a clear research direction. However, currently the software metrics features are extracted on the
file or commit level, which means that the prediction model cannot find the specific position where vulnerabilities are. Another problem is that the software prediction model using software metrics is totally
based on the statistical relationship between metrics and vulnerabilities, which may cause a high false-negative rate. Vulnerability recognition systems usually apply traditional program analysis techniques
to extract features from software source code or binary code. The feature extraction process is more complicated than the  vulnerability prediction model. The advantage of the vulnerability recognition system is that
it could find the vulnerabilities more precisely.

\subsection{Malware Detection}
\begin{figure}[]
\begin{centering}
\includegraphics[width=0.9\columnwidth]{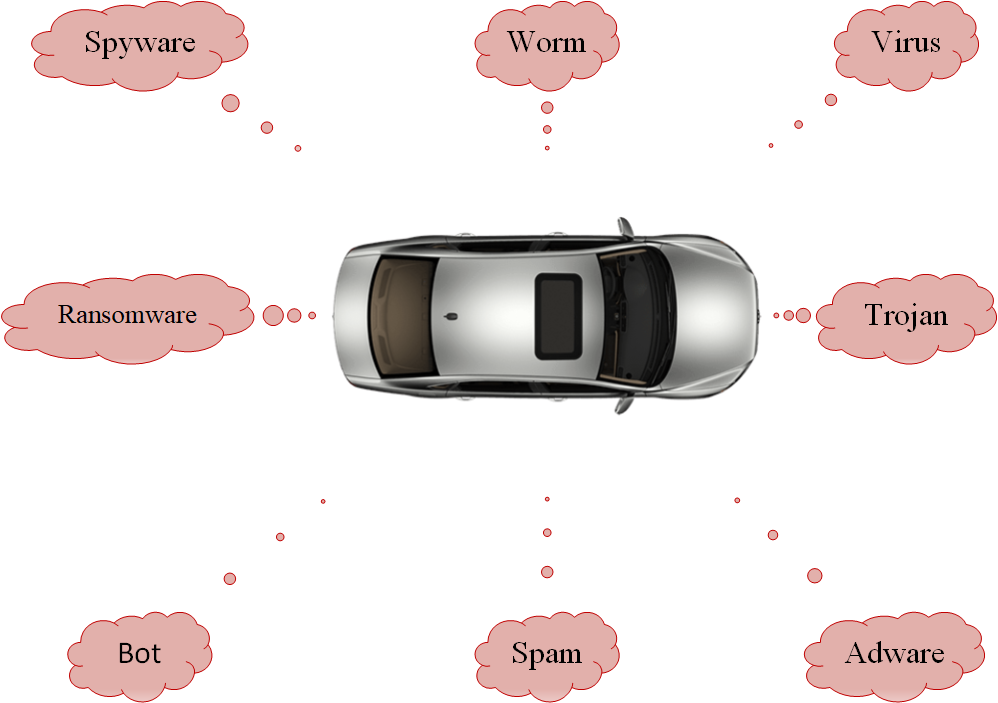}
\par\end{centering}
\caption{Various types of malware attacks on intelligent vehicles}
\label{malware}
\end{figure}

Whenever something new and as complex as an intelligent car or truck connects to the internet, it is exposed to the full force of malicious activity. As depicted in Figure \ref{malware}, leaving an attack surface unprotected will expose
vehicles to many
security risks including malware and trojans \cite{MCAFEE}.
 In this section, a number of significant and well-known machine learning approaches for malware detection are reviewed and compared with each other.

\begin{table*}[htbp] %
\caption{malware detection  defences  }
\centering
\begin{tabular}{p{3 cm} p{5 cm}  p{4cm} p{4.0cm}  }
\toprule
Method                                                                   & Key idea                                                              & Advantages                                                                            & Disadvantages   \\
\midrule%
MSPMD \cite{fan2016malicious}   & Uses a modified version of the k-nearest neighbor algorithm for malware detection      & High detection rate   &k-NN is a lazy learning algorithm   \\
MRMR-SVMS \cite{huda2016hybrids}   & Uses the combination of SVM wrapper with MRMR filter        & Low false positive rate, low false negative rate   &Collecting API calls is time-consuming   \\
Huda et al. \cite{huda2017defending} &Uses semi-supervised technique with unlabeled data for dynamic feature extraction &Significant performance improvement &Complexity of database update procedure \\
CloudIntell \cite{mirza2017cloudintell}                  &Computation offloading using SVM, decision tree and boosting on decision tree    & Energy efficiency, high detection rate      &Continuous connectivity is required  \\\bottomrule

\end{tabular}
\label{tab:malware}
\end{table*}

Fan et al. \cite{fan2016malicious}  proposed a malicious-sequential-pattern mining approach for automatic malware detection, called MSPMD. The major objective of MSPMD was to investigate a
malware-detection method to recognize new, unseen malicious executables. MSPMD emphasizes the order of extracted features before classification to discover malicious sequential patterns. At
first, PE files are disassembled to the machine instructions by the C32Asm (2011) disassembler, followed by parsing the instructions by ignoring the operands and encoding each operator to an
instruction ID. Next, in order to reduce the useless information, MSPMD selects instructions according to their frequency in malicious and benign files (feature selection). Then, the authors
proposed a Malicious Sequential Pattern Extraction (MSPE) algorithm for discovering malicious sequential patterns. In fact, MSPE is a modification of the Generalized Sequential Pattern (GPS) algorithm
\cite{srikant1996mining}. Last, MSPMD makes use of All Nearest Neighbor (ANN) as a machine-learning algorithm to identify malware automatically. All in all, the detection accuracy of MSPMD is 95\%.

Huda et al. \cite{huda2016hybrids} proposed a combination of SVM wrapper and filtering solution for malware detection. The main theoretical premise behind using
 an anomaly-based approach is that detecting unknown malware is not possible using signature-based techniques. In other words, signature-based methods have a high false-positive rate against
 code-obfuscation techniques such as encryption, packing, polymorphism, and metamorphism. This scheme is intended to extract application programming interface (API) calls from portable
 executable (PE) file formats such as (.EXE and .DLL) and then analyze them to understand their use for malicious purposes in order to identify malicious code. In this way, the authors have
 combined a Maximum Relevance Minimum Redundancy filter with SVM score (MRMR-SVMS). As a result, the MRMR-SVMS approach extracts more informative API features than either wrapper or filter alone.
 In other words, the proposed approaches integrate knowledge obtained by the filter from properties of data and the predictive accuracies of trained classifiers. In the final analysis, the
 accuracy of MRMR-SVMS is 96\%.

Huda et al. \cite{huda2017defending} have also investigated how to automatically integrate knowledge about unknown malware from unlabeled data by a semi-supervised approach. In fact, the proposed
 method is an integration of supervised and unsupervised learning. The authors have addressed the problem of antivirus (AV) engines which need to be updated regularly because most malware
 has inbuilt code that can generate new variants dynamically each time it is executed. Therefore, an unsupervised approach automatically extracts dynamic changes in malware variants using a sandbox
 environment and automatically updates the database. It should be noted that log files of the sandbox environment indicate API function calls of executable files dynamically. Furthermore, different
 supervised detection engines (Classifiers) such as Random forest, SVM, J48, Naive Bayes, and IB are trained with labeled data for detecting malware. In order to integrate unsupervised and supervised
 methods, the authors have used the term frequency (TF), inverse document frequency (IDF), and cosine similarity as measures of distance in clustering unlabeled files.

CloudIntell \cite{mirza2017cloudintell} is an intelligent malware detection system which has been designed using a combination of machine-learning algorithms for predicting any previously unknown malicious
activity and computation offloading for reducing the resource and power consumption of the client machine. As a case study, the authors have employed Amazon's EFS (Elastic File System), SQS
(Simple Queue Service) and EC2 (Elastic Compute Cloud) to store a large amount of benign and malicious files as malware repository, request-response queues, and detection engines respectively. Moreover,
malware repository files have been used for training and testing the machine-learning detection methods. After feature-extraction by an automated feature extracting tool, for the purpose of malware
detection, CloudIntell examines three types of machine-learning classifiers: Support Vector Machine (SVM), decision tree, and boosting on a decision tree. It must be remembered that continuous
connectivity between client and server is an important challenge of cloud computing. In the evaluation of CloudIntell, boosting on decision tree classifier indicated a better performance regarding ROC curve analysis.
A comparative analysis of malware detection defences is conducted in Table \ref {tab:malware}. It demonstrates the key features and points of malware detection defences, and also their merits and demerits.

\subsection{Comparision of Existing Security Defence Mechanisms}

Security defences in intelligent vehicles are developed to protect the in-vehicle communications and communications between vehicles. The evaluation of a solution is required to know whether a particular solution has achieved its aims or not. The evaluation will reveal the effectiveness of security defences against malicious attacks. The challenge here is the fact that
security defences are developed under different deployment configurations, which complicates the process of comparison. Herein, the experimental overview of above-mentioned existing security defences are compared
with one another.


\begin{table*}[htbp]  %
\label{tab:security solutions}
\caption{Experimental overview of main existing defence mechanisms}
\centering

\renewcommand{\arraystretch}{1.40}

\setlength{\tabcolsep}{12pt}
\begin{tabular}{c l c c c }

\toprule
\rowcolor{lightgray}
Category                                                                                &  Security solutions                                                                  &   Performance metrics  & Evaluation size & Evaluation environment \\
\midrule%

\multirow{15}{*}{\rotatebox[origin=c]{90}{Cryptography}}&2FLIP \cite{2flip}                      &  Delay, packet loss ratio, overhead                                                          &           0-90 vehicles         &         ProVerif         \\
&TESLA \cite{tesla}                                                                                             &              N/A                                              &            N/A          &       N/A         \\
&RAISE \cite{RAISE}                                                                                           &   Message loss, delay, overhead                                                          & 30-200 vehicles                   &     ns-2           \\
&PACP \cite{huang2011pacp}                                                                             &           Latency                                                  &      1-100 vehicles                &         C++     \\
&ECDSA \cite{ecdsa}                                                                                           &Latency                                                               &100000 test vectors              &Verilog\\
&SA-KMP \cite{sakmp}                                                                                                       &        Delay, overhead                    &      100-1000000 users                 &    C           \\
&GKMPAN \cite{gkmpan}                                                                                               &                Transmission overhead                                             &           75 nodes           &     N/A        \\
&PPAA \cite{ppaa}                                                                                                              &                  Fairness                                            &       N/A              &         N/A      \\
&Calandriello et al. \cite{Calandriello}                                                                              &           Overhead                                                  &         N/A             &  C              \\
&PPGCV \cite{ppgcv}                                                                                               &                             Transmission overhead                                &          5km highway            &      N/A       \\
&TACKs \cite{tack}                                                                                                   &                        Overhead, certificate update                                              &    $3000\text{m} \times $3000m     &     ns-2         \\
&GSIS \cite{GSIS}                                                                                                     &                                         Delay, message loss ratio                      &  $1000\text{m} \times $1000m      &        ns-2         \\
&SRAAC \cite{sraac}                                                                                                 &                     Overhead (bandwidth,  memory)                                        &         N/A               &     N/A             \\
&DABE \cite{att4}                                                                                                       &                                 Overhead                            & 10 nodes          &        VanetMobiSim        \\
&ABACS \cite{abacs}                                                                                                  &                                  Overhead, delay                            &       $1000 \text{m}\times $1000m                 &     ns-2           \\
&Xia et al. \cite{att5}                                                                                               &                                       Overhead                       &    50 vehicles                  &     Real vehicles      \\
&Bouabdellah et al. \cite{att6}                                                                                 &        N/A                                                      &             N/A         &         N/A       \\

\arrayrulecolor{gray}\hline
\multirow{13}{*}{\rotatebox[origin=c]{90}{Network Security} \newline \rotatebox[origin=c]{90}{}}
&Bi{\ss}meyer et al. \cite{bissmeyer2010intrusion}                                                 &                       DR                                 &       10 vehicles               &        VSimRTI, JiST/SWANS        \\
&REST-Net  \cite{tomandl2014rest}                                                                        &                              FPR, TPR, TNR                                &              8000 vehicles        &         VANET-Simulator       \\
&CIDS \cite{cho2016fingerprinting}                                                                         &                         FPR                                     &           3 vehicles           &            Real vehicles     \\
&Martynov et al.\cite{martynov2007design}                                                            &                             DR                                 &        5 nodes              &     Tmote Sky, MoteIv           \\
&IDFV \cite{sedjelmaci2014new}                                                                             &                            DR, FPR, DT                                  &        50-300 vehicles              &     NS-3           \\
&Song et al. \cite{song2016intrusion}                                                               &                             FPR                                 &            N/A          &   Real vehicles             \\
&Zaidi et al. \cite{zaidi2016host}                                                                     &                       DR, FPR, overhead                                       &             150, 300 vehicles         &       OMNET++, SUMO         \\
&OTIDS \cite{leeotids2017OTIDS}                                                             &               Reply ratio, correlation coefficient                                                 &            1 vehicle          &        Real vehicles        \\
&PES  \cite {yu2013detecting}                                                                           &                   DR, FPR                                    &  100 nodes           &      ns-2, GrooveNet          \\
 &AECFV \cite{sedjelmaci2015accurate}                                                                &                        DR, FPR, DT, overhead  & 50-300 nodes     &     NS-3              \\
&Markovitz et al. \cite{markovitz2017field}                                                           &                      FPR                                     &       10-500 packets       &              ECU simulator, Python      \\
&PML-CIDS \cite{zhang2018distributed}                                                              &     FPR, FNR                                                      & 4, 8, 16 nodes              &          N/A        \\
\arrayrulecolor{gray}\hline

\multirow{13}{*}{\rotatebox[origin=c]{90}{Software Vulnerability} \newline \rotatebox[origin=c]{90}{Detection}}&Tice et al. \cite{tice2014enforcing}       &     Overhead                                 &  142778 instructions          &     SPEC CPU, Chromium            \\
&Dahse and Holz \cite{dahse2014static}                                        &                  TPR, FPR, FNR                                                   &   143 KLOC            &   PHP                 \\
&PITTYPAT\cite{ding2017efficient}                                                &                     Overhead                              &   122 KLOC            &        SPEC CPU, LLVM 3.6.0               \\
&DFI \cite{castro2006securing}                                                      &                                  FPR, Overhead, throughput                &    80 clients        &          SPEC CPU, SPEC WEB           \\
&FindBugs \cite{ayewah2008using}                                               &           N/A                                       &       N/A                        &          Java         \\
&Generational Search \cite{godefroid2008automated}                  &       DT          &    197011 bytes          &           SAGE              \\
&TaintCheck \cite{newsome2005dynamic}                                      &                                 FPR, overhead                &     1KB-10MB page size       &        Valgring, DynamoRIO              \\
&Dytan \cite{clause2007dytan}                                                        &                     FPR                                       &    10640 attacks           &       C          \\
&GenProg \cite{le2012genprog}                                                      &                       FPR, FNR                                   &        1.25 mLOC       &       C          \\

&Shin and Williams \cite{shin2011evaluating}                                           &                           Complexity, code churn&    11 releases            &      SAS v9.1.3, Weka v3.6.0           \\
&Perl et al. \cite{perl2015vccfinder}                                                                       &                                      FPR, precision                     &           170860 commits    &          CVEs        \\
&Zhou and Sharma \cite{zhou2017automated}                                                      &                                          Precision, recall rate                 &        3000000 commits       &         GitHub, JIRA, Bugzilla     \\
&Shar et al. \cite{shar2015web}                                                                             &                                         FPR, recall rate                   &     7 php applications           &         php miner         \\
&VDiscover \cite{grieco2016toward}                                                                    &                            FPR, TPR                                &        138308 executions        &            VDiscover      \\

\arrayrulecolor{gray}\hline
\multirow{4}{*}{\rotatebox[origin=c]{90}{Malware} \newline \rotatebox[origin=c]{90}{  }\newline \rotatebox[origin=c]{90}{Detection}}&MSPMD \cite{fan2016malicious}     &                                 DR, FPR, FNR, accuracy               &        10307 files       &          Jave Development Kit          \\
&MRMR-SVMS \cite{huda2016hybrids}                                                                        &                                           Accuracy                  &              66703 files    &        IDA Pro., Python       \\
&Huda et al. \cite{huda2017defending}                                                                       &                                                    FPR, accuracy, ROC curve         &       1452 files          &      IDA Pro., Python           \\
&CloudIntell \cite{mirza2017cloudintell}                                                                      &               ROC curve                                              &              500 files      &     MATLAB             \\
\hline


\end{tabular}
\label{table:comparision}
\end{table*}

Generally, the challenges of the current main defence mechanisms are as follows:
\begin{itemize}

\item \textbf{Cryptography:} The main disadvantage of symmetric encryption is the problem of key transportation. This problem is solved in asymmetric encryption and exchanging keys are not required. Thus, asymmetric encryption reduces the overhead. Moreover, they can provide undeniable digital signatures. However, public-key encryption is not fast enough and it uses more computer resources. In the same vein, attribute-based encryption, that is a suitable mechanism for dynamic networks, is a type of public key encryption with the same drawbacks. Provided that, the key problem with presented cryptographic methods is that they do not meet the performance requirement of VANETs. In other words, vehicles need light and real-time data transmission out of cities because of their fast-movement nature. On the other hand, over any congested area of the city with a traffic load of more than 100 vehicles in communication range, storing and computation of encrypted messages is really challenging \cite{wang20162flip}.
\item \textbf{Network security:} The main disadvantage of signature-based detection mechanisms is that they cannot detect unknown attacks. In contrast, anomaly-based detection methods have an advantage over signature-based methods for detecting unknown attacks, but the difficulty of defining the rules is the main challenge of anomaly-based detection methods.
\item \textbf{Software vulnerability detection:} Static analysis refers to detecting possible defects in an early stage of software, before running the program. On the other hand, dynamic analysis is based on executing programs. Static analysis is usually time-consuming and finding trained professionals for dynamic testing is difficult. Automated machine learning techniques mitigate the problem. However, machine learning techniques for software vulnerability detection are not accurate enough and more accurate defences are required. In separate regard, since X-by-wire systems are highly safety-critical, they must comply with safety standards such as ISO 26262. Moreover, software testing techniques must guarantee the traceability of an artifact \cite{maro2018software}. Thus, a more intelligent learning defence for an accurate and precise software vulnerability detection is of great importance.
\item \textbf{Malware detection:} Although using machine learning techniques for malware detection have improved detection rate, it is still very difficult to detect all evasive malware by traditional anti-malware strategies and more intelligent strategies are required.
\end{itemize}

The above-mentioned security solutions have been proposed to mitigate malicious attacks and increase the security of vehicles. 
Table \ref{tab:attackdefence} associates identified attacks with the defence mechanisms. It is apparent from this table that cryptographic and intrusion detection techniques are recognized as well-known and popular defence mechanisms for protecting intelligent vehicles while not enough attention has been paid to other defences such as software vulnerability detection. As mentioned in this section, software size in intelligent vehicles are growing dramatically, and therefore software vulnerability detection and malware detection techniques for protecting software in vehicles require particular attention. 
In general, a specific defence mechanism is not adequate. For instance, DoS attack on CAN bus is very different from Dos attack on wireless vehicular communications. Therefore, a systematic approach is needed which integrates complimentary defence mechanisms. Cryptographic approaches are usually employed for protecting wireless communications between road-side units and vehicles in VANETs. Network security techniques are appropriate for protecting ECUs as well as intrusion detection in wireless communications. Software vulnerability detection techniques are suitable for testing and analysis of software before installation on the vehicle while malware detection techniques protect them against malware after installation.

\begin{table*}
\caption{Security defences against security attacks  }
\centering
\renewcommand{\arraystretch}{1.40}
\begin{tabular}{p{4 cm} c  c  c c  }
\toprule
                                                          &Cryptography          &Network Security             &Software Vulnerability Detection          &Malware Detection              \\
\midrule%
DoS                                                    &        \checkmark        &   \checkmark                   &         \checkmark                             &                                           \\
DDoS                                                  &\checkmark              & \checkmark                      &                                                        &                                        \\
Black-hole                                           & \checkmark             &   \checkmark                    &                                                             &                                               \\
Replay                                                &     \checkmark         &                                          &                                                            &                                                 \\
Sybil                                                    &     \checkmark         &    \checkmark                    &                                                             &                                                \\
Impersonation                                    &     \checkmark         &    \checkmark                     &                                                           &                                                   \\
Malware                                              &     \checkmark         &                                           &  \checkmark                                       &       \checkmark                         \\
Falsified Information                           &    \checkmark          &        \checkmark                  &                                                          &                                                    \\
Timing                                                  &    \checkmark          &                                           &                                                          &                                                     \\\hline
\end{tabular}
\label{tab:attackdefence}
\end{table*}

A comparative analysis of all security methods and strategies is briefly illustrated in Table \ref{table:comparision}. In order to compare these solutions, the following evaluation criteria have been used:
\begin{itemize}
\item {Performance metrics}
\item {Evaluation size}
\item {Evaluation environment}
\end{itemize}
The evaluations of the proposed approaches have been performed regarding performance metrics. The main performance metrics that are addressed for mitigating security attacks are as follow:
detection rate, detection time, false positive rate, false negative rate, true positive rate, true negative rate, precision, recall rate, packet delivery ratio, bit error rate, and overhead.

\begin{itemize}
\item {(a) Detection rate (DR):} This parameter indicates the ratio of detected malicious vehicles to the total number of malicious vehicles. As a matter of fact, detection rate plays an important role in mitigating cybersecurity attacks on vehicles.
\item {(b) Detection time (DT):} This metric indicates the required time of the security mechanism to detect malicious vehicles.
\item {(c) Packet delivery ratio (PDR):} It is defined as the ratio of the number of sent packets to the number of received packets. PDR is usually used to give an estimation of the link quality.
\item {(d) False positive rate (FPR):} False positive rate refers to the possibility of an alarm when there is no attack. False positive rate is one of the most widely used parameters in evaluating security solutions against attacks. False negative rate (FNR), true positive rate (TPR), and true negative rate (TNR) are other similar parameters. In addition, accuracy, precision and recall rate are defined based on  FPR, FNR, TPR, and TNR. Table \ref{formulas} indicates the formulas for calculating them \cite{sun2018data}. 
\item {(e) Throughput:} It is measured in bits per second (bps), megabits per second (Mbps) or gigabits per second (Gbps) and indicates the amount of data travel through a channel. Throughput is a common metric for evaluating medium access channel (MAC) protocols.
\item {(f) Bit error rate (BER):} This metric is defined as the ratio of the number of errors to the total number of bits sent. The possibility of errors in a data transmission channel due to noise or interference is measured by this parameter.
\item {(g) Overhead:} This parameter indicates the amount of resource that is required for performing a specific task or solution.  For instance, transmission overhead and storage overhead are among important parameters for performance evaluation of different defence mechanisms.
\end{itemize}

\begin{table}
\caption{formulas for calculating accuracy, precision, and recall rate}
\centering
\begin{tabular}{p{2cm} c}
\toprule
Performance metric                                                                                                                                     & Formula               \\
\midrule%
Accuracy                                                                                                                      &    (TP+TN)/(TP+TN+FP+ FN)                      \\
Precision                                                                                                                       &       TP/(TP+FN)                                  \\
Recall rate                                                                                                                     &          TP/(TP+FP)                             \\
\hline
\end{tabular}
\label{formulas}
\end{table}

Future research on securing intelligent vehicles against attacks should consider recent technologies and developments. The next section will focus on using lightweight authentication to improve cryptography, 3GPP and software defined security to improve network security, and deep learning to improve software vulnerability and malware detection. In fact, computational constraints and the requirement for real-time data transmission in intelligent vehicles are the main reasons for choosing lightweight authentication as a future direction. Besides, 3GPP as a new promising telecommunication standard for V2X security and software defined security (SDS) as an efficient, adaptable and dynamic method for detecting and mitigating security attacks are other main directions for future studies. Finally, deep learning techniques are introduced because they outperform machine learning solutions in terms of attack detection accuracy.  Therefore, we will consider these developments as future directions.


\section{Future Directions}
\label{sec:futureDirections}
The research to date in the field of securing vehicles against cybersecurity challenges has addressed a number of security issues and proposed many security solutions, however, there are still open challenges that need further investigation. Future studies on the current topic are therefore recommended. This section provides a discussion of open issues as well as available and possible methods and technologies to further secure intelligent vehicles. The aim of this part of the paper is to provide future directions for research and encourage future contributions. In this section, we outline four promising directions to further secure intelligent vehicle systems: lightweight authentication to improve cryptography, 3GPP and software defined security to improve network security, and deep learning to improve software vulnerability and malware detection.

Our main reasons for choosing these directions for future research are as follow:
\begin{itemize}
\item \textbf{Lightweight authentication:} In the modern inter-vehicle communications, the efficiency of authentication has become a central issue because fast-moving vehicles need to authenticate each other as quickly as possible before exchanging any information. Thus, we will introduce lightweight authentication as the first future direction.
\item \textbf{3GPP:} Resulting from the development of V2X communications, 3GPP Cellular-V2X (C-V2X) as an initial standard completed in early 2017 to provide reliable, scalable, and robust wireless communications for hazardous situations. With this in mind, C-V2X is the first step towards 5G and this area of study was chosen for its role in the development of network security in the future.
\item \textbf{Software defined security:} Software defined security is the automation of threat detection and the automatic mitigation of attacks. Therefore, SDS is a practical way for improving network security in vehicular networks. In general, design principles of vehicular software defined networking is an open issue for future research.
\item \textbf{Deep learning:} More recent attention has focused on using intelligent deep learning techniques in different applications such as self-driving cars and image recognition. Deep learning methods are accurate. They outperform not only machine learning techniques but also humans in many tasks. Future studies on utilizing neural networks in software vulnerability detection as well as mitigating malware attacks on vehicles are therefore recommended.
\end{itemize}


\subsection{Lightweight Authentication}
Achieving lightweight authentication is never a trivial task in intelligent vehicle systems. The reason is that the authentication in the systems should be secure and efficient, and it should be flexible to handle complicated transportation circumstances~\cite{WXZZZ2016}. As a future research direction, more attention should be paid to lightweight authenticated key generation protocols using communication-media signals.

In this part of the paper, three types of lightweight authentication protocols including key extraction protocols using the wireless fading channels, key establishment protocols using the keyless cryptography technology and key distribution protocols using the Li-Fi technology are reviewed.

\subsubsection{{Key Extraction from Wireless Fading Channels}}
 To date, various key-extraction protocols have utilized wireless fading channels' features at the physical layer~\cite{JPCKPK2009,SIS2011,XQHZZLZ2016,ZXNTLC2017,ZXHCA2018}. In practical multipath environments, the signals transmitted between two vehicles (say Alice and Bob) experience a time-varying and stochastic fading. It is investigated that the fading is invariant within the channel coherence time, whether the signals are transmitted from Alice to Bob or vice versa. Additionally, the fading decorrelates over distances of the order of half a wavelength, i.e., $\lambda/2$. These wireless fading channels' characteristics ensure that: (a). two vehicles extract a shared secret key as long as the transmitted signals are exchanged within the channel coherence time, and (b). the adversary cannot extract any useful information about the extracted key (shared between Alice and Bob) using the eavesdropped signals as long as it is more than $\lambda/2$ far away from the two vehicles. Take the 2.4 GHz WLAN, 624 MHz TV signals and 90 MHz FM Radio as examples~\cite{ZXHCA2018}, the half a wavelength $2/\lambda=6.25 $cm, 23.5 cm and 1.65 m, respectively.

Existing physical-layer key-extraction protocols were designed using the Received Signal Strength (RSS) or the Channel Impulse Response (CIR). Typically, these protocols consist of three phases, i.e., \emph{quantization}, \emph{reconciliation}, and \emph{privacy amplification}. In the quantization phase, two vehicles sample and quantize the transmitted signals, and obtain the initial bit sequences. Due to the imperfect reciprocity and noise, the initial bit sequences may have some mismatch bits. Thus, the reconciliation and privacy amplification phases are employed to remove the mismatch bits and ensure that the two vehicles can obtain a shared secret key with high entropy.

However, the physical layer key extraction protocols cannot be directly implemented in intelligent vehicle systems, because in practice, the channel coherence time in vehicular environments is very short (due to the rapid change of environment). Specifically, it is shown that the channel coherence time in vehicular environments is about a few hundred microseconds. Recalling that most of the existing devices operate in the half-duplex mode, the round-trip time of transmitted signals may be longer than the coherence time of vehicular environments. As a result, it is full of challenges for the two vehicles to sample packets within a duration of the short coherence time.

\subsubsection{{Key Establishment using Keyless Cryptography Technology}}
Alpern and Schneider designed a key-establishment protocol in~\cite{AS1983} using keyless cryptography technology, and it was improved by~\cite{CM2005,PO2015,ZXWWS2018}. In these protocols, the characteristics of the anonymous channel are utilized to establish secret keys. In the field of communication theory, the broadcast channel can be turned into the anonymous channel if the channel achieves source indistinguishability. Technically, source indistinguishability requires that the adversary cannot obtain a non-negligible advantage in identifying the source of the signals (transmitted over the channel) even using sophisticated signal-processing technologies.

These key-establishment protocols~\cite{AS1983,CM2005,PO2015,ZXWWS2018} can be abstracted via the following procedures:
\begin{itemize}
\item Alice uniformly chooses $\frac{|k|}{2}$ bits at random, i.e. $R_{a}\in \{0,1\}^{\frac{|k|}{2}}$. Similarly, Bob uniformly chooses $\frac{|k|}{2}$ bits at random , i.e. $R_{b}\in \{0,1\}^{\frac{|k|}{2}}$;
\item Alice builds $\frac{|k|}{2}$ messages $c_{A}^{1}, c_{A}^{2}, \ldots, c_{A}^{\frac{|k|}{2}}$ using the chosen bits $R_{a}$. For example, in~\cite{CM2005,PO2015}, the messages $c_{A}^{i}$s are built by following the pre-defined rule that the source identifier of the message $c_{A}^{i}$ is set to be ``Alice" if $R_{a}^{i}=1$. Otherwise, it is set to be ``Bob". Executing similar operations, Bob builds $\frac{|k|}{2}$ messages $c_{B}^{1}, c_{B}^{2}, \ldots, c_{B}^{\frac{k}{2}}$ using the chosen bits $R_{b}$; and
\item In the $i^{th}$ transmission round (where $i=1, 2, \ldots, |k|$), Alice or Bob sends a message ($c_{A}^{i}$ or $c_{B}^{i}$ with equal probability) at time $t_{i}$, where $t_{i}$ is uniformly chosen at random in the time interval $[(i-1)T, iT]$, where $T$ is a pre-defined system parameter.
\end{itemize}

Completing the above operations, two vehicles (say Alice and Bob) can establish a secret key with $|k|$ bits. To assess the efficiency of keyless cryptography, an analysis in~\cite{ZXWWS2018} was conducted and keyless cryptography was compared with Diffie-Hellman key exchange protocol (as a well-known asymmetric cryptography approach) in terms of energy and time consumption. The most interesting finding was that, using IEEE 802.15.4 CSMA/CA medium access control (MAC) protocol, keyless cryptography uses 159 times less energy than Diffie-Hellman protocol. Moreover, establishing a secret key with 112 bits in half-duplex mode takes 159 ms. As a matter of fact, keyless cryptography technology needs to be improved. Specifically, this type of key establishment protocols should be rigorously proved secure in order to ensure the security of the intelligent vehicle systems. Additionally, the protocols need to be implemented in intelligent vehicle systems in order to check their practicality.

\subsubsection{{Key Distribution using Li-Fi Technology}}
The rapid increase of wireless data communication makes the radio spectrum below 10 GHz become insufficient. Thus, researchers respond to this challenge by utilizing the radio spectrum above 10 GHz. Light-fidelity (Li-Fi) provides a promising perspective: it is demonstrated that Li-Fi can achieve high speed wireless communication, at over 3 Gb/s, from a single LED (which uses the optimized DCO-OFDM modulation)~\cite{TCRMV2014,HYWC2016}.


Compared with traditional Radio-Frequency (RF) based wireless communication technologies, Li-Fi has advantages~\cite{BYDZ2015} including:
\begin{itemize}
\item Li-Fi communication naturally provides a certain level of security, in term of avoiding interception and eavesdropping, due to the spatial confinement of the light beams.
\item Li-Fi signals have a much higher signal-to-noise ratio, the first reason is that Li-Fi technology uses high-brightness LEDs. Besides, the distance between the transmitter and receiver is limited in this technology.
\item Li-Fi infrastructure is easier to establish due to the wide utilization of LEDs and photodiodes. Taking the intelligent vehicle systems as an example, many LEDs and photodiodes have been deployed in them.
\end{itemize}

In recent years, there is an increasing interest in designing the intelligent vehicle systems using Li-Fi technology; the related work includes~\cite{AHALS2015,BMB2016,GS2016,DMBU2018}. However, the research is in its infancy, and more investigations need to be conducted. Specifically, it is critical and imperative to design key-distribution protocols using Li-Fi technology in order to ensure the security of Li-Fi communication in intelligent vehicle systems. 

\begin{figure}[h]
\begin{centering}
\includegraphics[width=0.9\columnwidth]{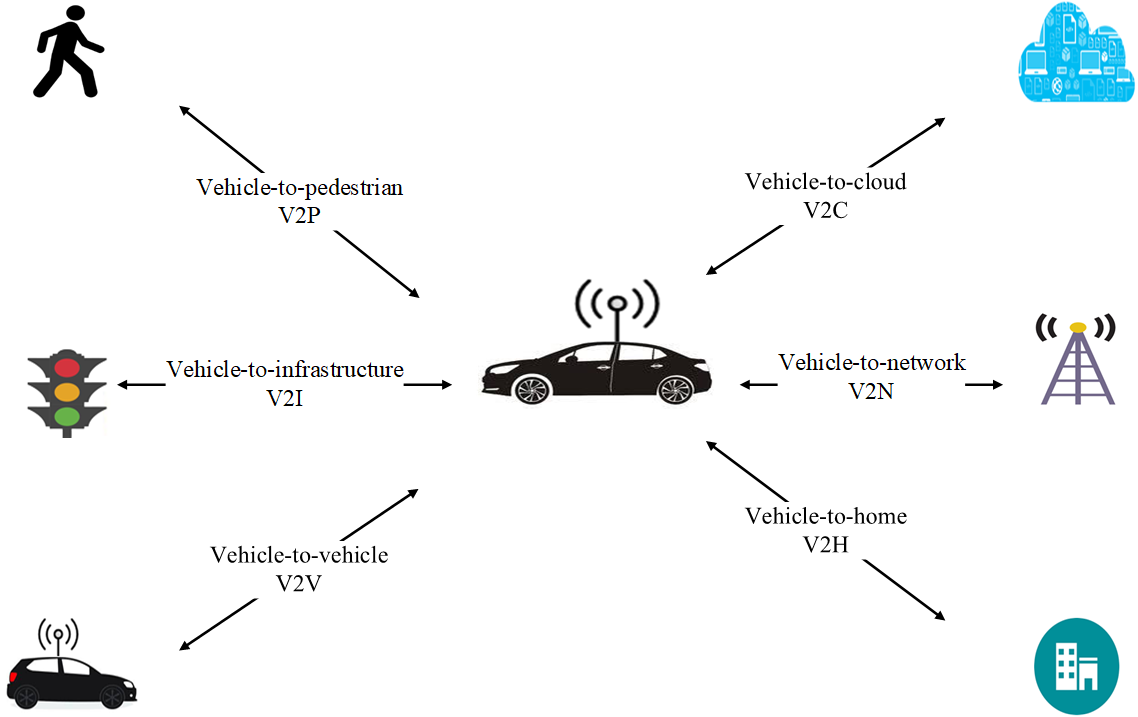}
\par\end{centering}
\caption{V2X models}
\label{V2X}
\end{figure}

\subsection{3GPP on V2X Security}
3GPP is assigned to create technical specification services for LTE support of V2X (3GPP TS33.185 V15.0.0 (2018-06)) \cite{version15}. The 3GPP V2X standard will develop the specifications for all aspects of LTE Advanced and 5G networks, including the protocols architecture, Vehicle-to-Vehicle (V2V), Vehicle-to-Infrastructure (V2I), Vehicle to Network (V2N), Vehicle to Pedestrian (V2P) and all related security concepts for all V2X models (Figure \ref{V2X}). An overview of the LTE enhancements is presented in \cite{version15} with an emphasis on the transport of V2X messages. This document also identifies some of the key threats to security in V2X networks as well as proposed mitigation methods. In addition, some preliminary security requirements are identified to enable safe and secure communication in V2X networks.

\begin{figure}[h]
\begin{centering}
\includegraphics[width=0.9\columnwidth]{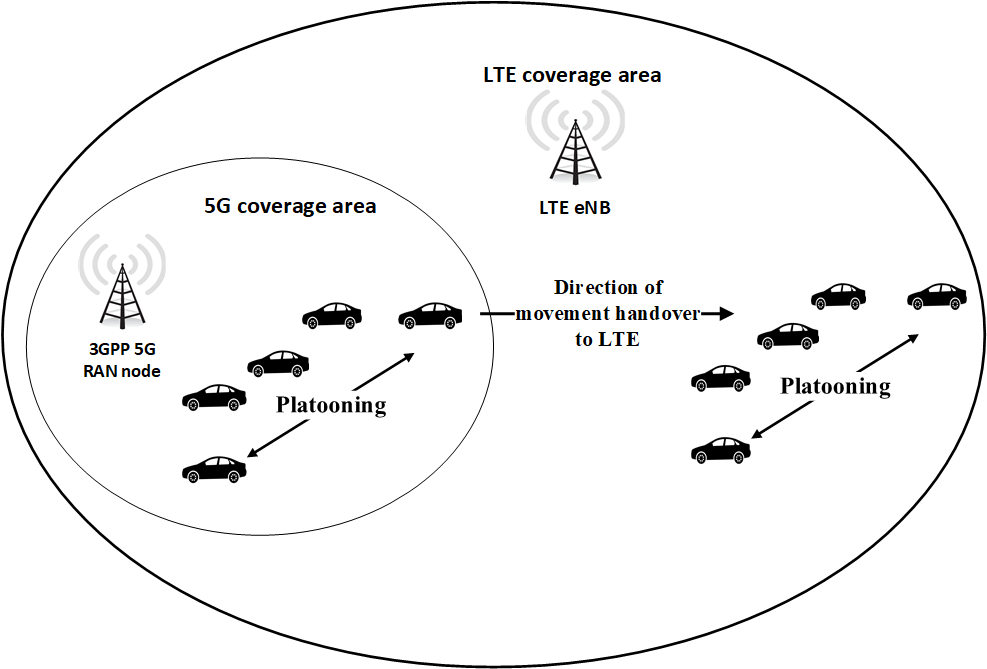}
\par\end{centering}
\caption{Use case out of 5G coverage}
\label{5G}
\end{figure}

\indent A key element in V2X communication is the ability for vehicles and roadside units to effectively and efficiently communicate. The 3GPP group outlines PC5 as the primary communication protocol used between two autonomous cars. To facilitate communication between the vehicle and the roadside unit a protocol called Uu is used. The Vehicle to road-side unit/server is carried over LTE-Uu in a payload of UDP/Ip packets. Emphasis on effective V2X messages is important. In order for vehicles and users to be safe, effective and efficient exchange of information between vehicles must be achieved. Constant communication and message exchange must be conducted in real time. Since vehicles are constantly evaluating their environment and their position, real-time communication is important \cite{version15}.
\newline
\indent Methods to safeguard information and authenticate users must be implemented in order to enable future V2X services. This includes safe and secure storage of credentials and other vital information, which must be protected from malicious users. Conditional information about the vehicle such as speed, location, heading, acceleration, and other dynamic characteristics must be securely  transmitted to roadside units and legitimate users. Failure to propagate this information could lead to serious incidents. In addition, compromising information sent out by a malicious user must be immediately identified \cite{version15}.
\newline
\indent Most V2X research is based on the LTE standard; this technology is currently being utilized and has been proven to be effective, with excellent performance, high bandwidth and low latency (up to 5 ms). 5G is the V2X enabling technology, and it provides ultra-low latency (as low as 1 ms). This allows a real-time response, which enables real-time warnings to be distributed to autonomous vehicles to avoid collisions in real time. This is key to ensuring that autonomous vehicles provide safe and reliable transport for their users. The end-to-end latency that is required for all real-time V2X transmissions is less than 5 ms for message sizes of about 1600 bytes with a probability of 99.999\%. This requirement must be guaranteed for all data traffic in real-time V2X communications. It must be remembered on high-speed roads, relative speeds are up to 500 km/h in high-speed scenarios \cite{luoto2017vehicle}.

\begin{figure}[h]
\begin{centering}
\includegraphics[width=0.9\columnwidth]{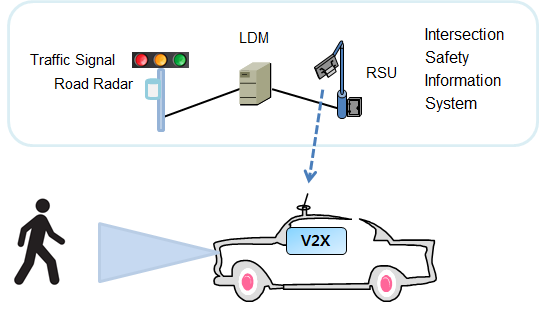}
\par\end{centering}
\caption{Concept of intersection safety information system}
\label{intersection}
\end{figure}

\indent Handover in and out of 5G coverage is illustrated in Figure \ref{5G}. It is an important aspect in supporting V2X application with multiple Radio Access Technology (RAT) modems \cite{version16}. UEs are grouped into platoons (clusters), as UEs move through the network it is important that platoon-related messages are transmitted between UEs with very low latency as per requirement. Thus, V2V messages needed to support platooning application are exchanged between the UEs in the target cell using device to device communication in 5G New RAT (NR), even though there is no 5G coverage in the target cell \cite{version16}. In \cite{version16}, intersection safety and provisioning for urban driving is discussed. Future applications lead to reduced traffic congestion as traffic is routed according to traffic incidents and conditions. An local dynamic map (LDM) is used to express traffic signal information, pedestrian and vehicle movement and direction and location information. 3GPP utilises the low latency capability of 5G communications to conduct real-time analysis of traffic conditions to reduce congestion. The concept of intersection safety information system is illustrated in Figure \ref{intersection}.

\subsubsection{{Proposal for Secure Software Update for Electronic Control Unit}}
\begin{figure}[h]
\begin{centering}
\includegraphics[width=0.9\columnwidth]{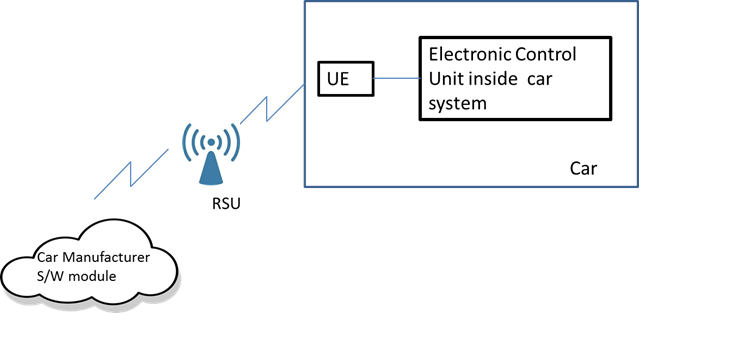}
\par\end{centering}
\caption{Secure software update for electronic control unit.}
\label{update}
\end{figure}

3GPP is proposing a car Electronic Control Unit (ECU) which is a software module able to control the car's system electronics; Examples are wheel steering and brakes, ECU has to be periodically software updatable. ECU software updates are very important for V2X and have to undergo major security testing.
Figure \ref{update} describes the ECU software secure update proposal:

\begin{itemize}
    \item UE is synchronized e.g. via Bluetooth to an ECU.
    \item Suppose a scenario where a car stops in a filling station and connects following a registration procedure to a nearby  roadside unit (RSU).
    \item When connected, the RSU detects the software module version of the ECU in the car via communication with the UE and detects that an update is needed. This is based on the list available to the RSU from a broadcast message from a car manufacturer cloud server.
    \item The RSU will notify the UE that an update is required to the ECU with a list of the updates required. The User will be able to choose the updates required from the list of updates for example. Also the user should be able to reject/defer the update required to the ECU.
    \item If the user chooses an update to the ECU, then additional security procedures should take place so the software download is definitely not from a wrong source and it’s actually the correct version.

\end{itemize}

\subsubsection{{Real-time Solution to Improve Security Thereat Mitigation}}
5G era will have trillions of things connected including autonomous cars and ITS (intelligent transport systems). we will need to get to the 5G future with intelligence, content should be distributed to the devices that make the wireless edge and that requires on-device AI capabilities for cars, smartphones, sensors, and other connected things, so they can autonomously. The 5G wireless edge will be able to reduce latency to less the 1 msec and will lead to improved transport safety and enhanced privacy and security.

Majority of security controls must have visibility to allow a decision as per policy at the security control used to mitigate the threat as at the edge, close to the source as possible, to minimize the damage of the threat.
Another future security solution is the Network slicing end to end (in Transport in the edge data center). 

\begin{figure}[h]
\begin{centering}
\includegraphics[width=0.9\columnwidth]{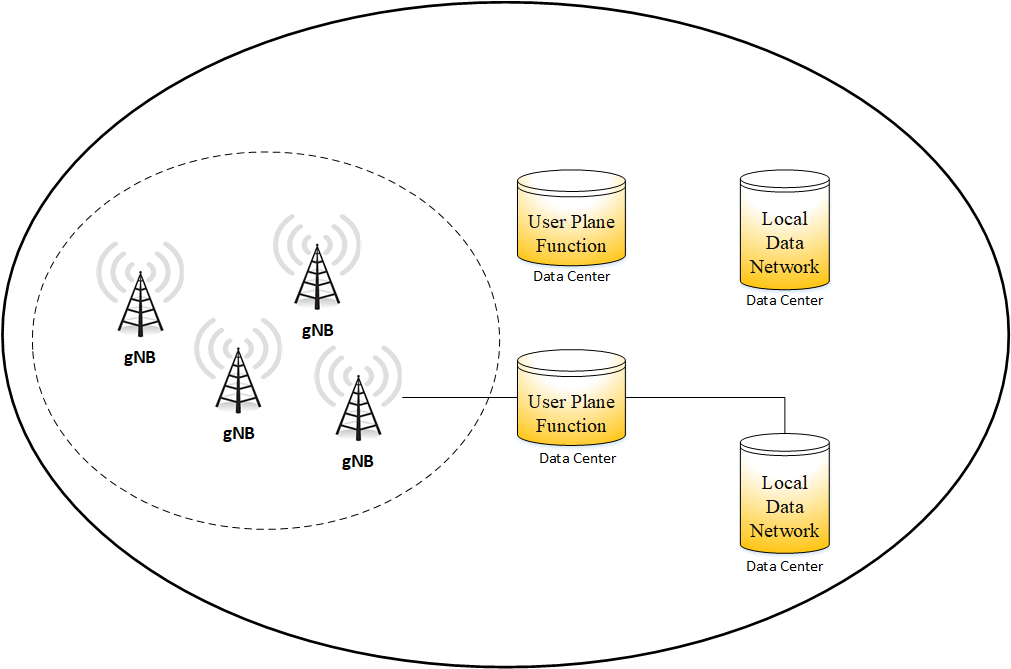}
\par\end{centering}
\caption{A typical edge computing network }
\label{edge}
\end{figure}

Figure \ref{edge} indicates a typical edge computing network deployment. It should be noted that in 5G terminology, gNB is the new name of the base station and User Plane Function (UPF) plays the role of the data plane. 

\subsection{Software-defined Security (SDS)}
Software-defined security refers to the automation of threat detection and automatic mitigation of threats using software defined platforms by adopting an open flow protocol, network function virtualization (NFV) and Software defined networking (SDN) which utilises the concept of a multi layered open virtual switch with programmatic extension to enable automation on a large scale, in addition to the open stack standard as a platform to manage the cloud and distributed data centers.
\newline
\indent Vehicle-to-everything (V2X) is the future of land-based transport. Autonomous vehicles look to eliminate human errors by continuously monitoring and adapting to environmental variations. In such networks, security management is key to ensuring safe and secure transport for occupants. The security for the system, which includes vehicle-to-vehicle, network and cloud security should be dynamic and adaptable. It must continuously apply SDN analytics to identify attacks and mitigate them quickly and effectively.  Threats can be mitigated by applying software defined security concepts that reinforce the security policy for analysis and mitigation.
\newline
\indent The most widely spread attacks on VANETs are DDoS attacks. DDoS attacks with high rates of more than 1 Tera bits/s compromise all kind of connected devices such as laptops, tablets, smartphones, printer’s cameras, Routers and IoT devices to part of a distributed botnet.
Redirection, Filtering, Blackholing, and Firewalls are the most widely used mechanisms to counter DDoS attacks. Redirection of traffic with an abnormal signature is done using forensic analysis. Access control lists (ACLs) are used to filter out ping traffic. However, it is too short to combat modern DDoS attacks. These solutions are ineffective as they handle legitimate traffic and traffic attacks the same way by dumping them to a black hole.
\newline
\indent Current systems take an unacceptably long time to recover from DDoS attacks because they require IT personal intervention and reconfigure major nodes settings and cannot be utilised in a V2X environment. The most promising solution is to automate the network configuration by applying software defined networking (SDN), which uses a central point of control and a decoupling control plan to automate the configuration and settings for major nodes.
\newline
\indent Based on SDN we can apply the concept of SDS, which would decouple the mitigation plan, from the detection plan to automate the security action and solution to mitigate attacks and threats to any node or component within the V2X system. SDS has to allow legitimate traffic to pass through to the designated destination and the redirection of traffic with an abnormal signature (Distribution, spike or abnormal emails) to submit to a forensic analysis to extract as much as possible information about the attack characteristics and traffic parameters in order to create patch file to mitigate the attack by pushing down the patch file along the network major nodes and devices.
\newline
\indent Dynamically changing environments of V2X will require dynamic software and hardware for practical implementation. SDNs are a key piece of the V2X architecture, enabling dynamic mitigation of security threats on the V2X network. 

\subsection{Deep Learning}

\begin{table*}[htbp] %
\caption{Deep learning defences}
\centering
\begin{tabular}{p{3cm} p{5cm}  p{4cm} p{4cm}  }

\toprule
Method                                                                   & Key deep learning idea                                                              & Advantages                                                                            & Disadvantages   \\
\midrule%
Loukas et al. \cite{loukas2017computation}           &Uses computation offloading method for a deep learning-based intrusion detection  &Energy efficiency, low detection latency               &Continuous connectivity is required              \\
DQN  \cite{xiao2017user}                                       &Uses combination of Q-learning and deep learning                          &Accelerates learning process                                               &Computational complexity, takes long time to make a decision\\
Loukas et al. \cite{loukas2018cloud}                     & Uses MLP and RNN for classification                             & Real test bed, high accuracy                                 &Not suitable for simple attacks or onboard intrusion detection   \\
Li et al. \cite   {li2018vuldeepecker}                      & Uses BLSTM for software vulnerability detection           &High accuracy  &Computational complexity \\
HaddadPajouh et al. \cite{haddadpajouh2018deep}                 &Uses LSTMs  to detect malware  & Learning information from previous steps      &LSTMs are not suitable for simple patterns  \\
CDT-DBN \& DDT-DBN \cite{Huda2018}                 &Uses DBN for training unlabeled data and an ANN for training labeled data   &Fast and simplified training model  &Computational complexity for learning RBMs   \\\bottomrule

\end{tabular}
\label{tab:deep}
\end{table*}

\begin{figure}[h]
\begin{centering}
\begin{tikzpicture}[font=\small]
\begin{axis}[width=7cm, x tick label style={
/pgf/number format/1000 sep=}, ylabel={}, ymin=0, ymax=30,xmin=2009.99, xmax=2018,
,axis lines=middle
,samples=32, grid, thick
,domain=2010:2020
,legend pos=outer north east
]
\addplot+ coordinates{(2010,28.0)(2011,26.0)(2012,16.0)(2013,11.0)(2014,6)(2015,4)(2016,3.5)(2017,3)};
\addlegendentry{Error rate}
\end{axis}
\end{tikzpicture}
\par\end{centering}
\caption{Classification error rates of ImageNet challenges}
\label{ImageNet}
\end{figure}
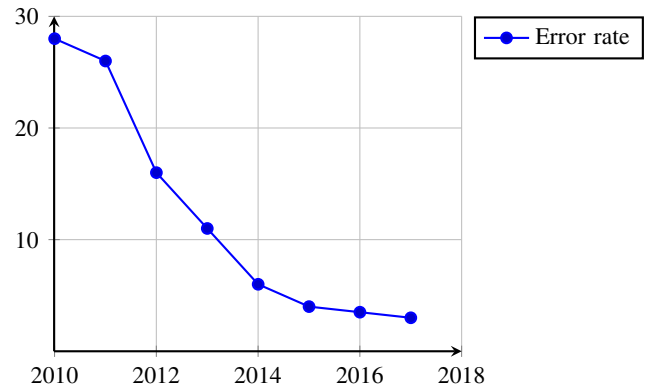

One of the most important events of ImageNet Large Scale Visual Recognition Challenge (ILSVRC) occurred in 2012 when deep learning methods outperformed shallow machine learning methods, therefore, deep learning algorithms attracted particular attention. Deep-learning approaches based on deep neural-network models have indicated amazing performance on various aspects of navigating self-driving cars using computer vision \cite{krizhevsky2012imagenet}. 
Figure \ref{ImageNet} illustrates the Image classification results for the ImageNet challenge where the most important progress occurred in the year 2012. The classification error rate dropped significantly (almost 10 percent) by using convolutional neural networks for image recognition.

%
%
Deep-learning models and techniques such as convolutional neural networks (CNN), long short-term memory (LSTM), deep belief networks (DBN), deep Boltzmann machines (DBM), multilayer perceptron (MLP), autoencoder-based methods, and sparse coding-based methods are based on training neural networks with a training set. After training the neural network, it recognizes the patterns and classifies a different set of examples called a test set \cite{lecun2015deep}. In deep-learning models, there are many layers between the input and output layers for finding features.
\begin{figure}[h]
\begin{centering}
\includegraphics[width=0.9\columnwidth]{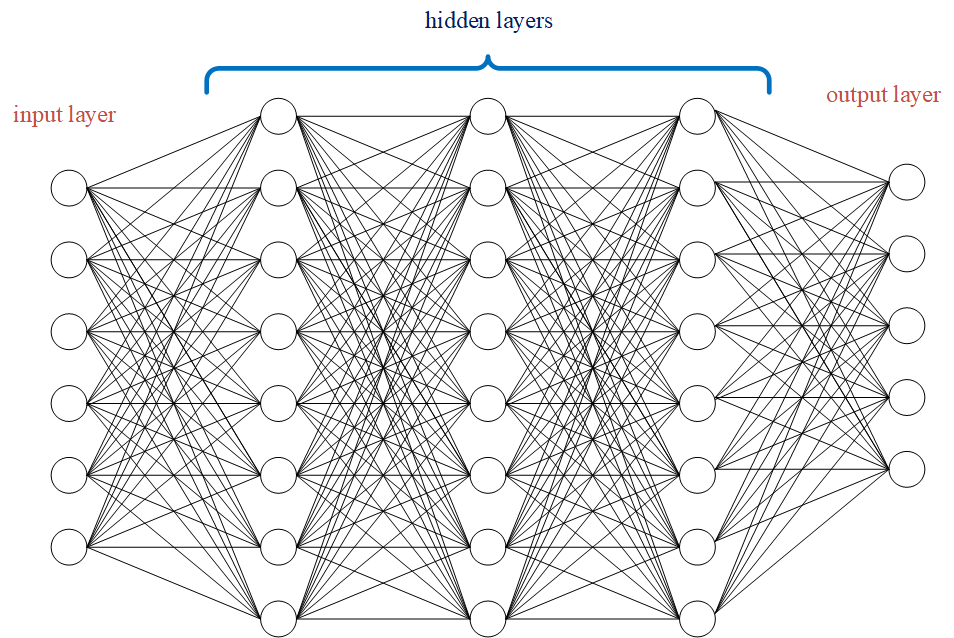}
\par\end{centering}
\caption{Input layer, hidden layers, and output layer in deep learning}
\label{deep}
\end{figure}

Along with this growth in deep learning, however, there is increasing concern over its computational complexity and traceability:
\begin{itemize}
\item Computational complexity: The most important limitation of deep neural networks lies in the fact that these approaches have a heavy computational load which prevents their use in many practical applications, especially on resource constrained vehicle nodes. Thus, recently many learning algorithms have been proposed for increasing the efficiency of deep-learning methods but, having said that, they often suffer from low classification accuracy. Therefore, this is an important issue for future research \cite{kim2018deep}.
\item Traceability: It is generally infeasible to know what features a deep neural network is using to make predictions on a big dataset. As shown in Figure \ref{deep}, the lack of traceability is due to using hidden factors for modeling the complicated relationship between the input variables and the output variables \cite{hinton2018deep}.
\end{itemize}

\subsubsection{Deep Learning Defences}
In this part of the paper, a number of significant and well-known deep-learning approaches for securing various aspects (intrusion detection, software vulnerability detection, malware detection) of intelligent vehicular systems are reviewed and compared with each other.

Loukas et al. \cite{loukas2017computation} proposed a cloud-based intrusion detection defence mechanism in order to achieve energy efficiency and performance. The authors addressed the shortcomings of existing methods for continuous tasks of vehicles such as intrusion detection in terms of energy consumption and processing limitations. Therefore, they proposed computational offloading for these tasks. This means, that the problem of computational complexity for vehicles deep-learning approaches can be solved by offloading computation. As a case study, the authors focused on deep-learning-based intrusion detection mechanism for a robotic vehicle. In this way, a Recurrent Neural Network (RNN) with Long Short Term Memory (LSTM) as the intrusion detection system has been employed. In order to address the inherent security and privacy risks in cloud computing, the HTTPS protocol has been selected for client-server communications; nevertheless, it should be noted that continuous connectivity between the vehicle and the server is another challenge of computation offloading, especially due to the bandwidth variability while vehicles move from one place to another.

Xiao et al. \cite{xiao2017user} have proposed a deep Q-network (DQN) based power allocation strategy. DQN combines Q-learning and deep learning against smart attacks on UAVs. Achieving the optimal power allocation is the main purpose of this scheme. To this end, a reinforcement-learning-based power allocation strategy was proposed. At first, the presented DQN-based method makes use of game theory to formulate interaction between Eve and the UAV transmission as a PT-based smart attack game, and the Nash Equilibrium (NE) of the game has been derived. Then, in order to find an optimal power-allocation strategy, the DQN learning-based power allocation method has been applied. In fact, the DQN is a convolutional neural network (CNN) which consists of two convolutional layers and two fully connected layers.

Loukas et al. \cite{loukas2018cloud} have proposed a deep learning defence mechanism against intrusion. Recurrent neural networks (RNN) and a multilayer perceptron (MLP) are proposed, both of which are considered as well-known artificial neural networks. The basic RNN is a neural network with feedback; hence, backpropagation is not suitable for training. The hidden layer of the RNN includes a Long Short-Term Memory (LSTM), a dense layer and a sigmoid activation function. Memorizing and forgetting the past in the learning process are two important characteristics of LSTMs. The hidden layer(s) of the proposed MLP makes use of leaky rectified linear units (LReLU) as the activation function. MLP is an example of feed-forwarding deep learning and has been proposed to make a comparison between RNN (with LSTM) and MLP. Furthermore, offloading infrastructure has been proposed to tackle the resource limitations of mobile vehicles. In order to evaluate the proposed approaches, denial of service (DoS), command injection and malware attacks were launched on a real robotic vehicle. The proposed methods provide a strong defence against intrusion but, having said that they are not suitable for simple attack classification or onboard intrusion detection.


Li et al. \cite{li2018vuldeepecker} applied deep-learning models to detect software vulnerabilities. First, a code gadget was defined as a composition of a number of relative program statements dependent on the same data or control flow. Then static analysis was used to extract features about library calls from the code gadget. A Bidirectional LSTM (BLSTM) network was selected as the classifier because LSTM could address the Vanishing Gradient (VG) problem and BLSTM could record the effect of code statements from two directions. Experimental results showed that the model achieved 5.7\% false-positive and 7.0\% false negative rate. The result indicates that a deep learning network may achieve better performance than traditional machine-learning models.

HaddadPajouh et al. \cite{haddadpajouh2018deep} have put forth a deep Recurrent Neural Network (RNN) based defence in order to detect malware attacks. Inasmuch as the majority of Unix System-V IoT devices use ARM processors, the authors have focused on analyzing ARM-based IoT applications’ execution operation codes (OpCodes). The proposed approach has used an ELF parser to extract the OpCodes of malware and benign files. After feature selection and removing redundant features (OpCodes), Long Short-Term Memory has been utilized for detecting IoT malware samples based on the sequences of OpCodes. In this way, three LSTM models with different configurations have been examined. It should be remembered that LSTMs are suitable for learning long-term dependencies due to their ability to forget, remember and update the information. In the final analysis, the detection accuracy of the proposed method is 98\%.

A recent study \cite{Huda2018} has attempted to design a defence model against malware attacks on the industrial cloud-assisted internet of things (CoT) using deep-belief networks (DBNs). The authors have argued that standard industrial-control system (ICS) malware detection systems are not able to defend against the unknown behavior of malware, and ICSs based on CoTs require advanced malware detection systems. The sandbox environment is proposed for detecting the run-time behavior of malware and dynamic feature extraction based on API calls. It should be mentioned that DBN based methods do not require feature reduction. The proposed DBN consists of one visible unit and two hidden units. The computed weights from the DBN have been transformed to an artificial neural network (ANN) for initialization, then the ANN is trained using only labeled data. Two detection models based on deep belief networks have been proposed. The first model makes use of disjoint training and testing sets (DDT-DBN). The second model trains a DBN with new unlabeled data to provide the DBN with additional knowledge about changes in the malicious attack patterns (CDT-DBN). The proposed defence mechanisms  have been tested against different types of Trojans, worms, and viruses. In the final analysis, the detection accuracy of the proposed models is 99\%.

In Table \ref{tab:deep}, the key idea, advantages, and disadvantages of deep learning defences are illustrated.
\subsubsection{{Real-time Simulation and Formal Verification}}
On a separate regard, the industry is adopting real-time simulation and formal verification as part of security compliance check for intelligent vehicles \cite{pan2017cyber}, \cite{zheng2014physically}, \cite{zheng2017perceptions}. In real-time simulation, computer models are used to accurately re-create repetitive and flexible test environment for vehicular systems \cite{zheng2017real} while formal verification provides security guarantee \cite{zheng2018efficient}, \cite{zheng2015braceassertion}. As a promising future direction, deep learning based models can be combined with real-time simulation and formal verification to provide more rigid yet accurate security assurance.

\subsubsection{{Deep Learning On Edge Computing}}
As illustrated earlier, one of the main drawbacks of deep learning approaches is computational complexity. Due to more datasets, cloud computing was a convenient solution for deep learning approaches. However, a major problem with this kind of application is huge data traffic and latency. In general, therefore, it seems that the distribution of computation between the nodes is a better idea. With this intention, in recent years edge computing or fog computing for the internet of things has been introduced \cite{sun2016edgeiot}. The term \textquotedblleft Fog Computing\textquotedblright{} the first time used by Cisco and generally understood to mean extending cloud computing to the edge of the network. It should be noted that low latency, geographical distribution, real-time interaction, support of mobility, and wireless access are the most significant characteristics of fog computing networks \cite{bonomi2012fog}.

In particular, exchanging safety critical information in the Internet of Vehicles (IoV) and its supporting platform \cite{zheng2018smartvm} between connected vehicles and roadside units need to minimize latency. With this in mind, mobile edge computing provides an important opportunity for deep learning applications to extend the connected car cloud to be close to the vehicles without sending data to distant servers \cite{hu2015mobile}.

\subsection{Summarising Future Directions}
 With the emerging and developing Internet of Things and Internet of Vehicles, the biggest challenge for intelligent vehicles in the future is security. 
By comparing the proposed directions for future security solutions it is obvious that they are usually light, fast, or intelligent. Therefore, they provide an appropriate environment for developing more adaptable and complicated security defences with high performance, meeting security requirements in the vehicles.


\section{Validity Discussion}
\label{sec:ValidityDiscuss}
There are always some threats to the validity of the results in any research study as to the findings of this survey paper. As a matter of fact, it is possible that some limitations have influenced this review paper. Herein, some of these limitations and threats have been discussed:
\begin{itemize}
\item \textbf{Limitation of the approach:} Due to brevity, we only selected a few interesting and major security attacks.  The scope of attacks is nowhere close to exhaustive.  But we hope the solutions provided for these attacks can provide some generalizable solutions for other attacks, which are not covered here.

\item \textbf{Solutions covered:} Once again, our solutions chosen are nowhere exhaustive, but we have chosen them based on our research capabilities from all possible collaborations for the manuscript. We hope the solutions, though inevitably limited, can provide some research insight into securing the intelligent vehicle systems.
\end{itemize}

\section{Conclusion}
\label{sec:concl}
In this review paper, we have surveyed securing state-of-the-art vehicles and discussed the architecture of intelligent vehicles. The purpose of this paper was to focus on security issues and the ways to deal with them accordingly. Therefore, we have stated the security requirements of vehicular networks, a number of security attacks on intelligent vehicle systems and challenges related to them. Moreover, the security defences have been classified regarding their effectiveness against these identified attacks. Our major purpose was to discover some advantages and shortcomings of the proposed defences. Finally, we comprehensively reviewed and discussed potential defences and directions for the future in order to further secure intelligent vehicle systems and their communications. This survey paper provides a good foundation for researchers who are interested to gain an insight into the security issues of intelligent vehicles and the proposed solutions. As future work, we plan to implement these proposed defence mechanisms for an international vehicle company and conduct some extensive experiments.


%

%
%

%
%
\ifCLASSOPTIONcaptionsoff
  \newpage
\fi



\bibliographystyle{IEEEtran}
\bibliography{vs_ref}

\begin{thebibliography}{100}
\providecommand{\url}[1]{#1}
\csname url@samestyle\endcsname
\providecommand{\newblock}{\relax}
\providecommand{\bibinfo}[2]{#2}
\providecommand{\BIBentrySTDinterwordspacing}{\spaceskip=0pt\relax}
\providecommand{\BIBentryALTinterwordstretchfactor}{4}
\providecommand{\BIBentryALTinterwordspacing}{\spaceskip=\fontdimen2\font plus
\BIBentryALTinterwordstretchfactor\fontdimen3\font minus
  \fontdimen4\font\relax}
\providecommand{\BIBforeignlanguage}[2]{{%
\expandafter\ifx\csname l@#1\endcsname\relax
\typeout{** WARNING: IEEEtran.bst: No hyphenation pattern has been}%
\typeout{** loaded for the language `#1'. Using the pattern for}%
\typeout{** the default language instead.}%
\else
\language=\csname l@#1\endcsname
\fi
#2}}
\providecommand{\BIBdecl}{\relax}
\BIBdecl

\bibitem{patsakis2014towards}
C.~Patsakis, K.~Dellios, and M.~Bouroche, ``Towards a distributed secure
  in-vehicle communication architecture for modern vehicles,'' \emph{Computers
  \& Security}, vol.~40, pp. 60--74, 2014.

\bibitem{cheah2018building}
M.~Cheah, S.~A. Shaikh, J.~Bryans, and P.~Wooderson, ``Building an automotive
  security assurance case using systematic security evaluations,''
  \emph{Computers \& Security}, vol.~77, pp. 360--379, 2018.

\bibitem{forbes}
\BIBentryALTinterwordspacing
A.~Greenberg. (2013) Hackers reveal nasty new car attacks--with me behind the
  wheel (video). [Online]. Available:
  \url{https://www.forbes.com/sites/andygreenberg/2013/07/24/hackers-reveal-nasty-new-car-attacks-with-me-behind-the-wheel-video/\#3901ab9a228c}
\BIBentrySTDinterwordspacing

\bibitem{wierd}
\BIBentryALTinterwordspacing
A.~GREENBERG. (2016) The jeep hackers are back to prove car hacking can get
  much worse. [Online]. Available:
  \url{https://www.wired.com/2016/08/jeep-hackers-return-high-speed-steering-acceleration-hacks/}
\BIBentrySTDinterwordspacing

\bibitem{guardian}
\BIBentryALTinterwordspacing
O.~Solon. (2016) Team of hackers take remote control of tesla model s from 12
  miles away. [Online]. Available:
  \url{https://www.theguardian.com/technology/2016/sep/20/tesla-model-s-chinese-hack-remote-control-brakes}
\BIBentrySTDinterwordspacing

\bibitem{helpnetsecurity}
\BIBentryALTinterwordspacing
C.~Miller. (2018) Researchers hack bmw cars, discover 14 vulnerabilities.
  [Online]. Available:
  \url{https://www.helpnetsecurity.com/2018/05/23/hack-bmw-cars/}
\BIBentrySTDinterwordspacing

\bibitem{MCAFEE}
\BIBentryALTinterwordspacing
C.~David and S.~Fry. (2016) Automotive security best practices. [Online].
  Available:
  \url{https://www.mcafee.com/enterprise/en-us/assets/white-papers/wp-automotive-security.pdf}
\BIBentrySTDinterwordspacing

\bibitem{mokhtar2015survey}
B.~Mokhtar and M.~Azab, ``Survey on security issues in vehicular ad hoc
  networks,'' \emph{Alexandria Engineering Journal}, vol.~54, no.~4, pp.
  1115--1126, 2015.

\bibitem{sakiz2017survey}
F.~Sakiz and S.~Sen, ``A survey of attacks and detection mechanisms on
  intelligent transportation systems: Vanets and iov,'' \emph{Ad Hoc Networks},
  vol.~61, pp. 33--50, 2017.

\bibitem{hasrouny2017vanet}
H.~Hasrouny, A.~E. Samhat, C.~Bassil, and A.~Laouiti, ``Vanet security
  challenges and solutions: A survey,'' \emph{Vehicular Communications},
  vol.~7, pp. 7--20, 2017.

\bibitem{bernardini2017security}
C.~Bernardini, M.~R. Asghar, and B.~Crispo, ``Security and privacy in vehicular
  communications: Challenges and opportunities,'' \emph{Vehicular
  Communications}, 2017.

\bibitem{haas2016cross}
W.~Haas and P.~Langjahr, ``Cross-domain vehicle control units in modern e/e
  architectures,'' in \emph{16. Internationales Stuttgarter Symposium}.\hskip
  1em plus 0.5em minus 0.4em\relax Springer, 2016, pp. 1619--1627.

\bibitem{brunner2017automotive}
S.~Brunner, J.~Roder, M.~Kucera, and T.~Waas, ``Automotive e/e-architecture
  enhancements by usage of ethernet tsn,'' in \emph{Intelligent Solutions in
  Embedded Systems (WISES), 2017 13th Workshop on}.\hskip 1em plus 0.5em minus
  0.4em\relax IEEE, 2017, pp. 9--13.

\bibitem{zeng2016vehicle}
W.~Zeng, M.~A. Khalid, and S.~Chowdhury, ``In-vehicle networks outlook:
  achievements and challenges,'' \emph{IEEE Communications Surveys \&
  Tutorials}, vol.~18, no.~3, pp. 1552--1571, 2016.

\bibitem{afsin2017c}
M.~E. Afsin, K.~W. Schmidt, and E.~G. Schmidt, ``C 3: configurable can fd
  controller: architecture, design and hardware implementation,'' in
  \emph{Industrial Embedded Systems (SIES), 2017 12th IEEE International
  Symposium on}.\hskip 1em plus 0.5em minus 0.4em\relax IEEE, 2017, pp. 1--9.

\bibitem{hartwich2012can}
F.~Hartwich \emph{et~al.}, ``Can with flexible data-rate,'' in \emph{Proc.
  iCC}.\hskip 1em plus 0.5em minus 0.4em\relax Citeseer, 2012, pp. 1--9.

\bibitem{bosch2012can}
R.~Bosch, ``Can with flexible data-rate specification,'' \emph{Robert Bosch
  GmbH, Stuttgart}, 2012.

\bibitem{flexray2010flexray}
F.~Consortium, ``Flexray communications system protocol specification,''
  \emph{Version}, vol. 3.0.1, no.~1, pp. 1--341, 2010.

\bibitem{Engelmann2010most}
S.~Khurshid, C.~S. P{\u{a}}s{\u{a}}reanu, and W.~Visser, ``Most150-development
  and production launch from an oems perstective.''\hskip 1em plus 0.5em minus
  0.4em\relax presented at the 11th MOST Intercon-nectivity conf. Asis, Seoul,
  Korea, 2010, pp. 553--568.

\bibitem{grzemba2011most}
A.~Grzemba, ``Most book from most25 to most150,'' \emph{MOST Cooperation,
  FRANZIS}, 2011.

\bibitem{zeeb2001optical}
E.~Zeeb, ``Optical data bus systems in cars: Current status and future
  challenges,'' in \emph{Optical Communication, 2001. ECOC'01. 27th European
  Conference on}, vol.~1.\hskip 1em plus 0.5em minus 0.4em\relax IEEE, 2001,
  pp. 70--71.

\bibitem{matheus2017automotive}
K.~Matheus and T.~K{\"o}nigseder, \emph{Automotive ethernet}.\hskip 1em plus
  0.5em minus 0.4em\relax Cambridge University Press, 2017.

\bibitem{hank2012automotive}
P.~Hank, T.~Suermann, and S.~Mueller, ``Automotive ethernet, a holistic
  approach for a next generation in-vehicle networking standard,'' in
  \emph{Advanced Microsystems for Automotive Applications 2012}.\hskip 1em plus
  0.5em minus 0.4em\relax Springer, 2012, pp. 79--89.

\bibitem{ABI2014ethernet}
``Ethernet in-vehicle networkingto feature in 40\% of vehicles shipping
  globally by 2020, london, u.k.'' \emph{[Online] Available:
  https://www.abiresearch.com/press/ethernet-in-vehiclenetworking-to-feature-in-40-of,
  accessed on Nov. 09, 2014}.

\bibitem{yang2018intelligent}
D.~Yang, K.~Jiang, D.~Zhao, C.~Yu, Z.~Cao, S.~Xie, Z.~Xiao, X.~Jiao, S.~Wang,
  and K.~Zhang, ``Intelligent and connected vehicles: Current status and future
  perspectives,'' \emph{Science China Technological Sciences}, pp. 1--26, 2018.

\bibitem{weiss2011v2x}
C.~Wei{\ss}, ``V2x communication in europe--from research projects towards
  standardization and field testing of vehicle communication technology,''
  \emph{Computer Networks}, vol.~55, no.~14, pp. 3103--3119, 2011.

\bibitem{obst2014multi}
M.~Obst, L.~Hobert, and P.~Reisdorf, ``Multi-sensor data fusion for checking
  plausibility of v2v communications by vision-based multiple-object
  tracking,'' in \emph{Vehicular Networking Conference (VNC), 2014 IEEE}.\hskip
  1em plus 0.5em minus 0.4em\relax IEEE, 2014, pp. 143--150.

\bibitem{kim2015multivehicle}
S.-W. Kim, B.~Qin, Z.~J. Chong, X.~Shen, W.~Liu, M.~H. Ang, E.~Frazzoli, and
  D.~Rus, ``Multivehicle cooperative driving using cooperative perception:
  Design and experimental validation,'' \emph{IEEE Transactions on Intelligent
  Transportation Systems}, vol.~16, no.~2, pp. 663--680, 2015.

\bibitem{liu2014vehicle}
W.~Liu, S.-W. Kim, K.~Marczuk, and M.~H. Ang, ``Vehicle motion intention
  reasoning using cooperative perception on urban road,'' in \emph{Intelligent
  Transportation Systems (ITSC), 2014 IEEE 17th International Conference
  on}.\hskip 1em plus 0.5em minus 0.4em\relax IEEE, 2014, pp. 424--430.

\bibitem{kim2015impact}
S.-W. Kim, W.~Liu, M.~H. Ang, E.~Frazzoli, and D.~Rus, ``The impact of
  cooperative perception on decision making and planning of autonomous
  vehicles,'' \emph{IEEE Intelligent Transportation Systems Magazine}, vol.~7,
  no.~3, pp. 39--50, 2015.

\bibitem{luthardt2017efficient}
S.~Luthardt, C.~Han, V.~Willert, and M.~Schreier, ``Efficient graph-based v2v
  free space fusion,'' in \emph{Intelligent Vehicles Symposium (IV), 2017
  IEEE}.\hskip 1em plus 0.5em minus 0.4em\relax IEEE, 2017, pp. 985--992.

\bibitem{leung2012decentralized}
K.~Y. Leung, T.~D. Barfoot, and H.~H. Liu, ``Decentralized cooperative slam for
  sparsely-communicating robot networks: A centralized-equivalent approach,''
  \emph{Journal of Intelligent \& Robotic Systems}, vol.~66, no.~3, pp.
  321--342, 2012.

\bibitem{shenjiang2010design}
L.~D.~W. Shenjiang, ``The design of the controller on automobile taillight
  based on at89s52 [j],'' \emph{Foreign Electronic Measurement Technology},
  vol.~8, p. 021, 2010.

\bibitem{Xu2014automotive}
J.~Xu and F.~M. Zhong, ``Automotive air conditioning control system based on
  stcl2c5a60s2 singlechip,'' \emph{Auto Electric Parts}, pp. 14--16, 2014.

\bibitem{Gan2004study}
H.~Y. Gan, J.~Z. Zhang, and Q.~C. Lu, ``Study on operating mode control of
  hybrid electric vehicle based on the high performance 32- bit scm mpc555,''
  \emph{Automobile Technology}, pp. 9--12, 2004.

\bibitem{Yu2011dsp}
X.~Q. Yu, B.~B. Chen, and T.~K. Ji, ``Dsp software design for eq effect of car
  multimedia system,'' \emph{Microcomputer \& Its Applications}, vol.~30, pp.
  47--50, 2011.

\bibitem{Yu2009dsp-based}
Y.~J. Yu, Z.~Z. Fu, L.~Rao \emph{et~al.}, ``Dsp-based advancecollision warning
  system,'' \emph{Process Automation Instrumentation}, vol.~30, no.~6, pp.
  11--13, 2009.

\bibitem{Yu2007design}
J.~Q. Yu, Z.~Z. Chen, and P.~Liang, ``The design and implementation signal
  processing system of the automotive collision. avoidance based on
  tms320vc5402,'' \emph{Microcomputer Information}, vol.~23, pp. 266--267,
  2007.

\bibitem{lindholm2008nvidia}
E.~Lindholm, J.~Nickolls, S.~Oberman, and J.~Montrym, ``Nvidia tesla: A unified
  graphics and computing architecture,'' \emph{IEEE micro}, vol.~28, no.~2,
  2008.

\bibitem{Liu2002osek}
B.~L. Liu and Y.~B. Sun, ``Osek/vdx--- an open-architectured platform of
  vehicle electronics system,'' \emph{Vehicle\& Power Technology}, pp. 61--64,
  2002.

\bibitem{guettier2016standardization}
C.~Guettier, B.~Bradai, F.~Hochart, P.~Resende, J.~Yelloz, and A.~Garnault,
  ``Standardization of generic architecture for autonomous driving: A reality
  check,'' in \emph{Energy Consumption and Autonomous Driving}.\hskip 1em plus
  0.5em minus 0.4em\relax Springer, 2016, pp. 57--68.

\bibitem{aly2017consolidating}
S.~Aly, ``Consolidating autosar with complex operating systems (autosar on
  linux),'' SAE Technical Paper, Tech. Rep., 2017.

\bibitem{furst2016autosar}
S.~F{\"u}rst and M.~Bechter, ``Autosar for connected and autonomous vehicles:
  The autosar adaptive platform,'' in \emph{Dependable Systems and Networks
  Workshop, 2016 46th Annual IEEE/IFIP International Conference on}.\hskip 1em
  plus 0.5em minus 0.4em\relax IEEE, 2016, pp. 215--217.

\bibitem{sagstetter2013security}
F.~Sagstetter, M.~Lukasiewycz, S.~Steinhorst, M.~Wolf, A.~Bouard, W.~R. Harris,
  S.~Jha, T.~Peyrin, A.~Poschmann, and S.~Chakraborty, ``Security challenges in
  automotive hardware/software architecture design,'' in \emph{Proceedings of
  the Conference on Design, Automation and Test in Europe}.\hskip 1em plus
  0.5em minus 0.4em\relax EDA Consortium, 2013, pp. 458--463.

\bibitem{jochem1995no}
T.~Jochem and D.~Pomerleau, ``No hands across america official press release,''
  \emph{Carnegie Mellon University}, 1995.

\bibitem{maurer1996compact}
M.~Maurer, R.~Behringer, S.~Furst, F.~Thomanek, and E.~D. Dickmanns, ``A
  compact vision system for road vehicle guidance,'' in \emph{Pattern
  Recognition, 1996., Proceedings of the 13th International Conference on},
  vol.~3.\hskip 1em plus 0.5em minus 0.4em\relax IEEE, 1996, pp. 313--317.

\bibitem{bertozzi1998vision}
M.~Bertozzi, A.~Broggi, G.~Conte, A.~Fascioli, and R.~Fascioli, ``Vision-based
  automated vehicle guidance: the experience of the argo vehicle,''
  \emph{Tecniche di Intelligenza Artificiale e Pattern Recognition per la
  Visione Artificiale}, pp. 35--40, 1998.

\bibitem{broggi2000architectural}
A.~Broggi, M.~Bertozzi, and A.~Fascioli, ``Architectural issues on vision-based
  automatic vehicle guidance: the experience of the argo project,''
  \emph{Real-Time Imaging}, vol.~6, no.~4, pp. 313--324, 2000.

\bibitem{campbell2010autonomous}
M.~Campbell, M.~Egerstedt, J.~P. How, and R.~M. Murray, ``Autonomous driving in
  urban environments: approaches, lessons and challenges,'' \emph{Philosophical
  Transactions of the Royal Society of London A: Mathematical, Physical and
  Engineering Sciences}, vol. 368, no. 1928, pp. 4649--4672, 2010.

\bibitem{johnson2008development}
D.~G. Johnson, ``Development of a high resolution mmw radar employing an
  antenna with combined frequency and mechanical scanning,'' in \emph{Radar
  Conference, 2008. RADAR'08. IEEE}.\hskip 1em plus 0.5em minus 0.4em\relax
  IEEE, 2008, pp. 1--5.

\bibitem{wang2014bionic}
X.~Wang, L.~Xu, H.~Sun, J.~Xin, and N.~Zheng, ``Bionic vision inspired on-road
  obstacle detection and tracking using radar and visual information,'' in
  \emph{Intelligent Transportation Systems (ITSC), 2014 IEEE 17th International
  Conference on}.\hskip 1em plus 0.5em minus 0.4em\relax IEEE, 2014, pp.
  39--44.

\bibitem{han2016frontal}
S.~Han, X.~Wang, L.~Xu, H.~Sun, and N.~Zheng, ``Frontal object perception for
  intelligent vehicles based on radar and camera fusion,'' in \emph{Control
  Conference (CCC), 2016 35th Chinese}.\hskip 1em plus 0.5em minus 0.4em\relax
  IEEE, 2016, pp. 4003--4008.

\bibitem{wang2016road}
X.~Wang, L.~Xu, H.~Sun, J.~Xin, and N.~Zheng, ``On-road vehicle detection and
  tracking using mmw radar and monovision fusion,'' \emph{IEEE Transactions on
  Intelligent Transportation Systems}, vol.~17, no.~7, pp. 2075--2084, 2016.

\bibitem{kato2002obstacle}
T.~Kato, Y.~Ninomiya, and I.~Masaki, ``An obstacle detection method by fusion
  of radar and motion stereo,'' \emph{IEEE Transactions on Intelligent
  Transportation Systems}, vol.~3, no.~3, pp. 182--188, 2002.

\bibitem{song2014robust}
S.~Song and M.~Chandraker, ``Robust scale estimation in real-time monocular sfm
  for autonomous driving,'' in \emph{Proceedings of the IEEE Conference on
  Computer Vision and Pattern Recognition}, 2014, pp. 1566--1573.

\bibitem{dagan2004forward}
E.~Dagan, O.~Mano, G.~P. Stein, and A.~Shashua, ``Forward collision warning
  with a single camera,'' in \emph{Intelligent Vehicles Symposium, 2004
  IEEE}.\hskip 1em plus 0.5em minus 0.4em\relax IEEE, 2004, pp. 37--42.

\bibitem{park2014robust}
K.-Y. Park and S.-Y. Hwang, ``Robust range estimation with a monocular camera
  for vision-based forward collision warning system,'' \emph{The Scientific
  World Journal}, vol. 2014, 2014.

\bibitem{dong2013driver}
Y.~Dong and Z.~Hu, ``Driver inattention monitoring system for intelligent
  vehicles,'' in \emph{Transportation Technologies for Sustainability}.\hskip
  1em plus 0.5em minus 0.4em\relax Springer, 2013, pp. 395--421.

\bibitem{bertozzi2000vision}
M.~Bertozzi, A.~Broggi, and A.~Fascioli, ``Vision-based intelligent vehicles:
  State of the art and perspectives,'' \emph{Robotics and Autonomous systems},
  vol.~32, no.~1, pp. 1--16, 2000.

\bibitem{tawari2014looking}
A.~Tawari, S.~Sivaraman, M.~M. Trivedi, T.~Shannon, and M.~Tippelhofer,
  ``Looking-in and looking-out vision for urban intelligent assistance:
  Estimation of driver attentive state and dynamic surround for safe merging
  and braking,'' in \emph{Intelligent Vehicles Symposium Proceedings, 2014
  IEEE}.\hskip 1em plus 0.5em minus 0.4em\relax IEEE, 2014, pp. 115--120.

\bibitem{WorldCar}
\BIBentryALTinterwordspacing
A.~Davies. (2018) The wired guide to self-driving cars. [Online]. Available:
  \url{https://www.wired.com/story/guide-self-driving-cars/}
\BIBentrySTDinterwordspacing

\bibitem{Europe}
\BIBentryALTinterwordspacing
P.~Campbell. (2018) Eu motors ahead with rules for self-driving cars. [Online].
  Available:
  \url{https://www.ft.com/content/f3a76e4c-5772-11e8-b8b2-d6ceb45fa9d0}
\BIBentrySTDinterwordspacing

\bibitem{crashOne}
\BIBentryALTinterwordspacing
J.~Stewart. (2018) Tesla's autopilot was involved in another deadly crash.
  [Online]. Available:
  \url{https://www.wired.com/story/tesla-autopilot-self-driving-crash-california/}
\BIBentrySTDinterwordspacing

\bibitem{crashTwo}
\BIBentryALTinterwordspacing
A.~Davies. (2018) The unavoidable folly of making humans train self-driving
  cars. [Online]. Available:
  \url{https://www.wired.com/story/uber-crash-arizona-human-train-self-driving-cars/}
\BIBentrySTDinterwordspacing

\bibitem{uber}
\BIBentryALTinterwordspacing
A.~news. (2018) Uber suspends self-driving car tests after vehicle hits and
  kills woman crossing the street in arizona. [Online]. Available:
  \url{http://www.abc.net.au/news/2018-03-20/uber-suspends-self-driving-car-tests-after-fatal-crash/9565586}
\BIBentrySTDinterwordspacing

\bibitem{tesla}
A.~Perrig, R.~Canetti, J.~D.~Tygar, and D.~Song, ``The tesla broadcast
  authentication protocol,'' vol.~5, 11 2002.

\bibitem{bmw}
\BIBentryALTinterwordspacing
Z.~Zorz. (2018) Researchers hack bmw cars, discover 14 vulnerabilities.
  [Online]. Available:
  \url{https://www.helpnetsecurity.com/2018/05/23/hack-bmw-cars/}
\BIBentrySTDinterwordspacing

\bibitem{hash2}
R.~Verdult, F.~Garcia, and J.~Balasch, ``Gone in 360 seconds: Hijacking with
  hitag2,'' pp. 237--252, 01 2012.

\bibitem{de2018driverless}
G.~De~La~Torre, P.~Rad, and K.-K.~R. Choo, ``Driverless vehicle security:
  Challenges and future research opportunities,'' \emph{Future Generation
  Computer Systems}, 2018.

\bibitem{stavrou2010survey}
E.~Stavrou and A.~Pitsillides, ``A survey on secure multipath routing protocols
  in wsns,'' \emph{Computer Networks}, vol.~54, no.~13, pp. 2215--2238, 2010.

\bibitem{safi2018cloud}
Q.~G.~K. Safi, S.~Luo, C.~Wei, L.~Pan, and G.~Yan, ``Cloud-based security and
  privacy-aware information dissemination over ubiquitous vanets,''
  \emph{Computer Standards \& Interfaces}, vol.~56, pp. 107--115, 2018.

\bibitem{silva2016geo}
F.~A. Silva, A.~Boukerche, T.~R.~B. Silva, L.~B. Ruiz, and A.~A. Loureiro,
  ``Geo-localized content availability in vanets,'' \emph{Ad Hoc Networks},
  vol.~36, pp. 425--434, 2016.

\bibitem{attacks}
M.~S. Al-kahtani, ``Survey on security attacks in vehicular ad hoc networks
  (vanets),'' in \emph{2012 6th International Conference on Signal Processing
  and Communication Systems}, Dec 2012, pp. 1--9.

\bibitem{sybilTwo}
L.~Wang and J.~Kangasharju, ``Measuring large-scale distributed systems: case
  of bittorrent mainline dht,'' in \emph{IEEE P2P 2013 Proceedings}, Sept 2013,
  pp. 1--10.

\bibitem{sybilThree}
------, ``Real-world sybil attacks in bittorrent mainline dht,'' in \emph{2012
  IEEE Global Communications Conference (GLOBECOM)}, Dec 2012, pp. 826--832.

\bibitem{TorAttack}
\BIBentryALTinterwordspacing
TorBlog. (2014) Tor security advisory: "relay early" traffic confirmation
  attack. [Online]. Available:
  \url{https://blog.torproject.org/tor-security-advisory-relay-early-traffic-confirmation-attack}
\BIBentrySTDinterwordspacing

\bibitem{firewall}
\BIBentryALTinterwordspacing
G.~Smith. (2003) The new threats to firewalls. [Online]. Available:
  \url{https://www.computerworld.com/article/2569753/security0/the-new-threats-to-firewalls.html}
\BIBentrySTDinterwordspacing

\bibitem{pan2017cyber}
L.~Pan, X.~Zheng, H.~Chen, T.~Luan, H.~Bootwala, and L.~Batten, ``Cyber
  security attacks to modern vehicular systems,'' \emph{Journal of Information
  Security and Applications}, vol.~36, pp. 90--100, 2017.

\bibitem{markovitz2017field}
M.~Markovitz and A.~Wool, ``Field classification, modeling and anomaly
  detection in unknown can bus networks,'' \emph{Vehicular Communications},
  vol.~9, pp. 43--52, 2017.

\bibitem{nilsson2009first}
D.~K. Nilsson, U.~E. Larson, F.~Picasso, and E.~Jonsson, ``A first simulation
  of attacks in the automotive network communications protocol flexray,'' in
  \emph{Proceedings of the International Workshop on Computational Intelligence
  in Security for Information Systems CISIS’08}.\hskip 1em plus 0.5em minus
  0.4em\relax Springer, 2009, pp. 84--91.

\bibitem{leszczyna2018review}
R.~Leszczyna, ``A review of standards with cybersecurity requirements for smart
  grid,'' \emph{Computers \& Security}, 2018.

\bibitem{dey2016vehicle}
K.~C. Dey, A.~Rayamajhi, M.~Chowdhury, P.~Bhavsar, and J.~Martin,
  ``Vehicle-to-vehicle (v2v) and vehicle-to-infrastructure (v2i) communication
  in a heterogeneous wireless network--performance evaluation,''
  \emph{Transportation Research Part C: Emerging Technologies}, vol.~68, pp.
  168--184, 2016.

\bibitem{zaidi2016host}
K.~Zaidi, M.~B. Milojevic, V.~Rakocevic, A.~Nallanathan, and M.~Rajarajan,
  ``Host-based intrusion detection for vanets: a statistical approach to rogue
  node detection,'' \emph{IEEE transactions on vehicular technology}, vol.~65,
  no.~8, pp. 6703--6714, 2016.

\bibitem{sedjelmaci2015accurate}
H.~Sedjelmaci and S.~M. Senouci, ``An accurate and efficient collaborative
  intrusion detection framework to secure vehicular networks,'' \emph{Computers
  \& Electrical Engineering}, vol.~43, pp. 33--47, 2015.

\bibitem{cui2018review}
J.~Cui, L.~S. Liew, G.~Sabaliauskaite, and F.~Zhou, ``A review on safety
  failures, security attacks, and available countermeasures for autonomous
  vehicles,'' \emph{Ad Hoc Networks}, 2018.

\bibitem{wei2018virus}
L.~Wei, H.~Qin, Y.~Wang, Z.~Zhang, and G.~Yu, ``Virus-traffic coupled dynamic
  model for virus propagation in vehicle-to-vehicle communication networks,''
  \emph{Vehicular communications}, vol.~14, pp. 26--38, 2018.

\bibitem{petit2015potential}
J.~Petit and S.~E. Shladover, ``Potential cyberattacks on automated vehicles.''
  \emph{IEEE Trans. Intelligent Transportation Systems}, vol.~16, no.~2, pp.
  546--556, 2015.

\bibitem{fussel2017birmingham}
\BIBentryALTinterwordspacing
S.~Fussel. (2017) Watch thieves hack keyless entry to steal a mercedes in less
  than a minute. [Online]. Available:
  \url{https://gizmodo.com/watch-thieves-hack-keyless-entry-to-steal-a-mercedes-in-1820767189}
\BIBentrySTDinterwordspacing

\bibitem{eiza2017driving}
M.~H. Eiza and Q.~Ni, ``Driving with sharks: Rethinking connected vehicles with
  vehicle cybersecurity,'' \emph{IEEE Vehicular Technology Magazine}, vol.~12,
  no.~2, pp. 45--51, 2017.

\bibitem{2flip}
F.~Wang, Y.~Xu, H.~Zhang, Y.~Zhang, and L.~Zhu, ``2flip: A two-factor
  lightweight privacy-preserving authentication scheme for vanet,'' \emph{IEEE
  Transactions on Vehicular Technology}, vol.~65, no.~2, pp. 896--911, Feb
  2016.

\bibitem{RAISE}
C.~Zhang, X.~Lin, R.~Lu, and P.~H. Ho, ``Raise: An efficient rsu-aided message
  authentication scheme in vehicular communication networks,'' in \emph{2008
  IEEE International Conference on Communications}, May 2008, pp. 1451--1457.

\bibitem{huang2011pacp}
D.~Huang, S.~Misra, M.~Verma, and G.~Xue, ``Pacp: An efficient pseudonymous
  authentication-based conditional privacy protocol for vanets,'' \emph{IEEE
  Transactions on Intelligent Transportation Systems}, vol.~12, no.~3, pp.
  736--746, 2011.

\bibitem{ecdsa}
M.~Kneževic, V.~Nikov, and P.~Rombouts, ``Low-latency ecdsa signature
  verification 2014;a road toward safer traffic,'' \emph{IEEE Transactions on
  Very Large Scale Integration (VLSI) Systems}, vol.~24, no.~11, pp.
  3257--3267, Nov 2016.

\bibitem{sakmp}
H.~Tan, M.~Ma, H.~Labiod, A.~Boudguiga, J.~Zhang, and P.~H.~J. Chong, ``A
  secure and authenticated key management protocol (sa-kmp) for vehicular
  networks,'' \emph{IEEE Transactions on Vehicular Technology}, vol.~65,
  no.~12, pp. 9570--9584, Dec 2016.

\bibitem{gkmpan}
S.~Zhu, S.~Setia, S.~Xu, and S.~Jajodia, ``Gkmpan: an efficient group rekeying
  scheme for secure multicast in ad-hoc networks,'' in \emph{The First Annual
  International Conference on Mobile and Ubiquitous Systems: Networking and
  Services, 2004. MOBIQUITOUS 2004.}, Aug 2004, pp. 42--51.

\bibitem{ppaa}
P.~P. Tsang and S.~W. Smith, ``Ppaa: Peer-to-peer anonymous authentication,''
  in \emph{Applied Cryptography and Network Security}, S.~M. Bellovin,
  R.~Gennaro, A.~Keromytis, and M.~Yung, Eds.\hskip 1em plus 0.5em minus
  0.4em\relax Berlin, Heidelberg: Springer Berlin Heidelberg, 2008, pp. 55--74.

\bibitem{Calandriello}
\BIBentryALTinterwordspacing
G.~Calandriello, P.~Papadimitratos, J.-P. Hubaux, and A.~Lioy, ``Efficient and
  robust pseudonymous authentication in vanet,'' in \emph{Proceedings of the
  Fourth ACM International Workshop on Vehicular Ad Hoc Networks}, ser. VANET
  '07.\hskip 1em plus 0.5em minus 0.4em\relax New York, NY, USA: ACM, 2007, pp.
  19--28. [Online]. Available: \url{http://doi.acm.org/10.1145/1287748.1287752}
\BIBentrySTDinterwordspacing

\bibitem{ppgcv}
A.~Wasef and X.~Shen, ``Ppgcv: Privacy preserving group communications protocol
  for vehicular ad hoc networks,'' in \emph{2008 IEEE International Conference
  on Communications}, May 2008, pp. 1458--1463.

\bibitem{tack}
A.~Studer, E.~Shi, F.~Bai, and A.~Perrig, ``Tacking together efficient
  authentication, revocation, and privacy in vanets,'' in \emph{2009 6th Annual
  IEEE Communications Society Conference on Sensor, Mesh and Ad Hoc
  Communications and Networks}, June 2009, pp. 1--9.

\bibitem{GSIS}
X.~Lin, X.~Sun, P.~H. Ho, and X.~Shen, ``Gsis: A secure and privacy-preserving
  protocol for vehicular communications,'' \emph{IEEE Transactions on Vehicular
  Technology}, vol.~56, no.~6, pp. 3442--3456, Nov 2007.

\bibitem{sraac}
L.~Fischer, A.~Aijaz, C.~Eckert, and D.~Vogt, ``Secure revocable anonymous
  authenticated inter-vehicle communication (sraac),'' 12 0002.

\bibitem{att4}
N.~Chen, M.~Gerla, D.~Huang, and X.~Hong, ``Secure, selective group broadcast
  in vehicular networks using dynamic attribute based encryption,'' in
  \emph{2010 The 9th IFIP Annual Mediterranean Ad Hoc Networking Workshop
  (Med-Hoc-Net)}, June 2010, pp. 1--8.

\bibitem{abacs}
L.-Y. Yeh, Y.-C. Chen, and J.-L. Huang, ``Abacs: An attribute-based access
  control system for emergency services over vehicular ad hoc networks,''
  vol.~29, pp. 630--643, 03 2011.

\bibitem{att5}
Y.~Xia, W.~Chen, X.~Liu, L.~Zhang, X.~Li, and Y.~Xiang, ``Adaptive multimedia
  data forwarding for privacy preservation in vehicular ad-hoc networks,''
  \emph{IEEE Transactions on Intelligent Transportation Systems}, vol.~18,
  no.~10, pp. 2629--2641, Oct 2017.

\bibitem{att6}
M.~Bouabdellah, F.~E. Bouanani, and H.~Ben-azza, ``A secure cooperative
  transmission model in vanet using attribute based encryption,'' in \emph{2016
  International Conference on Advanced Communication Systems and Information
  Security (ACOSIS)}, Oct 2016, pp. 1--6.

\bibitem{bissmeyer2010intrusion}
N.~Bi{\ss}meyer, C.~Stresing, and K.~M. Bayarou, ``Intrusion detection in
  vanets through verification of vehicle movement data,'' in \emph{Vehicular
  Networking Conference (VNC), 2010 IEEE}.\hskip 1em plus 0.5em minus
  0.4em\relax IEEE, 2010, pp. 166--173.

\bibitem{tomandl2014rest}
A.~Tomandl, K.-P. Fuchs, and H.~Federrath, ``Rest-net: A dynamic rule-based ids
  for vanets,'' in \emph{Wireless and Mobile Networking Conference (WMNC), 2014
  7th IFIP}.\hskip 1em plus 0.5em minus 0.4em\relax IEEE, 2014, pp. 1--8.

\bibitem{cho2016fingerprinting}
K.-T. Cho and K.~G. Shin, ``Fingerprinting electronic control units for vehicle
  intrusion detection.'' in \emph{USENIX Security Symposium}, 2016, pp.
  911--927.

\bibitem{martynov2007design}
D.~Martynov, J.~Roman, S.~Vaidya, and H.~Fu, ``Design and implementation of an
  intrusion detection system for wireless sensor networks,'' in
  \emph{Electro/Information Technology, 2007 IEEE International Conference
  on}.\hskip 1em plus 0.5em minus 0.4em\relax IEEE, 2007, pp. 507--512.

\bibitem{sedjelmaci2014new}
H.~Sedjelmaci and S.~M. Senouci, ``A new intrusion detection framework for
  vehicular networks,'' in \emph{Communications (ICC), 2014 IEEE International
  Conference on}.\hskip 1em plus 0.5em minus 0.4em\relax IEEE, 2014, pp.
  538--543.

\bibitem{song2016intrusion}
H.~M. Song, H.~R. Kim, and H.~K. Kim, ``Intrusion detection system based on the
  analysis of time intervals of can messages for in-vehicle network,'' in
  \emph{Information Networking (ICOIN), 2016 International Conference
  on}.\hskip 1em plus 0.5em minus 0.4em\relax IEEE, 2016, pp. 63--68.

\bibitem{leeotids2017OTIDS}
H.~Lee, S.~H. Jeong, and H.~K. Kim, ``Otids: A novel intrusion detection system
  for in-vehicle network by using remote frame.''

\bibitem{yu2013detecting}
B.~Yu, C.-Z. Xu, and B.~Xiao, ``Detecting sybil attacks in vanets,''
  \emph{Journal of Parallel and Distributed Computing}, vol.~73, no.~6, pp.
  746--756, 2013.

\bibitem{zhang2018distributed}
T.~Zhang and Q.~Zhu, ``Distributed privacy-preserving collaborative intrusion
  detection systems for vanets,'' \emph{IEEE Transactions on Signal and
  Information Processing over Networks}, vol.~4, no.~1, pp. 148--161, 2018.

\bibitem{tice2014enforcing}
C.~Tice, T.~Roeder, P.~Collingbourne, S.~Checkoway, {\'U}.~Erlingsson,
  L.~Lozano, and G.~Pike, ``Enforcing forward-edge control-flow integrity in
  gcc \& llvm.'' in \emph{USENIX Security Symposium}, 2014, pp. 941--955.

\bibitem{dahse2014static}
J.~Dahse and T.~Holz, ``Static detection of second-order vulnerabilities in web
  applications.'' in \emph{USENIX Security Symposium}, 2014, pp. 989--1003.

\bibitem{ding2017efficient}
R.~Ding, C.~Qian, C.~Song, B.~Harris, T.~Kim, and W.~Lee, ``Efficient
  protection of path-sensitive control security,'' in \emph{26th USENIX
  Security Symposium (USENIX Security 17). Vancouver, BC: USENIX Association},
  2017, pp. 131--148.

\bibitem{castro2006securing}
M.~Castro, M.~Costa, and T.~Harris, ``Securing software by enforcing data-flow
  integrity,'' in \emph{Proceedings of the 7th symposium on Operating systems
  design and implementation}.\hskip 1em plus 0.5em minus 0.4em\relax USENIX
  Association, 2006, pp. 147--160.

\bibitem{ayewah2008using}
N.~Ayewah, D.~Hovemeyer, J.~D. Morgenthaler, J.~Penix, and W.~Pugh, ``Using
  static analysis to find bugs,'' \emph{IEEE software}, vol.~25, no.~5, 2008.

\bibitem{godefroid2008automated}
P.~Godefroid, M.~Y. Levin, D.~A. Molnar \emph{et~al.}, ``Automated whitebox
  fuzz testing.'' in \emph{NDSS}, vol.~8, 2008, pp. 151--166.

\bibitem{newsome2005dynamic}
J.~Newsome and D.~X. Song, ``Dynamic taint analysis for automatic detection,
  analysis, and signaturegeneration of exploits on commodity software.'' in
  \emph{NDSS}, vol.~5.\hskip 1em plus 0.5em minus 0.4em\relax Citeseer, 2005,
  pp. 3--4.

\bibitem{clause2007dytan}
J.~Clause, W.~Li, and A.~Orso, ``Dytan: a generic dynamic taint analysis
  framework,'' in \emph{Proceedings of the 2007 international symposium on
  Software testing and analysis}.\hskip 1em plus 0.5em minus 0.4em\relax ACM,
  2007, pp. 196--206.

\bibitem{le2012genprog}
C.~Le~Goues, T.~Nguyen, S.~Forrest, and W.~Weimer, ``Genprog: A generic method
  for automatic software repair,'' \emph{Ieee transactions on software
  engineering}, vol.~38, no.~1, p.~54, 2012.

\bibitem{shin2011evaluating}
Y.~Shin, A.~Meneely, L.~Williams, and J.~A. Osborne, ``Evaluating complexity,
  code churn, and developer activity metrics as indicators of software
  vulnerabilities,'' \emph{IEEE Transactions on Software Engineering}, vol.~37,
  no.~6, pp. 772--787, 2011.

\bibitem{perl2015vccfinder}
H.~Perl, S.~Dechand, M.~Smith, D.~Arp, F.~Yamaguchi, K.~Rieck, S.~Fahl, and
  Y.~Acar, ``Vccfinder: Finding potential vulnerabilities in open-source
  projects to assist code audits,'' in \emph{Proceedings of the 22nd ACM SIGSAC
  Conference on Computer and Communications Security}.\hskip 1em plus 0.5em
  minus 0.4em\relax ACM, 2015, pp. 426--437.

\bibitem{zhou2017automated}
Y.~Zhou and A.~Sharma, ``Automated identification of security issues from
  commit messages and bug reports,'' in \emph{Proceedings of the 2017 11th
  Joint Meeting on Foundations of Software Engineering}.\hskip 1em plus 0.5em
  minus 0.4em\relax ACM, 2017, pp. 914--919.

\bibitem{shar2015web}
L.~K. Shar, L.~C. Briand, and H.~B.~K. Tan, ``Web application vulnerability
  prediction using hybrid program analysis and machine learning,'' \emph{IEEE
  Transactions on Dependable and Secure Computing}, vol.~12, no.~6, pp.
  688--707, 2015.

\bibitem{grieco2016toward}
G.~Grieco, G.~L. Grinblat, L.~Uzal, S.~Rawat, J.~Feist, and L.~Mounier,
  ``Toward large-scale vulnerability discovery using machine learning,'' in
  \emph{Proceedings of the Sixth ACM Conference on Data and Application
  Security and Privacy}.\hskip 1em plus 0.5em minus 0.4em\relax ACM, 2016, pp.
  85--96.

\bibitem{fan2016malicious}
Y.~Fan, Y.~Ye, and L.~Chen, ``Malicious sequential pattern mining for automatic
  malware detection,'' \emph{Expert Systems with Applications}, vol.~52, pp.
  16--25, 2016.

\bibitem{huda2016hybrids}
S.~Huda, J.~Abawajy, M.~Alazab, M.~Abdollalihian, R.~Islam, and J.~Yearwood,
  ``Hybrids of support vector machine wrapper and filter based framework for
  malware detection,'' \emph{Future Generation Computer Systems}, vol.~55, pp.
  376--390, 2016.

\bibitem{huda2017defending}
S.~Huda, S.~Miah, M.~M. Hassan, R.~Islam, J.~Yearwood, M.~Alrubaian, and
  A.~Almogren, ``Defending unknown attacks on cyber-physical systems by
  semi-supervised approach and available unlabeled data,'' \emph{Information
  Sciences}, vol. 379, pp. 211--228, 2017.

\bibitem{mirza2017cloudintell}
Q.~K.~A. Mirza, I.~Awan, and M.~Younas, ``Cloudintell: An intelligent malware
  detection system,'' \emph{Future Generation Computer Systems}, 2017.

\bibitem{teslaone}
A.~Studer, F.~Bai, B.~Bellur, and A.~Perrig, ``Flexible, extensible, and
  efficient vanet authentication,'' \emph{Journal of Communications and
  Networks}, vol.~11, no.~6, pp. 574--588, Dec 2009.

\bibitem{tari}
R.~Chen, D.~Ma, and A.~Regan, ``Tari: Meeting delay requirements in vanets with
  efficient authentication and revocation,'' 01 2009.

\bibitem{att1}
A.~Alston, ``Attribute-based encryption for attribute-based authentication,
  authorization, storage, and transmission in distributed storage systems,''
  \emph{CoRR}.

\bibitem{att2}
A.~Sahai and B.~Waters, ``Fuzzy identity-based encryption,'' in \emph{Advances
  in Cryptology -- EUROCRYPT 2005}, R.~Cramer, Ed.\hskip 1em plus 0.5em minus
  0.4em\relax Berlin, Heidelberg: Springer Berlin Heidelberg, 2005, pp.
  457--473.

\bibitem{att3}
\BIBentryALTinterwordspacing
V.~Goyal, O.~Pandey, A.~Sahai, and B.~Waters, ``Attribute-based encryption for
  fine-grained access control of encrypted data,'' in \emph{Proceedings of the
  13th ACM Conference on Computer and Communications Security}, ser. CCS
  '06.\hskip 1em plus 0.5em minus 0.4em\relax New York, NY, USA: ACM, 2006, pp.
  89--98. [Online]. Available: \url{http://doi.acm.org/10.1145/1180405.1180418}
\BIBentrySTDinterwordspacing

\bibitem{asim2014attribute}
M.~Asim, M.~Petkovic, and T.~Ignatenko, ``Attribute-based encryption with
  encryption and decryption outsourcing,'' 2014.

\bibitem{att1point1}
N.~Chen and M.~Gerla, ``Dynamic attributes design in attribute based
  encryption,'' 01 2009.

\bibitem{mamaghani2018security}
M.~T. Mamaghani and R.~Abbas, ``On the security and reliability performance of
  two-way wireless energy harvesting based untrusted relaying with cooperative
  jamming,'' \emph{IET Communications}, 2018.

\bibitem{att7}
L.~Nkenyereye, B.~A. Tama, Y.~Park, and K.~H. Rhee, ``A fine-grained privacy
  preserving protocol over attribute based access control for vanets,''
  \emph{JoWUA}, vol.~6, pp. 98--112, 2015.

\bibitem{block1}
N.~Kshetri, ``Can blockchain strengthen the internet of things?'' \emph{IT
  Professional}, vol.~19, no.~4, pp. 68--72, 2017.

\bibitem{block3}
A.~Dorri, S.~S. Kanhere, and R.~Jurdak, ``Towards an optimized blockchain for
  iot,'' in \emph{2017 IEEE/ACM Second International Conference on
  Internet-of-Things Design and Implementation (IoTDI)}, April 2017, pp.
  173--178.

\bibitem{block2}
A.~Dorri, S.~Kanhere, R.~Jurdak, and P.~Gauravaram, ``Blockchain for iot
  security and privacy: The case study of a smart home,'' 03 2017.

\bibitem{dhaliwal2018effective}
S.~Dhaliwal, A.-A. Nahid, and R.~Abbas, ``Effective intrusion detection system
  using xgboost,'' \emph{Information}, vol.~9, no.~7, p. 149, 2018.

\bibitem{butun2014survey}
I.~Butun, S.~D. Morgera, and R.~Sankar, ``A survey of intrusion detection
  systems in wireless sensor networks,'' \emph{IEEE communications surveys \&
  tutorials}, vol.~16, no.~1, pp. 266--282, 2014.

\bibitem{miller2014survey}
C.~Miller and C.~Valasek, ``A survey of remote automotive attack surfaces,''
  \emph{black hat USA}, vol. 2014, 2014.

\bibitem{sun2017attacks}
Y.~Sun, L.~Wu, S.~Wu, S.~Li, T.~Zhang, L.~Zhang, J.~Xu, Y.~Xiong, and X.~Cui,
  ``Attacks and countermeasures in the internet of vehicles,'' \emph{Annals of
  Telecommunications}, vol.~72, no. 5-6, pp. 283--295, 2017.

\bibitem{sharma2018survey}
S.~Sharma and A.~Kaul, ``A survey on intrusion detection systems and honeypot
  based proactive security mechanisms in vanets and vanet cloud,''
  \emph{Vehicular Communications}, 2018.

\bibitem{sedjelmaci2014efficient}
H.~Sedjelmaci, S.~M. Senouci, and M.~A. Abu-Rgheff, ``An efficient and
  lightweight intrusion detection mechanism for service-oriented vehicular
  networks,'' \emph{IEEE Internet of things journal}, vol.~1, no.~6, pp.
  570--577, 2014.

\bibitem{sharma2018hybrid}
S.~Sharma and A.~Kaul, ``Hybrid fuzzy multi-criteria decision making based
  multi cluster head dolphin swarm optimized ids for vanet,'' \emph{Vehicular
  Communications}, vol.~12, pp. 23--38, 2018.

\bibitem{wahab2016ceap}
O.~A. Wahab, A.~Mourad, H.~Otrok, and J.~Bentahar, ``Ceap: Svm-based
  intelligent detection model for clustered vehicular ad hoc networks,''
  \emph{Expert Systems with Applications}, vol.~50, pp. 40--54, 2016.

\bibitem{sedjelmaci2017hierarchical}
H.~Sedjelmaci, S.~M. Senouci, and N.~Ansari, ``A hierarchical detection and
  response system to enhance security against lethal cyber-attacks in uav
  networks,'' \emph{IEEE Transactions on Systems, Man, and Cybernetics:
  Systems}, 2017.

\bibitem{zhang2017dynamic}
T.~Zhang and Q.~Zhu, ``Dynamic differential privacy for admm-based distributed
  classification learning,'' \emph{IEEE Transactions on Information Forensics
  and Security}, vol.~12, no.~1, pp. 172--187, 2017.

\bibitem{emanuelsson2008comparative}
P.~Emanuelsson and U.~Nilsson, ``A comparative study of industrial static
  analysis tools,'' \emph{Electronic notes in theoretical computer science},
  vol. 217, pp. 5--21, 2008.

\bibitem{zheng2017security}
X.~Zheng, L.~Pan, and E.~Yilmaz, ``Security analysis of modern mission critical
  android mobile applications,'' in \emph{Proceedings of the Australasian
  Computer Science Week Multiconference}.\hskip 1em plus 0.5em minus
  0.4em\relax ACM, 2017, p.~2.

\bibitem{mcgraw2004software}
G.~McGraw, ``Software security,'' \emph{IEEE Security \& Privacy}, vol.~2,
  no.~2, pp. 80--83, 2004.

\bibitem{abadi2005control}
M.~Abadi, M.~Budiu, U.~Erlingsson, and J.~Ligatti, ``Control-flow integrity,''
  in \emph{Proceedings of the 12th ACM conference on Computer and
  communications security}.\hskip 1em plus 0.5em minus 0.4em\relax ACM, 2005,
  pp. 340--353.

\bibitem{wogerer2005survey}
W.~W{\"o}gerer, ``A survey of static program analysis techniques,'' Citeseer,
  Tech. Rep., 2005.

\bibitem{zheng2017real}
X.~Zheng, C.~Julien, H.~Chen, R.~Podorozhny, and F.~Cassez, ``Real-time
  simulation support for runtime verification of cyber-physical systems,''
  \emph{ACM Transactions on Embedded Computing Systems (TECS)}, vol.~16, no.~4,
  p. 106, 2017.

\bibitem{sutton2007fuzzing}
M.~Sutton, A.~Greene, and P.~Amini, \emph{Fuzzing: brute force vulnerability
  discovery}.\hskip 1em plus 0.5em minus 0.4em\relax Pearson Education, 2007.

\bibitem{ji2018coming}
T.~Ji, Y.~Wu, C.~Wang, X.~Zhang, and Z.~Wang, ``The coming era of
  alphahacking?: A survey of automatic software vulnerability detection,
  exploitation and patching techniques,'' in \emph{2018 IEEE Third
  International Conference on Data Science in Cyberspace (DSC)}.\hskip 1em plus
  0.5em minus 0.4em\relax IEEE, 2018.

\bibitem{takanen2008fuzzing}
A.~Takanen, J.~D. Demott, and C.~Miller, \emph{Fuzzing for software security
  testing and quality assurance}.\hskip 1em plus 0.5em minus 0.4em\relax Artech
  House, 2008.

\bibitem{schwartz2010all}
E.~J. Schwartz, T.~Avgerinos, and D.~Brumley, ``All you ever wanted to know
  about dynamic taint analysis and forward symbolic execution (but might have
  been afraid to ask),'' in \emph{Security and privacy (SP), 2010 IEEE
  symposium on}.\hskip 1em plus 0.5em minus 0.4em\relax IEEE, 2010, pp.
  317--331.

\bibitem{sinha2011architectural}
P.~Sinha, ``Architectural design and reliability analysis of a fail-operational
  brake-by-wire system from iso 26262 perspectives,'' \emph{Reliability
  Engineering \& System Safety}, vol.~96, no.~10, pp. 1349--1359, 2011.

\bibitem{maro2018software}
S.~Maro, J.-P. Stegh{\"o}fer, and M.~Staron, ``Software traceability in the
  automotive domain: Challenges and solutions,'' \emph{Journal of Systems and
  Software}, vol. 141, pp. 85--110, 2018.

\bibitem{deng2017mutation}
L.~Deng, J.~Offutt, P.~Ammann, and N.~Mirzaei, ``Mutation operators for testing
  android apps,'' \emph{Information and Software Technology}, vol.~81, pp.
  154--168, 2017.

\bibitem{li2018vuldeepecker}
Z.~Li, D.~Zou, S.~Xu, X.~Ou, H.~Jin, S.~Wang, Z.~Deng, and Y.~Zhong,
  ``Vuldeepecker: A deep learning-based system for vulnerability detection,''
  \emph{arXiv preprint arXiv:1801.01681}, 2018.

\bibitem{ghaffarian2017software}
S.~M. Ghaffarian and H.~R. Shahriari, ``Software vulnerability analysis and
  discovery using machine-learning and data-mining techniques: a survey,''
  \emph{ACM Computing Surveys (CSUR)}, vol.~50, no.~4, p.~56, 2017.

\bibitem{srikant1996mining}
R.~Srikant and R.~Agrawal, ``Mining sequential patterns: Generalizations and
  performance improvements,'' in \emph{International Conference on Extending
  Database Technology}.\hskip 1em plus 0.5em minus 0.4em\relax Springer, 1996,
  pp. 1--17.

\bibitem{wang20162flip}
F.~Wang, Y.~Xu, H.~Zhang, Y.~Zhang, and L.~Zhu, ``2flip: a two-factor
  lightweight privacy-preserving authentication scheme for vanet,'' \emph{IEEE
  Transactions on Vehicular Technology}, vol.~65, no.~2, pp. 896--911, 2016.

\bibitem{sun2018data}
N.~Sun, J.~Zhang, P.~Rimba, S.~Gao, Y.~Xiang, and L.~Y. Zhang, ``Data-driven
  cybersecurity incident prediction: A survey,'' \emph{IEEE Communications
  Surveys \& Tutorials}, 2018.

\bibitem{WXZZZ2016}
\BIBentryALTinterwordspacing
F.~Wang, Y.~Xu, H.~Zhang, Y.~Zhang, and L.~Zhu, ``{2FLIP}: {A} two-factor
  lightweight privacy-preserving authentication scheme for {VANET},''
  \emph{{IEEE} Transactions on Vehicular Technology}, vol.~65, no.~2, pp.
  896--911, 2016. [Online]. Available:
  \url{https://doi.org/10.1109/TVT.2015.2402166}
\BIBentrySTDinterwordspacing

\bibitem{JPCKPK2009}
\BIBentryALTinterwordspacing
S.~Jana, S.~N. Premnath, M.~Clark, S.~K. Kasera, N.~Patwari, and S.~V.
  Krishnamurthy, ``On the effectiveness of secret key extraction from wireless
  signal strength in real environments,'' in \emph{Proceedings of the 15th
  Annual International Conference on Mobile Computing and Networking, {MOBICOM}
  2009, Beijing, China, September 20-25, 2009}, K.~G. Shin, Y.~Zhang, R.~L.
  Bagrodia, and R.~Govindan, Eds.\hskip 1em plus 0.5em minus 0.4em\relax {ACM},
  2009, pp. 321--332. [Online]. Available:
  \url{http://doi.acm.org/10.1145/1614320.1614356}
\BIBentrySTDinterwordspacing

\bibitem{SIS2011}
\BIBentryALTinterwordspacing
T.~Shimizu, H.~Iwai, and H.~Sasaoka, ``Physical-layer secret key agreement in
  two-way wireless relaying systems,'' \emph{{IEEE} Transactions on Information
  Forensics and Security}, vol.~6, no. 3-1, pp. 650--660, 2011. [Online].
  Available: \url{https://doi.org/10.1109/TIFS.2011.2147314}
\BIBentrySTDinterwordspacing

\bibitem{XQHZZLZ2016}
\BIBentryALTinterwordspacing
W.~Xi, C.~Qian, J.~Han, K.~Zhao, S.~Zhong, X.~Li, and J.~Zhao, ``Instant and
  robust authentication and key agreement among mobile devices,'' in
  \emph{Proceedings of the 2016 {ACM} {SIGSAC} Conference on Computer and
  Communications Security, Vienna, Austria, October 24-28, 2016}, E.~R. Weippl,
  S.~Katzenbeisser, C.~Kruegel, A.~C. Myers, and S.~Halevi, Eds.\hskip 1em plus
  0.5em minus 0.4em\relax {ACM}, 2016, pp. 616--627. [Online]. Available:
  \url{http://doi.acm.org/10.1145/2976749.2978298}
\BIBentrySTDinterwordspacing

\bibitem{ZXNTLC2017}
\BIBentryALTinterwordspacing
X.~Zhu, F.~Xu, E.~Novak, C.~C. Tan, Q.~Li, and G.~Chen, ``Using wireless link
  dynamics to extract a secret key in vehicular scenarios,'' \emph{{IEEE}
  Transactions on Mobile Computing}, vol.~16, no.~7, pp. 2065--2078, 2017.
  [Online]. Available: \url{https://doi.org/10.1109/TMC.2016.2557784}
\BIBentrySTDinterwordspacing

\bibitem{ZXHCA2018}
\BIBentryALTinterwordspacing
Y.~Zhang, Y.~Xiang, X.~Huang, X.~Chen, and A.~Alelaiwi, ``A matrix-based
  cross-layer key establishment protocol for smart homes,'' \emph{Information
  Sciences}, vol. 429, pp. 390--405, 2018. [Online]. Available:
  \url{https://doi.org/10.1016/j.ins.2017.11.039}
\BIBentrySTDinterwordspacing

\bibitem{AS1983}
\BIBentryALTinterwordspacing
B.~Alpern and F.~B. Schneider, ``Key exchange using 'keyless cryptography',''
  \emph{Information Processing Letters}, vol.~16, no.~2, pp. 79--81, 1983.
  [Online]. Available: \url{https://doi.org/10.1016/0020-0190(83)90029-7}
\BIBentrySTDinterwordspacing

\bibitem{CM2005}
\BIBentryALTinterwordspacing
C.~Castelluccia and P.~Mutaf, ``Shake them up!: a movement-based pairing
  protocol for cpu-constrained devices,'' in \emph{Proceedings of the 3rd
  International Conference on Mobile Systems, Applications, and Services,
  MobiSys 2005, Seattle, Washington, USA, June 6-8, 2005}, K.~G. Shin, D.~Kotz,
  and B.~D. Noble, Eds.\hskip 1em plus 0.5em minus 0.4em\relax {ACM}, 2005, pp.
  51--64. [Online]. Available: \url{http://doi.acm.org/10.1145/1067170.1067177}
\BIBentrySTDinterwordspacing

\bibitem{PO2015}
\BIBentryALTinterwordspacing
R.~D. Pietro and G.~Oligeri, ``{ESC:} an efficient, scalable, and crypto-less
  solution to secure wireless networks,'' \emph{Computer Networks}, vol.~84,
  pp. 46--63, 2015. [Online]. Available:
  \url{https://doi.org/10.1016/j.comnet.2015.04.006}
\BIBentrySTDinterwordspacing

\bibitem{ZXWWS2018}
\BIBentryALTinterwordspacing
Y.~Zhang, Y.~Xiang, T.~Wang, W.~Wu, and J.~Shen, ``An over-the-air key
  establishment protocol using keyless cryptography,'' \emph{Future Generation
  Computer Systems}, vol.~79, pp. 284--294, 2018. [Online]. Available:
  \url{https://doi.org/10.1016/j.future.2016.12.013}
\BIBentrySTDinterwordspacing

\bibitem{TCRMV2014}
D.~Tsonev, H.~Chun, S.~Rajbhandari, J.~J. McKendry, S.~Videv, E.~Gu, M.~Haji,
  S.~Watson, A.~E. Kelly, G.~Faulkner, M.~D. Dawson, H.~Haas, and D.~O. Brien,
  ``A {3-Gb/s} single-{LED} {OFDM}-based wireless {VLC} link using a gallium
  nitride $\mu${LED},'' \emph{{IEEE} Photonics Technology Letters}, vol.~26,
  no.~7, pp. 637--640, 2014.

\bibitem{HYWC2016}
H.~Haas, L.~Yin, Y.~Wang, and C.~Chen, ``What is {LiFi}?'' \emph{Journal of
  Lightwave Technology}, vol.~34, no.~6, pp. 1533--1544, 2016.

\bibitem{BYDZ2015}
\BIBentryALTinterwordspacing
X.~Bao, G.~Yu, J.~Dai, and X.~Zhu, ``Li-fi: Light fidelity-a survey,''
  \emph{Wireless Networks}, vol.~21, no.~6, pp. 1879--1889, 2015. [Online].
  Available: \url{https://doi.org/10.1007/s11276-015-0889-0}
\BIBentrySTDinterwordspacing

\bibitem{AHALS2015}
N.~A. Abdulsalam, R.~A. Hajri, Z.~A. Abri, Z.~A. Lawati, and M.~M.
  Bait-Suwailam, ``Design and implementation of a vehicle to vehicle
  communication system using {Li-Fi} technology,'' in \emph{Information and
  Communication Technology Research (ICTRC), 2015 International Conference
  on}.\hskip 1em plus 0.5em minus 0.4em\relax IEEE, 2015, pp. 136--139.

\bibitem{BMB2016}
P.~Bhateley, R.~Mohindra, and S.~Balaji, ``Smart vehicular communication system
  using {LiFi} technology,'' in \emph{Computation of Power, Energy Information
  and Commuincation (ICCPEIC), 2016 International Conference on}.\hskip 1em
  plus 0.5em minus 0.4em\relax IEEE, 2016, pp. 222--226.

\bibitem{GS2016}
M.~Gupta and S.~Sharma, ``Infrastructure-less vehicular communication system
  using {Li-Fi} technology,'' \emph{International Journal of Computer (IJC)},
  vol.~23, no.~1, pp. 53--60, 2016.

\bibitem{DMBU2018}
\BIBentryALTinterwordspacing
F.~A. Dahri, H.~B. Mangrio, A.~Baqai, and F.~A. Umrani, ``Experimental
  evaluation of intelligent transport system with {VLC} vehicle-to-vehicle
  communication,'' \emph{Wireless Personal Communications}, Apr 2018. [Online].
  Available: \url{https://doi.org/10.1007/s11277-018-5727-0}
\BIBentrySTDinterwordspacing

\bibitem{version15}
3GPP, ``3rd generation partnership project; technical specification group
  services and system aspects; security aspect for lte support of
  vehicle-to-everything (v2x) services (release 15),'' \emph{3GPP TS 33.185},
  vol.~16, 2018.

\bibitem{luoto2017vehicle}
P.~Luoto, M.~Bennis, P.~Pirinen, S.~Samarakoon, K.~Horneman, and M.~Latva-aho,
  ``Vehicle clustering for improving enhanced lte-v2x network performance,'' in
  \emph{Networks and Communications (EuCNC), 2017 European Conference
  on}.\hskip 1em plus 0.5em minus 0.4em\relax IEEE, 2017, pp. 1--5.

\bibitem{version16}
3GPP, ``3rd generation partnership project; technical specification group
  services and system aspects; study on enhancement of 3gpp support for 5g v2x
  services (release 16), year={2018}, volume={16},,'' \emph{3GPP TS 33.185}.

\bibitem{loukas2017computation}
G.~Loukas, Y.~Yoon, G.~Sakellari, T.~Vuong, and R.~Heartfield, ``Computation
  offloading of a vehicle’s continuous intrusion detection workload for
  energy efficiency and performance,'' \emph{Simulation Modelling Practice and
  Theory}, vol.~73, pp. 83--94, 2017.

\bibitem{xiao2017user}
L.~Xiao, C.~Xie, M.~Min, and W.~Zhuang, ``User-centric view of unmanned aerial
  vehicle transmission against smart attacks,'' \emph{IEEE Transactions on
  Vehicular Technology}, 2017.

\bibitem{loukas2018cloud}
G.~Loukas, T.~Vuong, R.~Heartfield, G.~Sakellari, Y.~Yoon, and D.~Gan,
  ``Cloud-based cyber-physical intrusion detection for vehicles using deep
  learning,'' \emph{IEEE Access}, vol.~6, pp. 3491--3508, 2018.

\bibitem{haddadpajouh2018deep}
H.~HaddadPajouh, A.~Dehghantanha, R.~Khayami, and K.-K.~R. Choo, ``A deep
  recurrent neural network based approach for internet of things malware threat
  hunting,'' \emph{Future Generation Computer Systems}, vol.~85, pp. 88--96,
  2018.

\bibitem{Huda2018}
\BIBentryALTinterwordspacing
S.~Huda, S.~Miah, J.~Yearwood, S.~Alyahya, H.~Al-Dossari, and R.~Doss, ``A
  malicious threat detection model for cloud assisted internet of things (cot)
  based industrial control system (ics) networks using deep belief network,''
  \emph{Journal of Parallel and Distributed Computing}, pp.~--, 2018. [Online].
  Available:
  \url{https://www.sciencedirect.com/science/article/pii/S0743731518302442}
\BIBentrySTDinterwordspacing

\bibitem{krizhevsky2012imagenet}
A.~Krizhevsky, I.~Sutskever, and G.~E. Hinton, ``Imagenet classification with
  deep convolutional neural networks,'' in \emph{Advances in neural information
  processing systems}, 2012, pp. 1097--1105.

\bibitem{lecun2015deep}
Y.~LeCun, Y.~Bengio, and G.~Hinton, ``Deep learning,'' \emph{nature}, vol. 521,
  no. 7553, p. 436, 2015.

\bibitem{kim2018deep}
J.~Kim, H.~Kim, S.~Huh, J.~Lee, and K.~Choi, ``Deep neural networks with
  weighted spikes,'' \emph{Neurocomputing}, 2018.

\bibitem{hinton2018deep}
G.~Hinton, ``Deep learning—a technology with the potential to transform
  health care,'' \emph{JAMA}, 2018.

\bibitem{zheng2014physically}
X.~Zheng, ``Physically informed assertions for cyber physical systems
  development and debugging,'' in \emph{Pervasive Computing and Communications
  Workshops (PERCOM Workshops), 2014 IEEE International Conference on}.\hskip
  1em plus 0.5em minus 0.4em\relax IEEE, 2014, pp. 181--183.

\bibitem{zheng2017perceptions}
X.~Zheng, C.~Julien, M.~Kim, and S.~Khurshid, ``Perceptions on the state of the
  art in verification and validation in cyber-physical systems,'' \emph{IEEE
  Systems Journal}, vol.~11, no.~4, pp. 2614--2627, 2017.

\bibitem{zheng2018efficient}
X.~Zheng, C.~Julien, R.~Podorozhny, F.~Cassez, and T.~Rakotoarivelo,
  ``Efficient and scalable runtime monitoring for cyber--physical system,''
  \emph{IEEE Systems Journal}, vol.~12, no.~2, pp. 1667--1678, 2018.

\bibitem{zheng2015braceassertion}
X.~Zheng, C.~Julien, R.~Podorozhny, and F.~Cassez, ``Braceassertion: Runtime
  verification of cyber-physical systems,'' in \emph{Mobile Ad Hoc and Sensor
  Systems (MASS), 2015 IEEE 12th International Conference on}.\hskip 1em plus
  0.5em minus 0.4em\relax IEEE, 2015, pp. 298--306.

\bibitem{sun2016edgeiot}
X.~Sun and N.~Ansari, ``Edgeiot: Mobile edge computing for the internet of
  things,'' \emph{IEEE Communications Magazine}, vol.~54, no.~12, pp. 22--29,
  2016.

\bibitem{bonomi2012fog}
F.~Bonomi, R.~Milito, J.~Zhu, and S.~Addepalli, ``Fog computing and its role in
  the internet of things,'' in \emph{Proceedings of the first edition of the
  MCC workshop on Mobile cloud computing}.\hskip 1em plus 0.5em minus
  0.4em\relax ACM, 2012, pp. 13--16.

\bibitem{zheng2018smartvm}
T.~Zheng, X.~Zheng, Y.~Zhang, Y.~Deng, E.~Dong, R.~Zhang, and X.~Liu,
  ``Smartvm: a sla-aware microservice deployment framework,'' \emph{World Wide
  Web}, pp. 1--19, 2018.

\bibitem{hu2015mobile}
Y.~C. Hu, M.~Patel, D.~Sabella, N.~Sprecher, and V.~Young, ``Mobile edge
  computing—a key technology towards 5g,'' \emph{ETSI white paper}, vol.~11,
  no.~11, pp. 1--16, 2015.

\end{thebibliography}
%

%
%
%
%
%




\end{document}